\documentclass[journal]{IEEEtran}

\ifCLASSINFOpdf
\else
\fi

\usepackage{cite}
\usepackage{hyperref}
\usepackage{url}
\usepackage{amsmath}
\usepackage{amsfonts}
\usepackage{graphicx}
\usepackage{multirow}
\usepackage{float}
\usepackage{wrapfig}
\usepackage{hyperref}
\usepackage{graphicx}
\usepackage{caption}
\usepackage{subcaption}
\usepackage{epstopdf}
\usepackage{color}
\usepackage{longtable}

\usepackage[labelsep=period]{caption}

\begin{document}
\title{A Survey on Legacy and Emerging Technologies for Public Safety Communications}
\author{\IEEEauthorblockN{Abhaykumar Kumbhar$^{1,2}$, Farshad Koohifar$^1$, \.{I}smail~G\"uven\c{c}$^1$}, and  Bruce Mueller$^2$\\
\IEEEauthorblockA{$^1$Department of Electrical and Computer Engineering, Florida International University, Miami, FL 33174\\
$^2$Motorola Solutions, Inc., Plantation, FL 33322 USA\\
Email: {\tt \{akumb004, fkooh001, iguvenc\}@fiu.edu, bruce.d.mueller@motorolasolutions.com}}
}

\maketitle

\begin{abstract}

Effective emergency and natural disaster management depend on the efficient mission-critical voice and data communication between first responders and victims. Land  Mobile Radio System (LMRS) is a legacy narrowband technology used for critical voice communications with limited use for data applications. Recently Long Term Evolution (LTE) emerged as a broadband communication technology that has a potential to transform the capabilities of public safety technologies by providing broadband, ubiquitous, and mission-critical voice and data support. For example, in the United States, FirstNet is building a nationwide coast-to-coast public safety network based of LTE broadband technology. This paper presents a comparative survey of legacy and the LTE-based public safety networks, and discusses the LMRS-LTE convergence as well as mission-critical push-to-talk over LTE. A simulation study of LMRS and LTE band class~14 technologies is provided using the NS-3 open source tool. An experimental study of APCO-25 and LTE band class 14 is also conducted using software-defined radio, to enhance the understanding of the public safety systems. Finally, emerging technologies that may have strong potential for use in public safety networks are reviewed.

\end{abstract}

\begin{IEEEkeywords}
APCO-25, FirstNet, heterogeneous networks, LMRS-LTE convergence, mission-critical push-to-talk, P-25, public safety communications, software-defined radio (SDR), TETRA, unmanned aerial vehicle (UAV), 3GPP, 5G.
\end{IEEEkeywords}

\let\thefootnote\relax\footnotetext{This research was supported in part by the U.S. National Science Foundation under the grant AST-1443999 and CNS-1453678.}

\section{Introduction}

Public safety organizations protect the well-being of the public in case of natural and man-made disasters, and are tasked with preparing for, planning for, and responding to emergencies. The emergency management agencies include law enforcement agencies, fire departments, rescue squads, emergency medical services (EMS), and other entities that are referred to as emergency first responders (EFR). The ability of EFR to communicate amongst themselves and seamlessly share critical information directly affects their ability to save lives. The communication technologies such as legacy radio system, commercial network (2G/3G), and broadband (LTE/WiFi) are largely used by the public safety organizations.

Over the recent years, there has been increasing interest in improving the capabilities of public safety communications (PSC) systems. For example, in \cite{spectrum1, spectrum2, spectrum3}, efficient spectrum management techniques, allocation models, and infrastructure options are introduced for PSC scenarios. Reforms on PSC policy have been discussed in \cite{CommPolicy} and \cite{CommPolicy1}, which study the decoupling of spectrum licenses for spectrum access, a new nationwide system built on open standards with consistent architecture, and fund raising approach for the transition to a new nationwide system. As explained in \cite{ER}, communication of time critical information is an important factor  for emergency response. In \cite{cognitive} and \cite{cognitive1}, insights on cognitive radio technology are presented, which plays a significant role in making the best use of scarce spectrum in public safety scenarios. Integration of other wireless technologies into PSC is studied in \cite{integrating}, with a goal to provide faster and reliable communication capability in a challenging environment where infrastructure is impacted by the unplanned emergency events.

\begin{table*}[!htbp]
\centering
\caption{Literature overview on existing and emerging PSC technologies and open research directions.}
\begin{tabular}{|p{1.5cm}|p{1.77cm}|p{1.3cm}|p{1.75cm}|p{1.49cm}|p{1.35cm}|p{6.2cm}|}
\hline \textbf{PSC technologies} &\textbf{Spectrum policy, management, and regulation} &\textbf{Role in PSC} &\textbf{Survey, modeling, and simulation} &\textbf{3GPP Rel. 12 and Rel. 13 support} & \textbf{Field study, devices and frameworks} &\textbf{Summary and possible research directions related to PSC}\\
\hline

\hline LMRS  & \cite{spectrum1, spectrum2, spectrum3, CommPolicy, CommPolicy1,SpectrumScarcity1,SpectrumScarcityAfter,SpectrumScarcityProb,merwaday2014incentivizing,yuksel2013fostering,Auction73,DblockChutiyapa,hallahan2013enabling,fnprm,npsbn,Ref1,Ref1.1,VHF_UHF,Ref3,freq-db0,freq-db,SpectrumChart,UHF-TV,4.9GHz,4.9GHz1, moto49Ghz,Interoperability,Interoperability1} & \cite{previousWork1,previousWork2,doumi2013lte,Congestion,tetra3,tetra4} &\cite{previousWork1,previousWork2,HauptPublicSafety,DAQ,desourdis2002emerging,dataRate,Congestion,talkaround,cai,C4FM,P25ControlChannel,P25RadioWiki}  &-& \cite{4.9GHz,4.9GHz1,msitetra,sepura,hytera,kenwood,harris}
& \textbf{Summary:} LMRS was designed to provide mission-critical voice and extensive coverage.\\
\cline{7-7}
 & & & & & &\textbf{Research areas:} Design and optimization of converged LMRS-LTE devices, and tailoring MBMS/eMBMS into them.\\

\hline
LTE & \cite{FirstNetWebpage,fnprm,FirstNet2, Auction73, 700mhzChutiyapa, hallahan2013enabling, DblockSigning, DblockChutiyapa,ituFreq,WRC15,EuBbPsc,CanadaBbPsc,700Band14Interoperability}& \cite{previousWork1,previousWork2,HauptPublicSafety,doumi2013lte,LTEVehicle,FirstNet,coverage,MC-PTT2,MC-PTT4,desourdis2002emerging,FirstNetModel}& \cite{solanki2013lte,desourdis2002emerging,simsek2013device,babun2015extended,babun2014intercell,MC-PTT1,FirstNet,LENA,ici1,ici2, ici3} & \cite{MC-PTT1,MC-PTT2,MC-PTT3,MC-PTT4,MC-PTT5} & \cite{converge0, VALR, converge1,P25TxPower}
& \textbf{Summary:} LTE for PSC can deliver mission-critical broadband data with minimum latency.\\
\cline{7-7}
 & & & & & &\textbf{Research areas:} Integration and optimization of 3GPP Rel.~12/13 enhancements for PSC. \\

\hline SDR &\cite{wang2016software}& \cite{sdr5,sdrpsc1,sdrpsc} & \cite{sdr1,sdr3,ferraos2015mobile} & \cite{wyglinski2016revolutionizing}& \cite{sdr4,sdr41,sdr6,hackrf,rtl2832u} 
 & \textbf{Summary:} SDR provides low-cost infrastructure for public safety experiments and test activities.\\
 \cline{7-7}
 & & & & & &\textbf{Research areas:} Development of SDR prototypes and vendor-compatible solutions for PSC. \\

\hline MBMS/eMBMS & - & \cite{mbmsCite, mbms2, mbms3} & \cite{monserrat2012joint} & \cite{3gpp.22.246} & 
& \textbf{Summary:} MBMS/eMBMS provides the PSN with an ability to carry out multicast/broadcast of emergency messages and data efficiently.\\ 
\cline{7-7}
 & & & & & &\textbf{Research areas:} Design and optimization of flexible MBSFN resource structures that can accommodate different user distributions. Integration of D2D/ProSe into MBMS/eMBMS. \\

\hline mmWave & \cite{mmWave} &\cite{bleicher2013millimeter,mmWave2} & \cite{sulyman2014radio,niu2015survey,thomas2015millimeter, kuttybeamforming} & - & \cite{akdeniz2014millimeter} 
& \textbf{Summary:} mmWave technology for PSC can reduce spectrum scarcity, network congestion, and provide broadband communication.\\
\cline{7-7}
 & & & & & &\textbf{Research areas:} Efficient interference management and spatial reuse. mmWave channel propagation measurements in PSC scenarios. \\

\hline Massive MIMO &\cite{mopidevi2011compact}& \cite{mopidevi2011compact} & \cite{marzetta2010noncooperative,jose2011channel,gopalakrishnan2011analysis,elijahcomprehensive} &-& \cite{larsson2013massiveMimo} 
& \textbf{Summary:} Massive MIMO can achieve high throughput with reduced communication errors, which can be decisive during the exchange of mission-critical data.\\
\cline{7-7}
 & & & & & &\textbf{Research areas:} Symbiotic convergence of mmWave and massive MIMO for higher capacity gains and better spectral efficiency. \\

\hline Small Cells & \cite{smallCell1} & \cite{merwaday2015uav, babun2014intercell,LTEvsWifi2} & \cite{smallCell1,smallcell2} & \cite{smallCell3GPP,LAASmallCell} & \cite{sui2015interference}
&\textbf{Summary:} Small cell deployment boosts coverage and capacity gains, which can enhance PSC between EFR during an emergency.\\
\cline{7-7}
 & & & & & &\textbf{Research areas:} Systematic convergence of mmWave and massive MIMO with small cells. Addressing interference/mobility challenges with moving small cells such as within firetrucks. \\

\hline UAVs & \cite{bennett2014civilian,UAVintoPSLTE} &\cite{merwaday2015uav,DronePS,gomez2013realistic,UAVintoPSLTE,wolfefeasibility,gomez2013performance,arvindUavPS,gomez2015capacity} & \cite{merwaday2015uav, guptasurvey, saleem2015integration,WahabUavUWB,NadisankaUav} & - & \cite{UAVLTEVideo,daniel2009airshield} 
& \textbf{Summary:} UAVs as a deployable system can be crucial for reducing coverage gaps and network congestion for PSC.\\
\cline{7-7}
 & & & & & &\textbf{Research areas:} Introducing autonomy to UAVs in PSC scenarios. Developing new UAV propagation models for PSC, such as in mmWave bands. \\

\hline LTE-based V2X &-& \cite{alazawi2011intelligent,eltoweissy2010towards,babun2015extended,firstNetVechicular} & \cite{V2X_nokia,chen2015vehicular} & \cite{V2X1,V2X2,V2X3} & - 
& \textbf{Summary:} LTE-based V2X communication can assist EFR to be more efficient during disaster management and rescue operations.\\
\cline{7-7}
 & & & & & &\textbf{Research areas:} Evolving eMBMS to LTE-based V2X needs. Improved privacy preservation schemes for V2X participants. Interference and mobility management. \\

\hline LAA & \cite{LAA2} & \cite{ferraos2015mobile} & \cite{ULS,rupasinghe2015reinforcement, rupasinghe2014licensed, LTEvsWifi1, LTEvsWifi2,DusanUavPsc} & \cite{LAA_1,LAA1,LAA3,LAA4,LAA_9,LAA_8} & & \textbf{Summary:} LAA can complement PSCthat are deployed in licensed bands and avoid any possibility of network congestion.\\
\cline{7-7}
 & & & & & &\textbf{Research areas:} Policies to ensure fair access to all technologies while coexisting in the unlicensed spectrum. Protocols for carrier aggregation of licensed and unlicensed bands. \\

\hline Cognitive radio & \cite{SpectrumScarcityProb,merwaday2014incentivizing,yuksel2013fostering,guvencSpectrum2016} &\cite{ghafoor2014cognitive} & \cite{ghafoor2014cognitive,akhtar2016white,sakr2015cognitive,hossain2009dynamic,hossain2007cognitive} &-& \cite{SpectrumScarcity1} 
&\textbf{Summary:} Cognitive radio technology is a viable solution for efficiently using public safety spectrum.\\ 
\cline{7-7}
 & & & & & &\textbf{Research areas:} Spectral/energy efficient spectrum sensing and sharing.
Database assisted spectrum sharing. Prioritized spectrum access.\\

\hline WSNs &-&\cite{gomez2009secure,gomez2010secure} & \cite{akan2009cognitive,hansurvey,gu2015evolution,bukhari2015survey} & - & - 
& \textbf{Summary:} Deployment of large-scale WSN into PSN can increase situational awareness of EFR and assist in evading any potential disaster.\\
\cline{7-7}
 & & & & & &\textbf{Research areas:} Robust models for multi-hop synchronization. Tethering wireless sensor data attached to EFR equipment.  \\

\hline IoT &-& \cite{iotps,iotps2,iotPsSec} & \cite{iot1,iot2,iot3} &-& \cite{iotps1} 
& \textbf{Summary:} Intelligent analysis of real-time data from IoT devices can enhance decision-making ability of EFR.\\
\cline{7-7}
 & & & & & &\textbf{Research areas:} Tailoring public safety wearables into IoT, considering also openness, security, interoperability, and cost. Formulating policies and regulations to strike right balance between privacy and security. \\

\hline Cybersecurity enhancements&-& \cite{mcgee2012public,Hastings2015,ghafghazi2014classification,EdwinUavSecurity} & \cite{wangbig} &-&-
& \textbf{Summary:} Securing mission-critical information has become critical with real-time data readily flowing through PSN. Concrete techniques and policies can help secure mission-critical data over PSN.\\
\cline{7-7}
 & & & & & &\textbf{Research areas:} Securing emergency medical services and law enforcement data operating across the LTE-based PSN. \\
\hline
\end{tabular}
\label{Table::SummaryReferences}
\end{table*}

\subsection{Related Works and Contributions}
There have been relatively limited studies in the literature on PSC that present a comprehensive survey on public safety LMRS and LTE systems. In \cite{previousWork1}, authors present a discussion on voice over LTE as an important aspect of PSC and then provide a high-level overview of LMRS and LTE technologies for their use in PSC scenario. In~\cite{previousWork2,HauptPublicSafety}, authors survey the status of various wireless technologies in public safety network (PSN), current regulatory standards, and the research activities that address the challenges in PSC. The ability of LTE to meet the PSN requirements, and classifying possible future developments to LTE that could  enhance its capacity to provide the PSC is discussed in \cite{doumi2013lte,FirstNet}. 

In this paper our focus is more on the comparative analysis of legacy and emerging technologies for PSC, when compared with the contributions in \cite{previousWork1,previousWork2,doumi2013lte}. We take up the public safety spectrum allocation in the United States as a case study, and present an overview of spectrum allocation  in VHF, UHF, 700 MHz, 800, MHz, 900 MHz, and 4.9 GHz bands for various public safety entities. We also provide a holistic view on the current status of the broadband PSN in other regions such as the European Union, the United Kingdom, and Canada.

We review the LTE-based FirstNet architecture in the United States, the convergence of LTE-LMR technologies, and support for mission-critical PTT over LTE, which are not addressed in  earlier survey articles such as \cite{previousWork1,previousWork2}. In addition, a unified comparison between LMRS and LTE-based PSN is undertaken in Section~\ref{SectionVI} and Section~\ref{SectionVII}, which is not available in existing survey articles on PSC to our best knowledge. Study of LMRS and LTE band class 14 is carried out using NS-3 simulations, and software-defined radio (SDR) measurement campaigns for LMRS and LTE band class 14 technologies are reported. Different than existing literature, we also provide a comprehensive perspective on how emerging wireless technologies can shape PSN  and discuss open research problems.

\subsection{Outline of the Paper}
This survey article is organized as follows. Section~\ref{SectionIII} provides classification of PSN, related standards, and existing challenges in PSN. Section~\ref{SectionII} summarizes public safety spectrum allocation in the United States, while Section~\ref{firstNet} studies the FirstNet as an example for LTE-based broadband PSN. Section~\ref{SectionIV}  gives a synopsis of LTE and LMR convergence in terms of available equipment and framework. Subsequently, Section~\ref{SectionV} provides a brief overview of mission-critical communication over LTE based on Release 12 and 13 enhancements. Section~\ref{SectionVI} introduces a comparison between LMRS and LTE systems explaining the various attributes of the spectrum, Section~\ref{SectionVII} provides a simulation study of LMRS and LTE, whereas Section~\ref{SectionVIII} furnishes the experimental study using software-defined radios. The potential of various emerging technologies for enhancing PSC capabilities are discussed in Section~\ref{SectionIX}. Current status of broadband PSNs in different regions of the world is discussed in Section~\ref{section:wrc15}. Section~\ref{issuesFutureResearchDirections} discusses the issues and possible research areas for different public safety technologies and Section~\ref{SectionXII} concludes the paper. Some of the key references to be studied in this survey related to LMRS, LTE, SDR, and emerging PSC technologies are classified in Table~\ref{Table::SummaryReferences}, along with possible research directions.

\section{Public Safety Networks}
\label{SectionIII}
A PSN is a dedicated wireless network used by emergency services such as police, fire rescue, and EMS. This network gives better situational awareness, quicker response time to the EFRs, and speed up the disaster response. The scope of a PSN can span over a large geographical area, with vital data flowing into broadband wireless mesh network such as Wi-Fi and LTE. The PSNs are also networked with mobile computing applications to improve the efficiency of the EFR and public well-being. In this article, we broadly classify PSNs into two categories: LMRS and  broadband networks. APCO-25 and TETRA suite of standards falls under LMRS network, while LTE-based broadband PSC network falls under broadband network.

\begin{figure} [!htbp]
\centering
\includegraphics[width=1\linewidth]{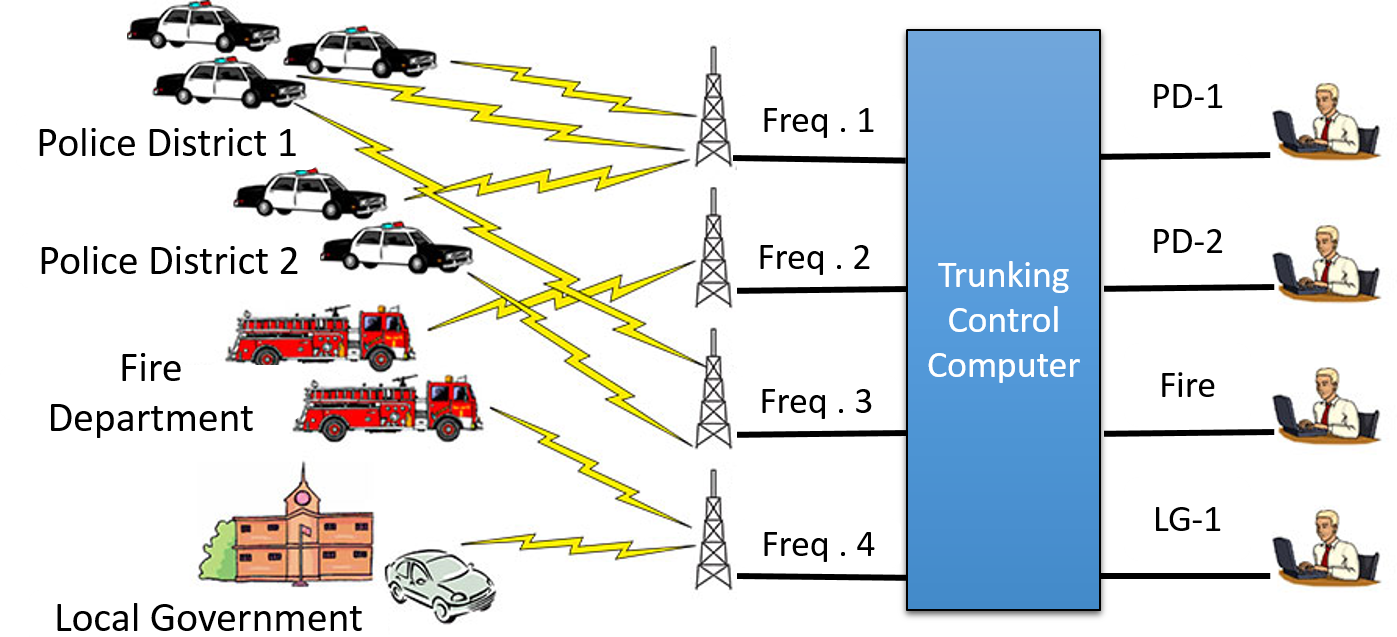}
\caption{An example for a LMRS network \cite{Fig1}.}
\label{fig:LMRSNetwork}
\end{figure}

\subsection{LMRS Network}
LMRS is a wireless communication system intended for terrestrial users comprised of portables and mobiles, such as two-way digital radios or walkie-talkies. LMRS networks and equipment are being used in military, commercial, and EFR applications as shown in Fig.~\ref{fig:LMRSNetwork}. The main goal of LMRS systems are to provide mission critical communications, to enable integrated voice and data communications for emergency response, and to maintain ruggedness, reliability, and interoperability.

\begin{figure} [!htbp]
\centering
\includegraphics[width=1\linewidth]{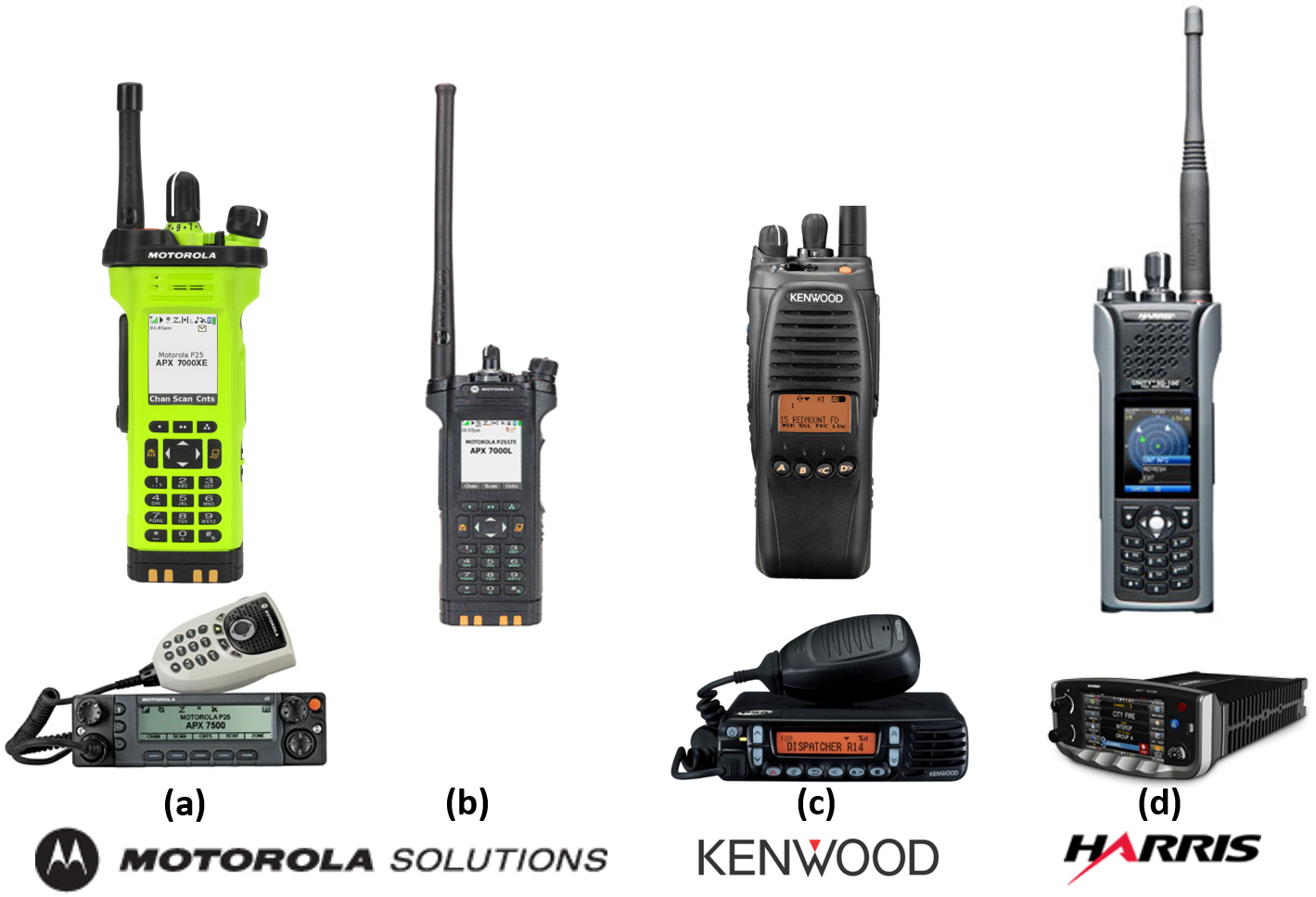}
\caption{APCO-25 portable and mobile radios in the United States. The radios are from (a), (b) Motorola Solutions \cite{P25TxPower}, (c) Kenwood \cite{kenwood}, and (d) Harris \cite{harris}. The radios (a), (c) and (d) are LMRS capable whereas radio (b) is a hybrid equipment with LMRS and LTE capabilities}
\label{fig:APCO-25Radios}
\end{figure}

\begin{figure} [!htbp]
\centering
\includegraphics[width=1\linewidth]{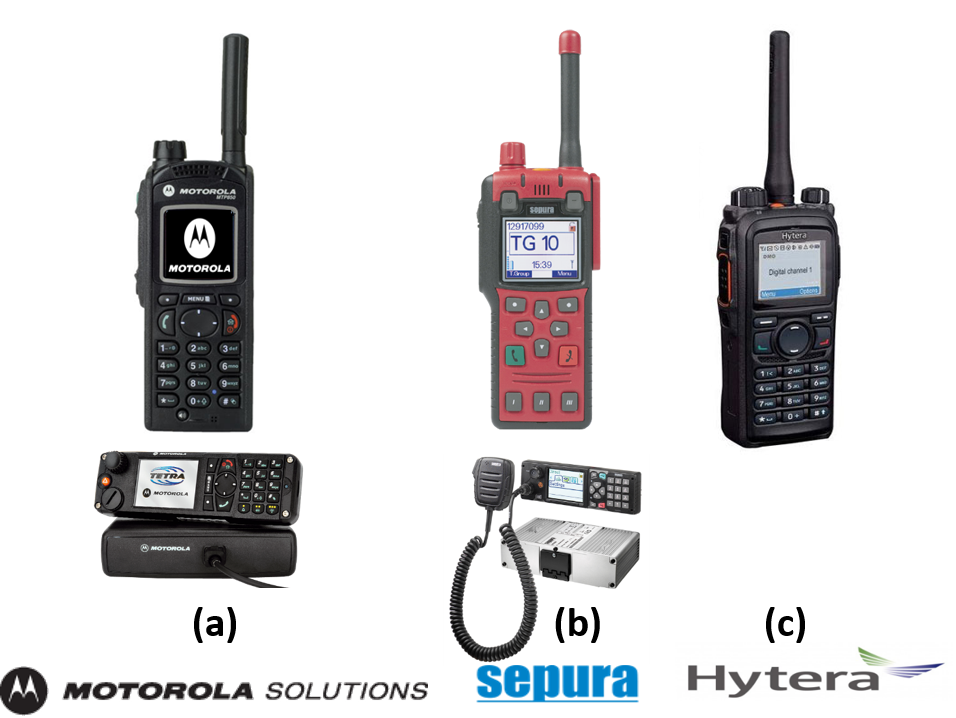}
\caption{TETRA portable and mobile radios from (a) Motorola Solutions \cite{msitetra}, (b) Sepura \cite{sepura}, and (c) Hytera \cite{hytera}.}
\label{fig:TETRAradios}
\end{figure}

APCO-25 and European Terrestrial Trunked Radio (TETRA) are widely used suite of standards for LMRS based digital radio communications. APCO-25 also known as Project~25 (P-25), and it is widely used by federal, state/province, and local public safety agencies in North America. The APCO-25 technology enables public safety agencies enabling them to communicate with other agencies and mutual aid response teams in emergencies. On the other hand, TETRA fulfills the same role for European and Asian countries. However, these two standards are not interoperable. Fig.~\ref{fig:APCO-25Radios} showcases some of the available APCO-25 portable and mobile digital radios in the market, while Fig.~\ref{fig:TETRAradios} showcases TETRA portable and mobile digital radios, procured from the product sheet and websites of the respective companies.

\subsubsection{APCO-25}
Public safety radios have been upgraded from analog to digital since the 1990s due to the limitations of analog transmission, and implement technological advances by expanding the capabilities of digital radio. Radios can communicate in analog mode with legacy radios, and in either digital or analog mode with other APCO-25 radios. Additionally, the deployment of APCO-25 compliant systems will allow for a high degree of equipment interoperability and compatibility. APCO-25 compliant technology has been deployed in three different phases,  where advancements have been gradually introduced \cite{P25Phase}.\\

\begin{enumerate}
\item[\bf Ph.~1]
In Phase 1, radio systems operate in 12.5 KHz band analog, digital, or mixed mode. Phase 1 radios use continuous 4-level frequency modulation (C4FM) technique, which is a non-linear modulation for digital transmissions.  C4FM is a special type of 4FSK modulation as explained in \cite{C4FM} and  was developed for the Telecommunications Industry Association (TIA) 102 standard for digital transmission in 12.5 KHz. C4FM results in a bit rate of 9600 bits/s. Phase~1 P25-compliant systems are backward compatible and interoperable with legacy systems. P25-compliant systems also provide an open interface to the radio frequency (RF) subsystem to facilitate interlinking between different vendor systems.

\item[\bf Ph.~2]
With a goal of improving the spectrum utilization, Phase 2 was implemented. Phase~2 introduces a 2~-slot TDMA system which provides two voice traffic channels in a 12.5 KHz band allocation, and doubles the call capacity. It also lays emphasis on interoperability with legacy equipment, interfacing between repeaters and other subsystems, roaming capacity, and spectral efficiency/channel reuse.

\item[\bf Ph.~3]
Project MESA (Mobility for Emergency and Safety Applications) is a collaboration between the European Telecommunications Standards Institute (ETSI) and the TIA to define a unified set of requirements for APCO~-~25 Phase 3. Initial agreement for project MESA  was ratified in year 2000. Phase 3 planning activities address the need for high-speed data for public-safety use \cite{P25Phase}. Project MESA also aims to facilitate effective, efficient, advanced specifications, and applications that will address public safety broadband communication needs \cite{mesa}.
\end{enumerate}

\subsubsection{TETRA}
TETRA is a digital mobile radio system, which is essentially confined to layers 1-3 of the OSI model. The TETRA system is intended to operate in existing VHF and UHF professional mobile radio frequencies \cite{TCCA}, and it has been developed by the European Telecommunications Standards Institute (ETSI). The user needs and technological innovations have led the TETRA standard to evolve with release 1 and release 2 \cite{TCCA}.

\begin{enumerate}
\item[\bf Rel.~1]
Release 1 is the original TETRA standard, which was known as TETRA V+D standard. Under this release TETRA radio system supports three modes of operation which are voice plus data (V+D), direct operation mode (DMO), and packet data optimized (PDO) \cite{tetrabook}. 

The V+D is the most commonly used mode, which allows switching between voice and data transmission. Voice and data can be transmitted on the same channel using different slots. The main characteristics of V+D are: 1) support for independent multiple concurrent bearer services, 2) support for transmitter preemption, 3) support for several grades of handover, 4) crossed calls are minimized by labeling an event on the air interface, 5) support for slot stealing during voice/data transmission, 6) support for different simultaneous access priorities \cite{tetrabook}.

The DMO, on the other hand, supports direct voice and data transmission between the subscriber units without the base stations, especially when the users are in the outside coverage area. Calls in the DMO can be either clear or encrypted, and full duplex radio communication is not supported under DMO~\cite{tetra2}.

As a third mode of operation, the PDO standard has been created for occasional data-only, to cater the demands for high volume of data in near future. Location based services and voice are necessary for mission-critical communications and need high volume of data, which can be the beneficiaries of the PDO standard~\cite{tetra2}.

\item[\bf Rel.~2]
Release~2 provided additional functions and improvements to already existing functionality of TETRA. The major enhancement provided by TETRA release 2 is TETRA enhanced data services, which provides more flexibility and greater levels of data capacity \cite{tetra3}. With adaptive selection of modulation schemes, RF channel bandwidths, and coding user bit rates can vary between $10$ Kbits/s to $500$ Kbits/s.

Data rate plays a important role in relaying mission-critical information during the emergency situation in timely manner. For example, an emergency scenario monitoring a remote victim would require data rate to support real-time duplex voice/video communication and telemetry. In such a mission-critical scenario, TETRA enhanced data service would play an important role by supporting the applications that need high data rate such as mulitmedia and location services. Other TETRA improvements also include adaptive multiple rate voice codec, mixed excitation linear predictive enhanced voice codec, and trunked mode operation range extension, which extended the range for air-ground-air services to $83$ kilometers when compared to 58 kilometers in TETRA release 1~\cite{tetra4}.
\end{enumerate}

\begin{table}[!htbp]
\centering
\caption{ Emerging wireless broadband communication technologies for creating a PSN\cite{LTEVehicle}. QoS stands for Quality of service, whereas CQI for channel quality indicator.}
\begin{tabular}{|p{1.7cm}|p{1.8cm}|p{1.8cm}|p{1.8cm}|}
\hline \textbf{Feature} & \textbf{Wi-Fi} &\textbf{UMTS} & \textbf{LTE}\\
\hline Channel width & 20 MHz & 5 MHz & 1.4, 3, 5, 10, 15 and 20 MHz \\
\hline Frequency bands & 2.4 - 2.483 GHz, 5.15 - 5.25 GHz, 5.25 - 5.35 GHz & 700 - 2600 MHz & 700 - 2690 MHz \\
\hline Max. data rate & 54 Mbits/s & 42 Mbits/s & Up to 300 Mbits/s \\
\hline Range & Up to 100 m & Up to 10 km & Up to 30 km \\
\hline Data capacity & Medium & Medium & High \\
\hline Coverage & Intermittent & Ubiquitous & Ubiquitous \\
\hline Mobility support & Low & High & Up to 350 km/h \\
\hline QoS support & Enhanced distributed channel access & QoS classes and bearer selection & QCI and bearer selection\\
\hline
\end{tabular}
\label{Table:EmergingWirelessBroadbandCommunicationTechnologiesPSN}
\end{table}

\subsection{LTE Broadband Network}
LTE is a broadband technology that will allow high data rate applications that are not possible to support with LMRS. LTE will enable unprecedented broadband service to public safety agencies and will bring the benefits of lower costs, consumer-driven economies of scale, and rapid evolution of advanced communication capabilities \cite{FirstNetWebpage}.

The Table~\ref{Table:EmergingWirelessBroadbandCommunicationTechnologiesPSN} , showcases various potential wireless communication standards, that can be used for public safety broadband networks. LTE standard, is developed by the 3GPP as a 4G broadband mobile communication technology. The main goal of LTE is to increase the capacity and high-speed data over wireless data networks. The technological advances has brought LTE to a performance level close to Shannon's capacity bound \cite{previousWork2}. Due to limited code block length in LTE, full SNR efficiency is not feasible. More specifically, LTE performance is less than 1.6 dB~-~2 dB off from the Shannon capacity bound as discussed in \cite{LTEShannon}. With steady increase of investment in broadband services, the LTE technology is soon expected to become most widely deployed broadband communication technology ever. An LTE-based broadband PSN, dedicated solely for the use of PSC can deliver the much needed advanced communication and data capabilities. As an example for LTE-based PSN, we briefly study \textit{FirstNet} deployment in the United States in Section~\ref{firstNet}.

\subsection{Major Challenges in PSNs}
\label{subsection::challenges}
Even with several technical advancements in PSN, there are major challenges that can hinder efficient operation of EFR. For instance, during a large-scale disaster, different public safety organizations are bound to use different communications technologies and infrastructure at the same time. This can cause major challenges such as, network congestion, low data rates, interoperability problems, spectrum scarcity, and security problems.
\subsubsection{Network Congestion}
In the case of an emergency, network activity spikes up, causing traffic congestion in public safety and commercial networks. Many EFR agencies using IP-based the commercial data services in case of emergency would result in a competition of air time with the general public \cite{Congestion}. The probability of any natural or man-made emergency can be assumed as a discrete random event. In case of such events, providing reliable communication becomes critical. Therefore, ensuring that network congestion will not happen in public safety networks during emergency situations is becoming increasingly more important. Emerging technologies such as licensed assisted access (LAA) of LTE to the unlicensed bands, small cells, and UAVs can be used as supplementary solutions during emergency situations to reduce the impact of network congestion \cite{ULS,rupasinghe2015reinforcement,rupasinghe2014licensed}. Furthermore, addressing the open challenges in the field of mmWave, massive MIMO, and vehicular network system can further help to lower the network congestion. Nevertheless, effective mitigation of network congestion problems during emergency situations remains an open issue.
\subsubsection{Maintaining Ubiquitous Throughput and Connectivity}
The data rate plays a critical role in relaying the information (i.e., voice and data) during the emergency situation in a timely manner. For example, an emergency scenario monitoring a remote victim would require data rate to support real-time duplex voice/video communication, telemetry, and so forth. Another example needing a real-time situational awareness is during fires for firefighters, through the use of streaming video and mission critical voice/data. These examples of mission-critical scenarios as per \cite{CommPolicy} would need higher data rates and require broadband allocation of spectrum. Lower data rate limits the usage of data applications, such as multimedia services which have a great potential to improve the efficiency of disaster recovery operations \cite{dataRate,galloix2014public}. The traditional public safety systems was designed to provide better coverage, mission-critical voice, but not peak data capacity. With the advent of 3GPP specifications, the data rates have been steadily increasing and will help in enhancing the capabilities of PSC systems. These data rates can be further increased by making technologies such as mmWave and massive MIMO commercially available. However, 3GPP systems deliver less coverage area when compared to LMRS. Furthermore, they can also experience relatively low data rates and dropped communications at the cell-edge. The coverage and fringe condition in LTE-based PSN can be addressed by the deployment of UAVs or vehicular network system to set up a small cell or virtual cell site. Regardless of the advancements in PSC, maintaining ubiquitously high throughput and connectivity during the emergency situation is still a challenge.
\subsubsection{Interoperability}
Traditionally, interoperability of a radio network means coordinating operating parameters and pre-defined procedures between the intended operators in the network. As discussed in \cite{Interoperability}, the interoperability failure that occurred during the September 11, 2001 incident was due to the presence of various independent public safety agencies, which resulted in entanglement of systems that were not interoperable. This led to communication failures between various EFRs, and posed a risk to public lives.  A possible solution for avoiding any potential interoperability issue includes EFR carrying multiple devices to be effective, which is an expensive strategy and an inefficient use of spectrum resources. Alternatively, communication infrastructure and policies can be evolved for EFR which is a more efficient approach. As per \cite{Interoperability1}, FCC is taking steps in resolving the lack of interoperability in 700 MHz band in a cost-effective manner by leveraging on commercial technologies and infrastructure.

\section{public safety spectrum allocation in the united states }
\label{SectionII}

Public safety operations can be classified mainly based on the applications and usage scenarios, such as mission-critical, military, transportation, utilities, and corporations with large geographical footprint. In this section, we provide a comprehensive overview of VHF, UHF, 700 MHz, 800 MHz, 900 MHz, and 4.9 GHz public safety band plans and spectrum allocation in the United States. The Federal Communications Commission (FCC) in the United States is a regulatory body for regulating communication by radios, television, wire, satellite, and cable. The FCC has taken steps to ensure that 911 and other critical communication remain operational when a disaster strikes. A major goal of FCC is to enable an operable and interoperable public safety communications (PSC) system in narrowband and broadband spectrum \cite{Ref1}.

\begin{figure}
        \centering
        \begin{subfigure}[b]{0.5\textwidth}
                \includegraphics[width=1\linewidth]{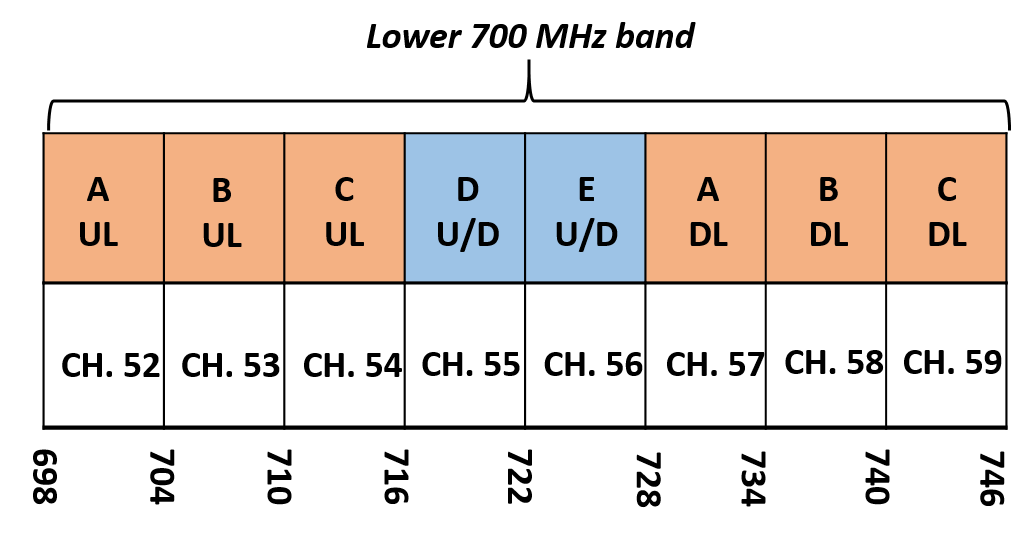}
                \caption{700 MHz lower band plan. Bands A, B, and C represent narrowband public safety bands, which are used for voice communications.}
                \label{fig:700MHzBandLowerPlan}
        \end{subfigure}

        \begin{subfigure}[b]{0.51\textwidth}
                \includegraphics[width=1\linewidth]{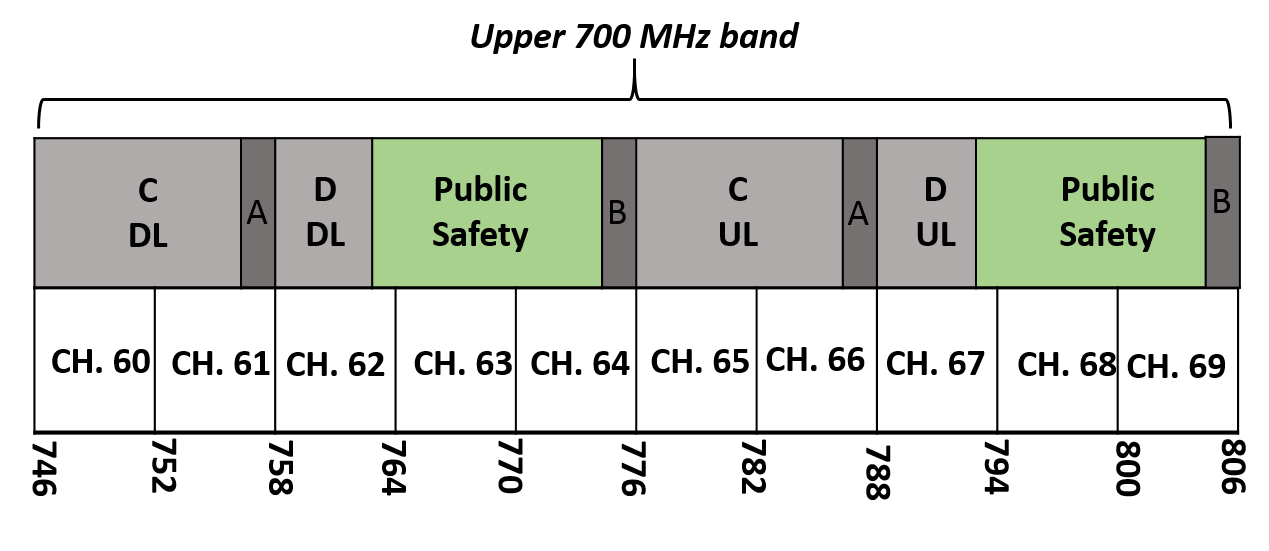}
                \caption{700 MHz upper band plan. Broadband public safety spectrum is explicitly illustrated, which dedicated for LTE communications.}
                \label{fig:700MHzBandUpperPlan}
        \end{subfigure}
        \caption{700 MHz band plan for public safety services in the United States. A, B, C, and D denote different bands in the 700 MHz spectrum.}
        \label{fig:700MHzBandPlan}
\end{figure}

\begin{figure*} [!htbp]
\centering
\includegraphics[width=1\linewidth]{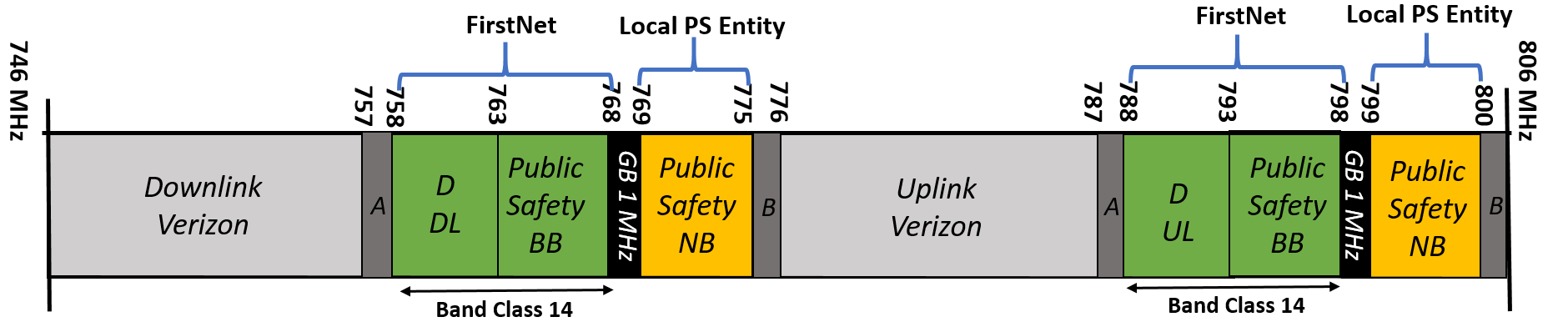}
\caption{Band class 14 plan for public safety services. D block will be reallocated for use by public safety entities as directed by Congressional mandate \cite{Ref1}.
758 MHz - 768 MHz would be the downlink and 788 MHz - 798 MHz would be uplink public safety frequency allocation in band class 14. Bands A and B are the guard bands of 1 MHz each.}
\label{fig:BandClass14Plan}
\end{figure*}

\subsection{700 MHz Public Safety Spectrum}
\label{subsubsection:700Chutiyapa}
The 700 MHz spectrum auction by FCC is officially know as \textit{Auction 73} \cite{Auction73}. The signals in 700 MHz spectrum travel longer distances than other typical cellular bands, and penetrate well. Therefore the 700 MHz band is an appealing spectrum to build systems for both commercial and PSN networks. Fig.~\ref{fig:700MHzBandLowerPlan}, Fig.~\ref{fig:700MHzBandUpperPlan}, Fig.~\ref{fig:BandClass14Plan}, and along with Table~\ref{Table:SpectrumAllocation700MHz} depict the spectrum allocated in the 700 MHz band in the United States. The lower 700 MHz range covers channels (CH) 52-59 and 698-746 MHz frequency, while the upper 700 MHz range covers CH 60-69 and 746-806 MHz frequency. The 3GPP standards have created four different band classes within 700 MHz band, i.e., band class 12, 13, 14, and 17.

\begin{table}[!htbp]
\centering
\caption{Spectrum allocation of 700 MHz: band class 12, 13, 14, and 17 \cite{bandGap}.}
\begin{tabular}{|c|c|c|c|c|}
\hline \textbf{Band } & \textbf{Spectrum} &\textbf{Uplink } & \textbf{Downlink } & \textbf{Band gap}  \\
\textbf{class} & \textbf{block} &\textbf{(MHz)} & \textbf{(MHz)} & \textbf{(MHz)}  \\
\hline 12  & Lower block A & 698 - 716 & 728 - 746 & 12\\
& Lower block B & & &\\
& Lower block C & & &\\
\hline 13 & Upper C block & 777 - 787 & 746 - 756 & 41\\
\hline 14 & Upper D Block and & 788 - 798 & 758 - 768 & 40\\
 & Public safety allocation & & &\\
\hline 17 & Lower B block & 704 - 716 & 734 - 746 & 18\\
 & Lower C block & & &\\
\hline
\end{tabular}
\label{Table:SpectrumAllocation700MHz}
\end{table}

Under the current framework, 20 MHz of dedicated spectrum is allocated to PSN in the 700 MHz band as shown in Fig. 2. Green color blocks represent public safety broadband ranging from 758 MHz to 768 MHz with 763 MHz as the center frequency and 788 MHz to 798 MHz with 793 MHz as the center frequency. Narrowband spectrum is represented in orange color blocks ranging from 769 MHz to 775 MHz and 799 MHz to 805 MHz. This part of 700 MHz public safety band is available for local public safety entity for voice communication. Guard bands (GB) of 1 MHz are placed between broadband, narrowband and commercial carrier spectrum to prevent any interference. A part of the 700 MHz public safety narrowband spectrum is also available for nationwide interoperable communications. In particular, interoperability is required to enable different governmental agencies to communicate across jurisdictions and with each other, and it is administered at state level by an executive agency \cite{Ref2}.

The deployment strategy of EFR varies depending on the magnitude of the emergency. As the emergency grows, the demand for additional tactical channels would grow too. These tactical channels would render necessary short-range communications at the emergency scene, without taxing the main dispatch channels. This implies the need for additional bandwidth allocation to EFR. However, before reallocation of D~block, there was none to little bandwidth available for PSC in 700 MHz spectrum. The only option for the EFR would be to use commercial network for PSC. However, commercial networks are not fully capable of mission-critical communication. Therefore, the full allocation of D Block i.e., 20 MHz of 700 MHz was critical to public safety \cite{DblockChutiyapa}. With reallocation of the D Block to public safety, LTE-based PSN in the 700 MHz would be beneficial to both commercial and PSN. Conversely, the interoperability and roaming between a commercial and PSN during the emergencies without seriously compromising quality of service for commercial users is an ongoing discussion \cite{hallahan2013enabling}.

\subsection{800 MHz Public Safety Spectrum}
\begin{figure}[!htbp]
\centering
\includegraphics[width=1\linewidth]{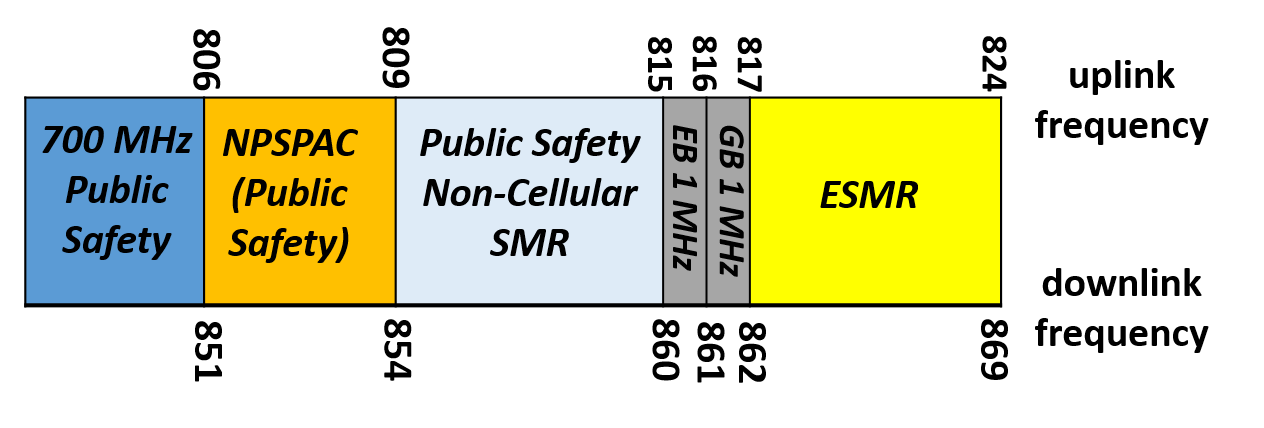}
\caption{Current 800 MHz band plan for public safety services \cite{800band}, \cite{SpectrumChart}.}
\label{800MHz}
\end{figure}

The 800 MHz public safety band is currently configured as illustrated in Fig.~\ref{800MHz}. It houses public safety spectrum allocation, commercial wireless carriers, and private radio systems. National Public Safety Planning Advisory Committee (NPSPAC) has come up with guidelines for 800 MHz band and has uplink channels allocated in 806 MHz - 809 MHz and downlink in 851 MHz - 854 MHz.  Various regional public safety planning committees administer 800 MHz NPSPAC spectrum. Specialized mobile radio (SMR) and enhanced specialized mobile radio (ESMR) can be either analog or digital trunked two-way radio system. SMR channels are allocated in 809~MHz - 815~MHz and downlink in 854~MHz~-~860~MHz. Expansion band (EB) and GB are of 1 MHz each, which are allocated between 815 MHz -  817 MHz. The ESMR channels are allocated in the 817 MHz - 824 MHz uplink and 862 MHz - 869 MHz for the downlink \cite{800band}, \cite{SpectrumChart}.

\begin{figure}[!htbp]
\centering
\includegraphics[width=1\linewidth]{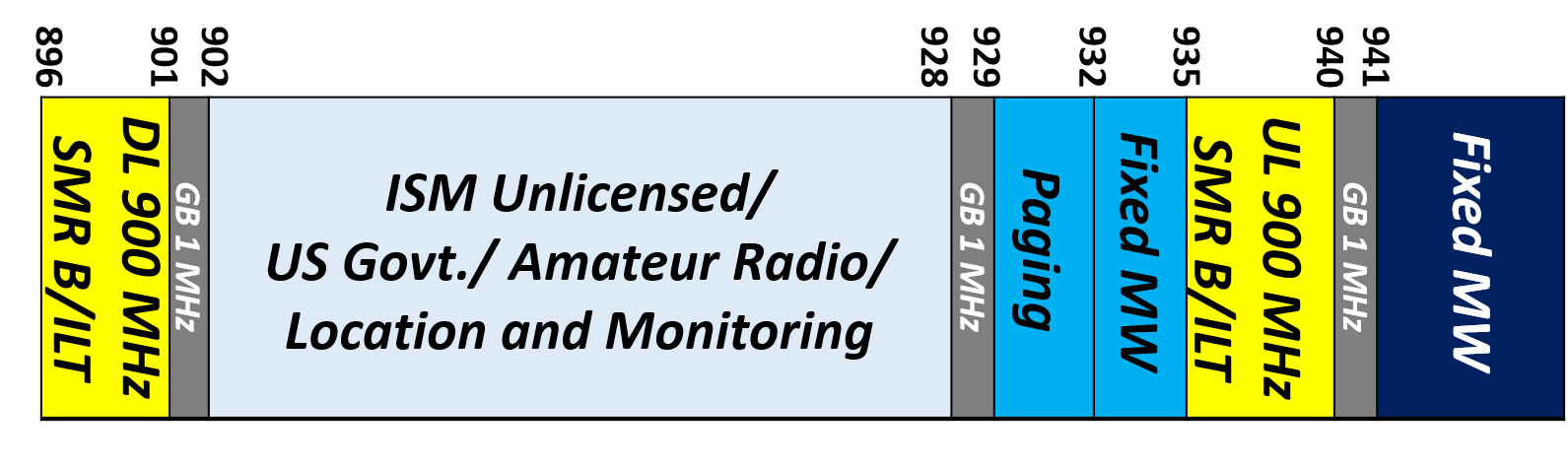}
\caption{SMR, industrial, scientific and medical (ISM), paging, and fixed microwave are the radio bands in 900 MHz spectrum plan \cite{SpectrumChart}, \cite{900}.}
\label{900Mhz}
\end{figure}

\subsection{900 MHz Public Safety Spectrum}
The FCC allows utilities and other commercial entities to file licenses in 900 MHz business and industrial land transport (B/ILT) and the 900 MHz spectrum band plan is shown in Fig.~\ref{900Mhz}. The new licenses in the 900 MHz B/ILT band are now allowed on a site-by-site basis for base mobile operator for various commercial (manufacturing, utility, and transportation) and non-commercial (medical, and educational) activities \cite{Ref1}, \cite{Ref1.1}.

\subsection{VHF and UHF Public Safety Spectrum}
All public safety and business industrial LMRS are narrowbands and operate using 12.5 KHz technology \cite{Ref1}. The operating frequencies for VHF low band ranges between 25~MHz - 50~MHz, VHF high band between 150~MHz - 174~MHz, and UHF band between 421~MHz - 512~MHz \cite{Ref1}. Licensees in the private land mobile VHF and UHF bands have previously operated on channel bandwidth of 25~KHz. As of January 1, 2013 FCC has mandated that all existing licenses in VHF and UHF bands must operate on 12.5~KHz channel bandwidth. Narrowbanding of VHF and UHF bands has ensured channel availability, and creation of additional channel capacity within the same radio spectrum. Narrowbanding will result in efficient use of spectrum, better spectrum access for public safety and non-public safety users, and support more users \cite{Ref1.1,VHF_UHF,Ref3}.

Previously, the frequencies from 470~MHz - 512~MHz were designated as UHF-TV sharing frequencies and were available in certain limited areas of the United States. Public safety in urbanized area such as Boston (MA), Chicago (IL), Cleveland (OH), Dallas/Fort Worth (TX), Detroit (MI), Houston (TX), Los Angeles (CA), Miami (FL), New York (NY), Philadelphia~(PA), San Francisco/Oakland (CA), and  Washington DC were allowed to share UHF-TV frequencies and were governed by FCC rules 90.301 through 90.317. However, as of February 22, 2012 legislation was enacted to reallocate spectrum in the D Block within the 700 MHz band for public safety broadband operation. The legislative act prompted FCC to impose a freeze on new T-Band licenses or any modifications to existing licenses. Furthermore, public safety operation is needed to vacate the T-Band spectrum by the year 2023 \cite{UHF-TV}.

\begin{table}[!htbp]
\centering
\caption{Public safety spectrum bands, frequency, and the respective users in the state of Florida. Primary control channels are denoted by (c) and alternate control channels by (a) \cite{freq-db0, freq-db, UHF-TV}.}
\begin{tabular}{|c|p{2.4cm}|p{4.1cm}|}
\hline \textbf{Spectrum} & \textbf{Frequency (MHz)}  & \textbf{Description}\\

\hline {\multirow{3}{*}{VHF}} & 155.70000              & Law Dispatch, City of Homestead, FL.   \\
							  &                         & Police main channel 1.                 \\
\cline{2-3}
           					  & 151.07000, 154.81000, & Law Tactical, City of Homestead, FL.     \\
	           				  & 155.01000, 155.65500, & Police channel 2 - 6.                    \\
	           				  & and 156.21000        &                                          \\

\hline {\multirow{3}{*}{UHF}} &           & Fire-Talk, and local government entity. Miami-Dade county, FL. \\
                              &           &                                                              \\
\cline{2-3}
                              & 453.13750 & Digital-A channel.\\
							  & 453.45000 & Digital-B channel.\\

\hline UHF-TV & 470 - 476 & Miami public safety operations. TV channel 14 is designated for LMRS use in Miami urban area. \\

\hline {\multirow{3}{*}{700 MHz}} & & FL Statewide, Statewide Law Enforcement Project 25 Radio System (SLERS P25). System type APCO-25 Phase I, supporting voice over common air interface.\\
\cline{2-3}
                                  & 773.59375 (c), 773.84375 (a) and 774.09375 (a),  & Orange county, Fort White, FL.\\

\hline 800 MHz & 851.05000,~853.87500, 855.11250, 856.16250, and 858.81250 (c)& Broward county, Coral Springs FL has Coral Springs public safety (Project-25). System type APCO-25 Phase I, supporting voice over common air interface.\\

\hline 900 MHz & 902 - 928 &  Radio spectrum is allocated to amateur radio i.e., amateur 33 centimeters band. It is also used by ISM and low power unlicensed devices. Example Motorola DTR650 FHSS (frequency-hopping spread spectrum) digital two‑way radio.\\
\cline{2-3}
               & 932 - 935 & Government and private shared operation, fixed system. \\

\hline 4.9 GHz & 4940 - 4990& Mission Critical Solutions, has designed a dual radio Wi-Fi mesh network for the city of Hollywood, FL called as "Wireless Hollywood". Each of the strategically located internet access points is a two-radio system that uses a 4.9 GHz for law enforcement and emergency first responders. The mesh network showed a 96.3 percent street level coverage on the 4.9 GHz safety band, and each access point had a RSSI value above -79 dBm \cite{4.9GHz1}.\\
\hline
\end{tabular}
\label{Table:SoFloSpectrumUsersAndExamples}
\end{table}

\subsection{4.9 GHz Public Safety Spectrum}
In year 2002, FCC allocated 50 MHz of spectrum in the 4940~MHz~-~4990~MHz band (i.e., 4.9 GHz band) for fixed and mobile services. This band supports a wide variety of broadband applications such as Wireless LAN for incident scene management, mesh network, WiFi hotspots, VoIP, temporary fixed communications, and permanent fixed point-to-point video surveillance \cite{Ref1.1,4.9GHz,4.9GHz1, moto49Ghz}.

Table~\ref{Table:SoFloSpectrumUsersAndExamples}, illustrates the public safety spectrum bands, frequencies and the respective users as an example in the state of Florida.

\section{LTE-based FirstNet PSN}
\label{firstNet}
In January 26, 2011, the FCC in the United States adopted a Third Report and Order and Fourth Further Notice of Proposed Rulemaking (FNPRM). In the third report and order, the FCC mandates the 700~MHz public safety broadband network operators to adopt LTE (based of 3GPP Release 8 specifications) as the broadband technology for nationwide public safety broadband network (NPSBN). The Fourth FNPRM focuses on the overall architecture and proposes additional requirements to promote and enable nationwide interoperability among public safety
broadband networks operating in the 700 MHz band \cite{fnprm}.

\begin{figure}[!htbp]
\centering
\includegraphics[width=9.05cm,height=12cm,keepaspectratio]{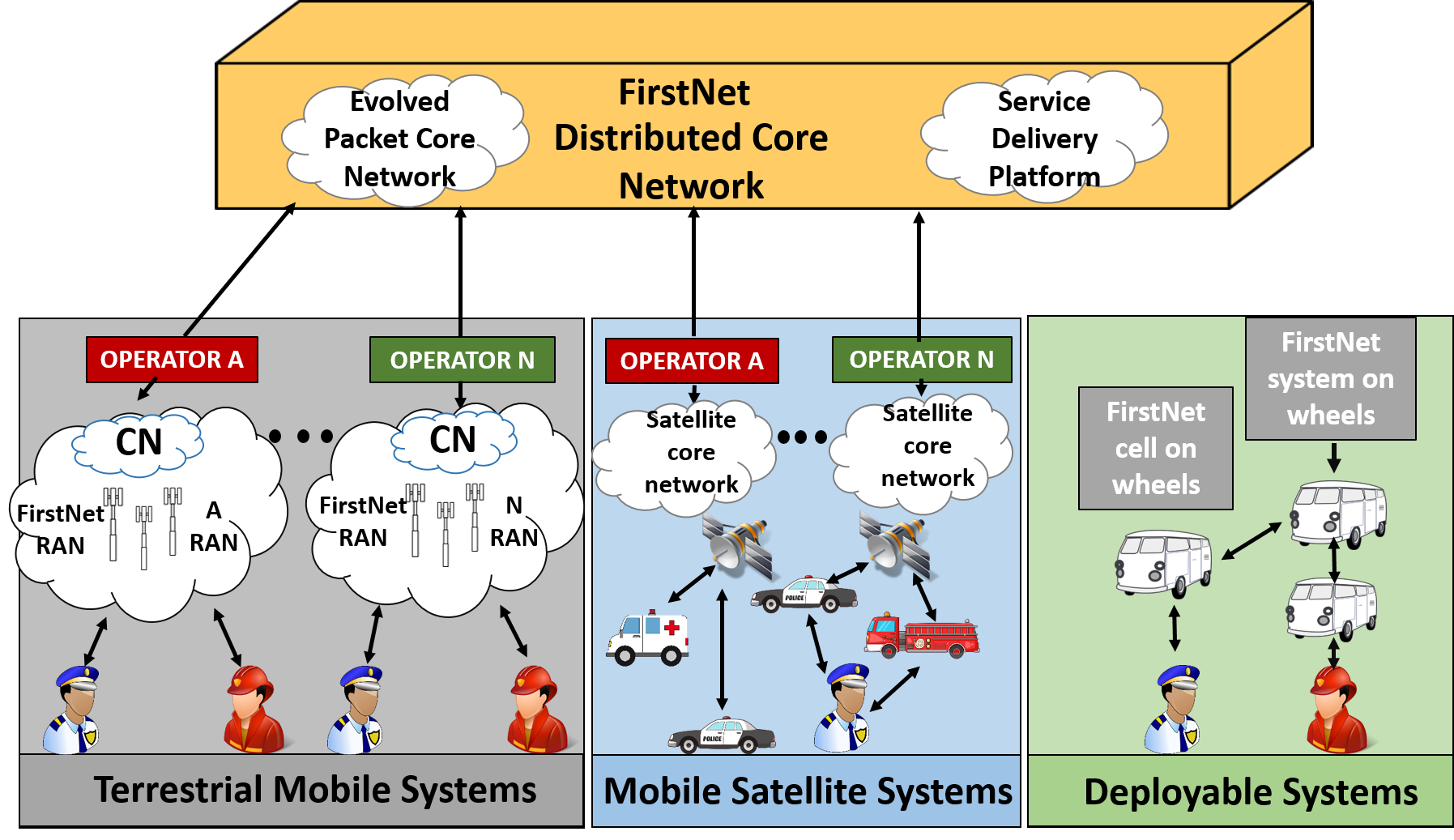}
\caption{FirstNet PSC architecture proposed by \cite{FirstNetModel}. }
\label{fig:ProposedFirstNet}
\end{figure}

\begin{table*}[!htbp]
\centering
\caption{ A variety of deployable technologies for FirstNet \cite{firstNetVechicular}.}
\begin{tabular}{|p{2.2cm}|p{1.8cm}|p{3cm}|p{3cm}|p{3cm}|p{2.9cm}|}
\hline \textbf{Characteristics } & \textbf{VNS} &\textbf{COLTS} & \textbf{COW} & \textbf{SOW} &\textbf{DACA} \\
\hline Capacity  & Low/medium & Medium & High & High & Low/medium \\
\hline Coverage  &  Small cells such as pico and femto & Cell size can be either macro, or micro, or pico, or femto &  Cell size can be either macro, or micro, or pico, or femto &  Cell size can be either macro, or micro, or pico, or femto & Small cells such as pico and femto\\
\hline Band class 14 radio  & Yes & Yes & Yes & Yes & Yes\\
\hline Standalone  &  Yes & No &  No & Yes & No\\
\hline Availability  &  Immediate & Vehicular drive time & Vehicular drive time & Vehicular drive time and system deployment time & Aerial launch time\\
\hline Power  &  Limited to vehicle battery & Generator & Generator & Generator & Limited to the airframe\\
\hline Deployment time  & Zero to low & Low & Medium & Long & Long\\
\hline Deployment nature  & EFR vehicles & Dedicated Trunks & Dedicated Trailers & Dedicated Trucks with Trailer & Aerial such as UAVs and balloons\\
\hline Deployment quantity  & Thousands	& Hundreds & Hundreds & Dozens & Dozens (based on experimentations and simulations) \\
\hline Incident duration & Low & Medium & Medium & Long & Long \\
\hline
\end{tabular}
\label{Table:FirstNetDeployTech}
\end{table*}

\subsection{FirstNet Architecture}
As per \cite{FirstNetWebpage}, Fig.~\ref{fig:ProposedFirstNet} shows the proposed FirstNet architecture. The FirstNet is a LTE-based broadband network dedicated to public safety services. Core features of FirstNet will include direct communication mode, push-to-talk (PTT), full duplex voice system, group calls, talker identification, emergency alerting, and audio quality. Band class 14 (D Block and Public Safety Block) shown in Fig.~\ref{fig:BandClass14Plan} is the dedicated band and will be used by FirstNet PSN~\cite{FirstNet2}. The FirstNet is composed of distributed core, terrestrial mobile system, mobile satellite system, and deployable systems. Distributed core consists of an evolved packet core (EPC) network and a service delivery platform to provide various services to end users such as EFR. The terrestrial mobile system comprises terrestrial based communication, while mobile satellite system will use a satellite communication link to communicate with distributed core network for services. Deployable systems are cells on wheels, providing services in network congestion areas or filling coverage holes.

In \cite{firstNetVechicular}, FirstNet vehicular PSNs were divided into 5 categories: vehicle network system (VNS), cells on light trucks (COLTS), cells on wheels (COW), system on wheels (SOW), and deployable aerial communications architecture (DACA). These different vehicular FirstNet PSN systems are expected to  play a significant role in providing coverage extension for NPSBN deployment. Thus, these deployments will deliver greater coverage, capacity, connectivity, and flexibility in regions which are outside of terrestrial coverage, or where traditional coverage become unavailable due to a natural or man-made disaster \cite{previousWork1, FirstNet}. Table \ref{Table:FirstNetDeployTech} discusses the characteristics of the five different vehicular PSN architectures in FirstNet.

\begin{figure}[!htbp]
\centering
\includegraphics[width=8cm,height=10cm,keepaspectratio]{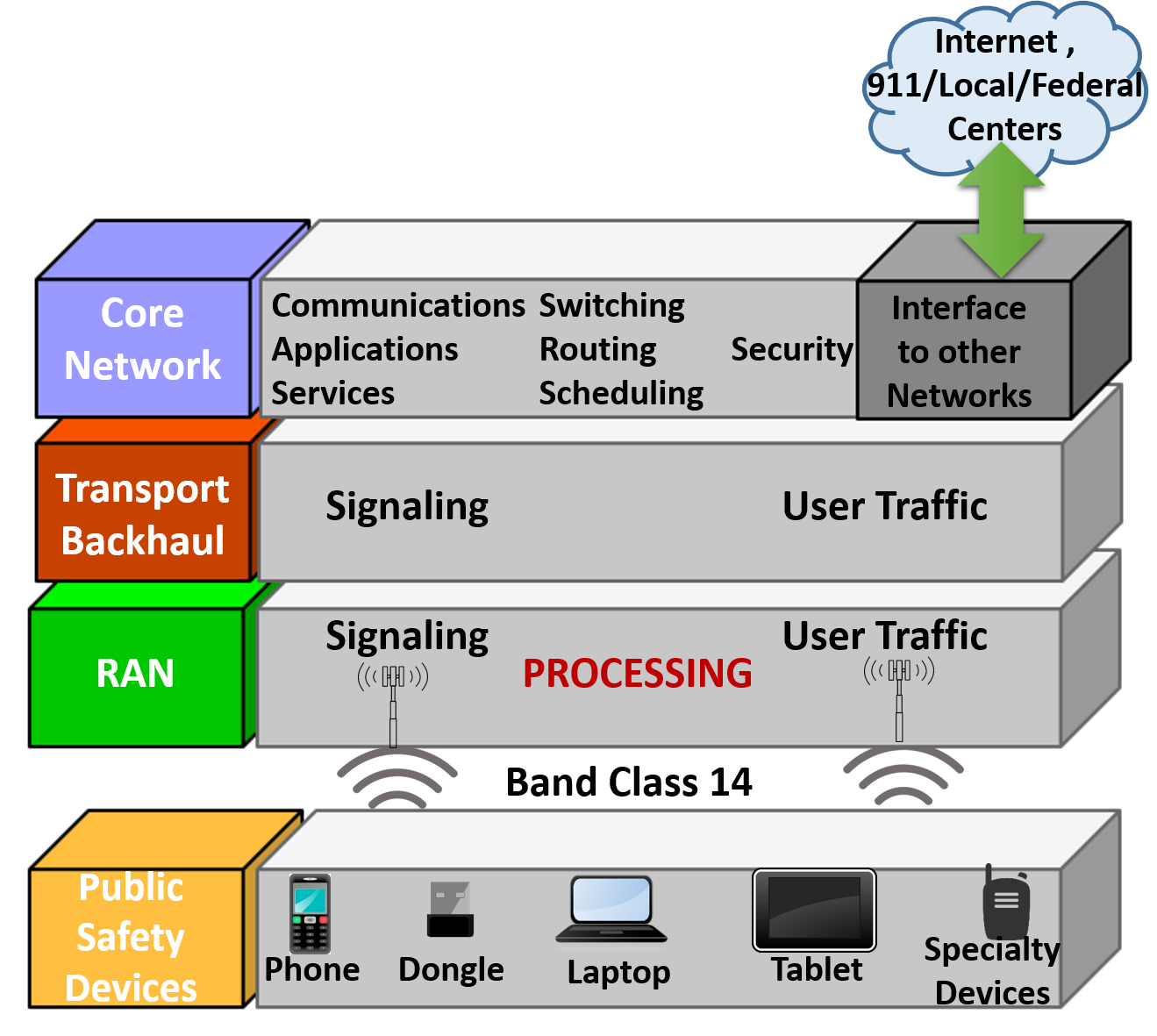}
\caption{LTE band class 14 architecture for FirstNet \cite{FirstNetWebpage}. }
\label{fig:LTEbandclass14Infrastructure}
\end{figure}

As per \cite{FirstNetWebpage}, FirstNet broadly defines LTE network in distinct layers: Core network, transport backhaul, radio access network (RAN) and public safety devices as seen in Fig.~\ref{fig:LTEbandclass14Infrastructure}. Public safety devices can be smartphones, laptops, tablets or any other LTE-based user equipment. RAN implements radio access technology, which conceptually resides between public safety devices and provides a connection to its CN. The transport backhaul comprises intermediate CN and subnetworks, wherein CN is the central part of a telecommunication network that provides various services to public safety devices across access network. 

The band class 14 communication will be based on the LTE commercial standards and is solely dedicated to PSC in North America region. Table~\ref{Table:BandClass14FreqParam} shows the available downlink and  uplink frequencies and the E-UTRA Absolute Radio Frequency Channel Number (Earfcn) for a band class 14 system. As per 3GPP specifications,  the carrier frequency is specified  by an absolute radio-frequency channel number (ARFCN). The Earfcn is a designated code pair for physical radio carrier for both transmission and reception of the LTE system i.e., one Earfcn for the uplink and one for the downlink. The Earfcn also reflects the center frequency of an LTE carriers, and falls within range of 0 to 65535. The bandwidth allocated for band class~14 is 10~MHz.

\begin{table}[!hb]
\centering
\caption{Band class 14 parameters.}
\begin{tabular}{|l|l|l|l|l|l|l|l|}
\hline
\multicolumn{1}{|c|}{ } &\multicolumn{3}{|c|}{\textbf{Downlink (MHz)}}
& \multicolumn{3}{|c|}{\textbf{Uplink (MHz)}} \\
\hline
 & Low &Middle & High &Low & Middle & High\\
\hline
Frequency&758&763&768&788&793&798 \\
\hline
Earfcn&5280&5330&5379&23280&23330&23379 \\
\hline
\end{tabular}
\label{Table:BandClass14FreqParam}
\vspace{-2mm}
\end{table}

\section{LMRS and LTE Convergence}
\label{SectionIV}
Since data is just as important as voice in PSC scenario, mission-critical voice and data communication together can deliver the much required reliable intelligence and support to EFR during any emergency \cite{converge0}. For example, in the United States, FCC has been taking steps to achieve this LMRS-LTE convergence by setting various policies and a framework for a NPSBN. A private network like NPSBN would include FirstNet (a single licensed operator), 700~MHz band class 14 (single frequency band), and LTE (a common operating technology). With large investment in LMRS over the last decade or so, LMRS has gained a national footprint in the United States \cite{npsbn}. During this time the United States has relied on LMRS technology to provide mission critical communication to the EFR. LTE deployments for replacing 2G and 3G infrastructure had no issues in relative measures, whereas the integration of LTE with LMRS is going to be a challenging task. As the complete roll out of the NPSBN by FirstNet would take few years to meet all mission-critical EFR requirements, it is inevitable for LMRS and LTE to coexist for some time in the near future.

There are many deployment models available today for converging LMRS and LTE in PSC. This includes the combination of mission-critical radios on PSN, broadband devices on dedicated broadband networks, and consumer grade devices on commercial carriers. The major challenges would be 1)~providing an efficient end-user experience to EFR using a multitude of different device, networks, and infrastructure and 2)~sharing mission-critical information regardless of device or network, without compromising security.

\begin{figure}[!htbp]
\centering
\includegraphics[width=6cm,height=6cm,keepaspectratio]{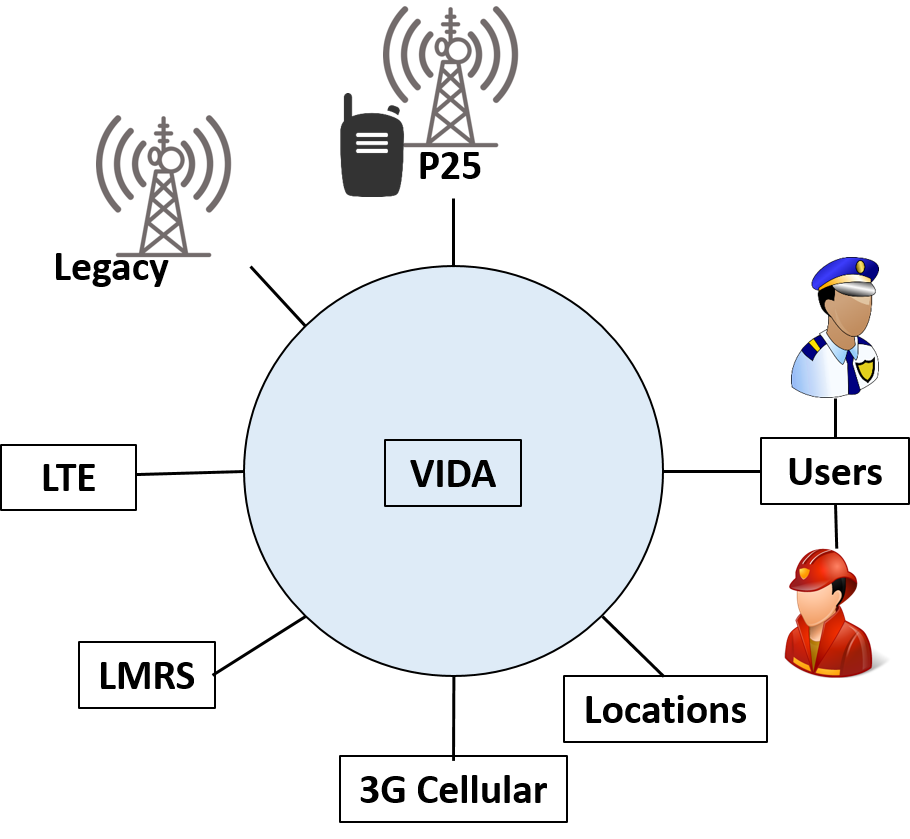}
\caption{VIDA from Harris Corporation is a converge platform that integrates legacy and broadband PSC into a core network~\cite{converge1}.}
\label{fig:VIDA}
\end{figure}

As a step towards the next generation public safety networks, Motorola Solutions has envisioned LMRS taking advantage of public and private LTE broadband networks. APX7000L, as shown in Fig. 6(b), is a portable APCO-25 digital radio converged with LTE data capabilities in a public-safety-grade LMR handheld device \cite{P25TxPower}. In order to have a seamless transition between LMRS and broadband networks, Motorola VALR mission critical architecture is utilized \cite{VALR}. Harris Corporation's VIDA mission-critical platform provides an ability to support narrowband and broadband PSC technologies \cite{converge1} as illustrated in Fig.~\ref{fig:VIDA}. Similarly, Etherstack's LTE25 solution bridges LTE with existing APCO-25 narrowband networks  with integrated push-to-talk solutions and interoperability in multi-vendor APCO-25/LTE networks \cite{LTE25}. These developments build flexibly on technological advances, without compromising requirements, security, and coverage need for mission-critical communication. The converged solutions and frameworks have a common goal to provide mission-critical solutions, location services, and priority/emergency call support.

These devices, solutions, and frameworks would provide EFR, the most need continuous connection to mission-critical voice, and data communications, and would be a first step in providing a necessary infrastructure for NPSBN. The LMRS and LTE both  have individual strengths, and their convergence would result in a nationwide PSN and therefore enable a better collaboration amongst EFR, interoperability, data capability, security, robust coverage, and provide mission critical communications.

\section{Mission-Critical PTT over LTE}
\label{SectionV}
The main goal of mission-critical voice is to enable reliable communication between various EFR. The essential attributes of mission~-critical voice is to enable fast-call setup and group communication amongst the EFR, push-to-talk for various group talks, high audio quality, emergency alerting, and support secure/encrypted voice communication. The mission~-critical encryption uses data encryption standard (DES) and advanced encryption standard (AES) for over-the-air-rekeying application \cite{MC-PTT1,MC-PTT2}. To support mission-critical voice, LTE needs to incorporate all the essential attributes mentioned in this section. The objective of mission-critical PTT over LTE would be to emulate the functions by LMRS.

The 3GPP group is actively working to produce technical enhancement for LTE that support public safety applications. The main objective of 3GPP is to protect the quality of LTE while including the features needed for public safety. In LTE Release 12, 3GPP has focused on two main areas to address public safety applications. These focus areas include proximity services (ProSe) that enable optimized communication between the mobiles in physical proximity and group call system that support operations such as one-to-many calling and dispatcher services which enable efficient and dynamic group communication \cite{MC-PTT4}.

\begin{figure}[!htbp]
\centering
\includegraphics[width=8cm,height=14cm,keepaspectratio]{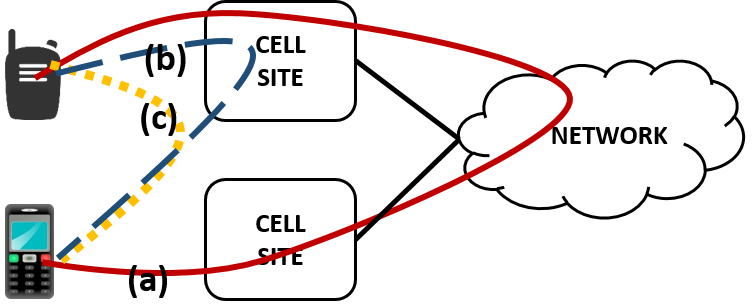}
\caption{Proximity service examples as proposed in 3GPP LTE Release 12\cite{MC-PTT4}. Path (a) shows current conventional LTE communication path, (b) shows direct communication with proximity services and (c) shows locally routed communication with proximity services.}
\label{Proximity}
\end{figure}

Path (a) in Fig.~\ref{Proximity} shows the current conventional LTE communication whereas a network assisted discovery of users in close physical proximity is shown in Fig.~\ref{Proximity} path (c). The direct communication, between these users with or without the supervision of the network is illustrated in Fig.~\ref{Proximity} as path (b). In direct communication a radio can establish a communication link between the mobile users without going through a network, thus saving network resources and also enable mission-critical communication amongst the EFR even outside the network coverage \cite{MC-PTT4,babun2014intercell,babun2015extended}.

The requirement definitions of mission-critical PTT over LTE are ongoing and will be finalized in LTE Release 13, which will enhance the D2D/ProSe framework standardized in Release 12 and support more advanced ProSe for public safety applications \cite{MC-PTT3, MC-PTT5,simsek2013device}. Release 12 added basic discovery and group communications functionality specific to D2D/ProSe. ProSe direct discovery, ProSe UE-to-network relay, ProSe UE-to-UE relay, and group communication among members via network, and a ProSe UE-to-network relay for public safety application will be included as part of Release 13 \cite{MC-PTT3}. Isolated E-UTRAN operation for public safety is an another public safety feature included in Release 13 specifications. It would support locally routed communications in E-UTRAN for nomadic eNodeBs operating without a backhaul for critical communication \cite{MC-PTT3}, as shown in Fig.~\ref{Proximity}.

\section{Comparison of LMRS and LTE PSC}
\label{SectionVI}
\begin{figure} [!htbp]
\centering
\includegraphics[width=1\linewidth]{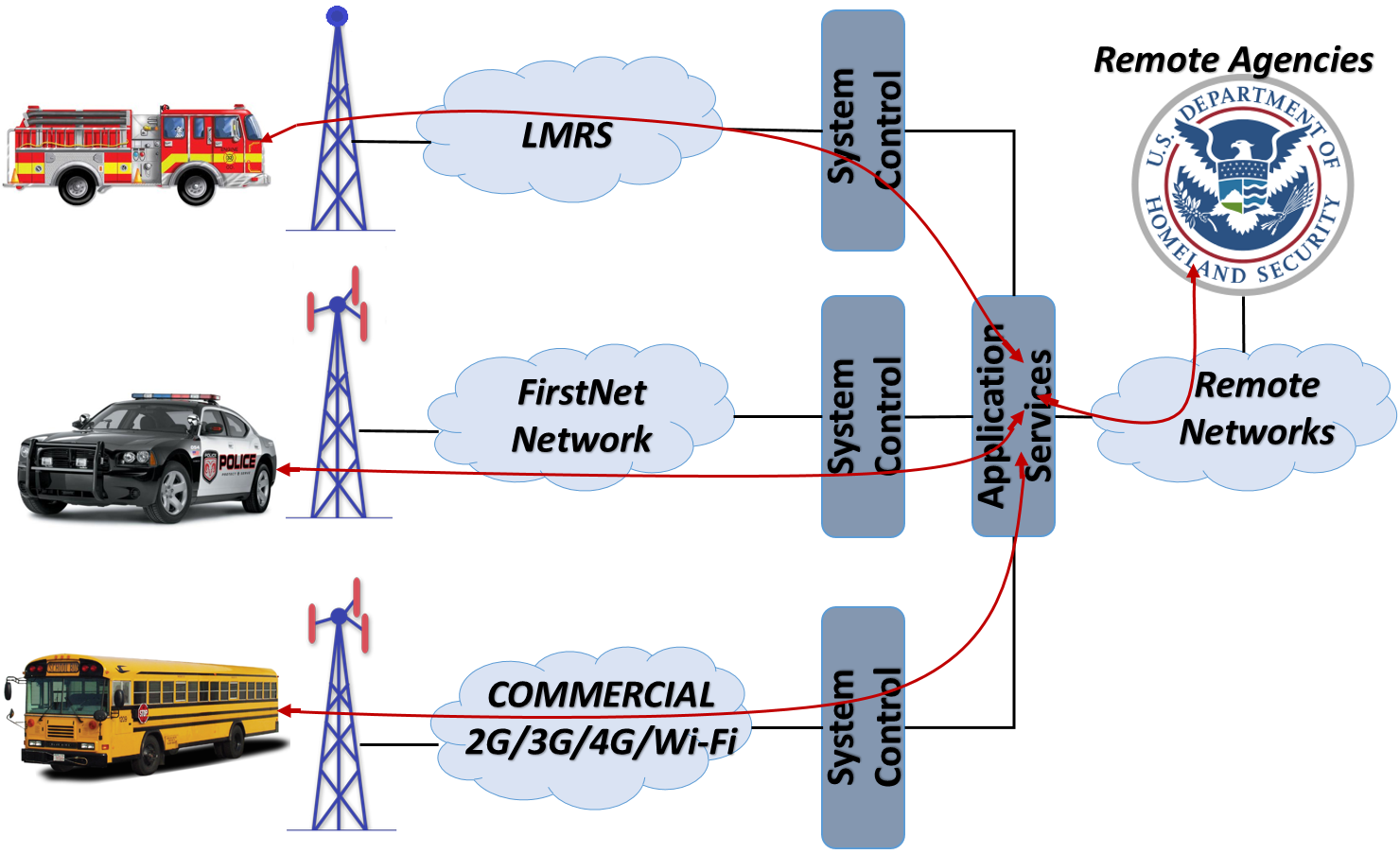}
\caption{Current PSN structure in the United States.}
\label{fig:CurrentPSNstructure}
\end{figure}

The LMRS technology has been widely used in PSC for over a decade with significant technological advancements. On the other hand, with the emerging of the LTE push-to-talk technology, focus has been shifting towards high speed low latency LTE technology.  With various advancements in commercial networks, it can be another possible viable solution for PSC.  Consider an emergency scenario, where different public safety organization using different communications technologies arrive at the incident scene. As shown in Fig.~\ref{fig:CurrentPSNstructure}, these public safety organizations can be using different types of PSNs. In particular, the PSN deployed in the United States is a mixture of LMRS, FirstNet, and commercial networks. The LMRS provides push-to-talk mission critical voice capability over narrowband, with restricted data rates for voice/data services. However, LMRS has a wide area coverage, which can be achieved through the deployment of high-power base stations and handsets, portable base stations, repeaters, and/or peer-to-peer communication \cite{coverage}. FirstNet provides high-speed network and can transform EFR capabilities with real-time updates. EFR can still rely on LMRS for mission-critical voice communication alongside FirstNet for high-speed data.

\subsection{Network Architecture}
Commercial networks are IP-based technologies such as 2G/3G/4G/Wi-Fi. These technologies need adequate signal coverage and network scaling to support PSC services between EFR and victims. Due to existing commercial infrastructure, these networks are beneficial for providing PSC services. However, commercial networks are vulnerable to network congestion due to higher intensity of inbound and outbound communication at the incident scene by the public safety organizations and victims. Such vulnerability can result in disrupted communications services and bring harm to victims.

\begin{figure} [!htbp]
\centering
\includegraphics[width=1\linewidth]{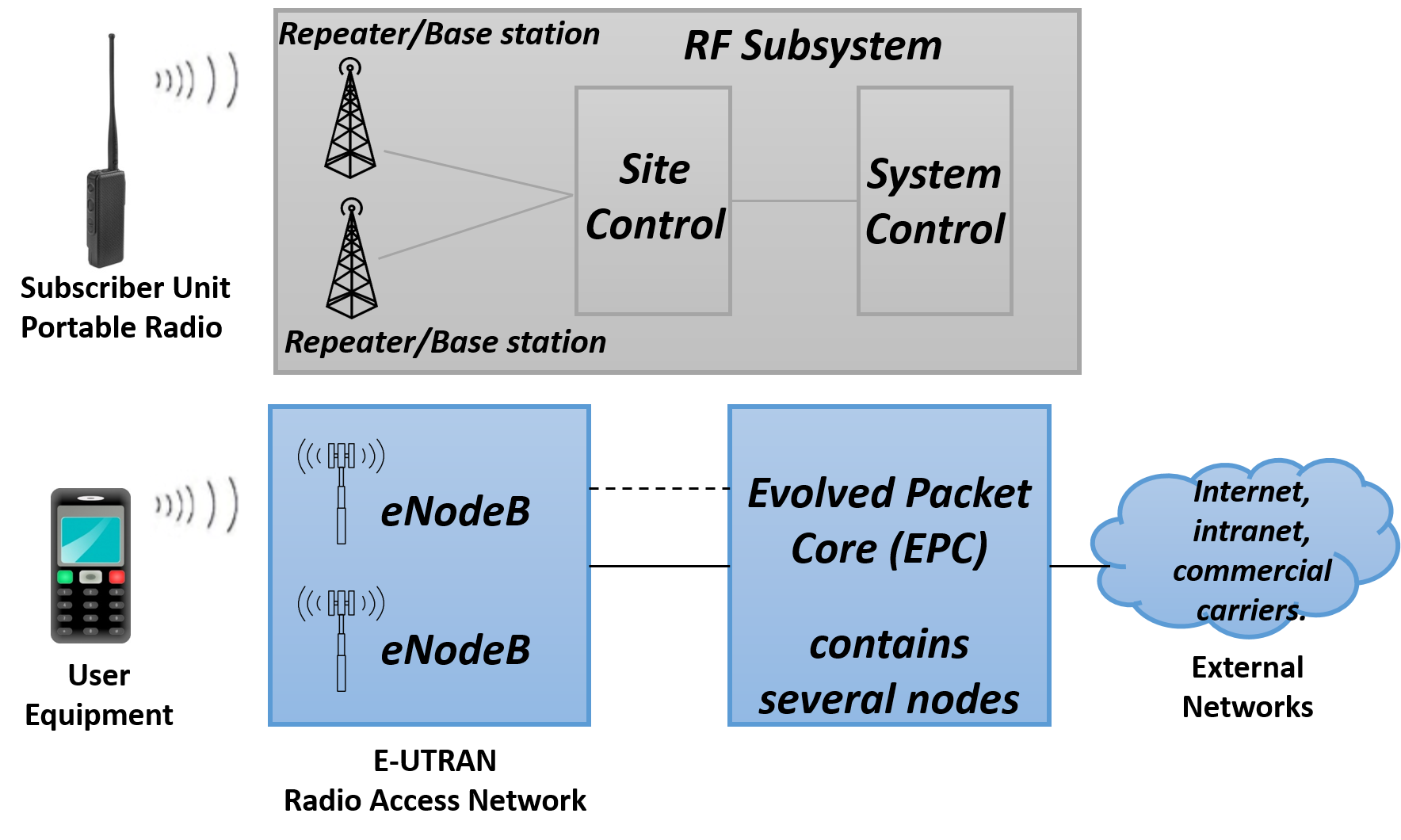}
\caption{System level representation of the LMRS and the LTE networks for PSC.}
\label{fig:SystemLevelComparison}
\end{figure}

System level block representation of the LMRS and LTE system are as shown in Fig.~\ref{fig:SystemLevelComparison}. The LMRS comprises of a mobile/portable subscriber units, repeater or a base station which is connected to an RF subsystem and then to a customer enterprise network. On other hand the LTE system includes user equipment (UE), eNodeB, and EPC which is connected to external Network. EPC, also known as system architecture evolution (SAE), may also contain several nodes. Key components of EPC are mobility management entity (MME), serving gateway (SGW), public data network gateway (PGW), and policy and charging rules function (PCRF).

\begin{figure} [!htbp]
\centering
\includegraphics[width=1\linewidth]{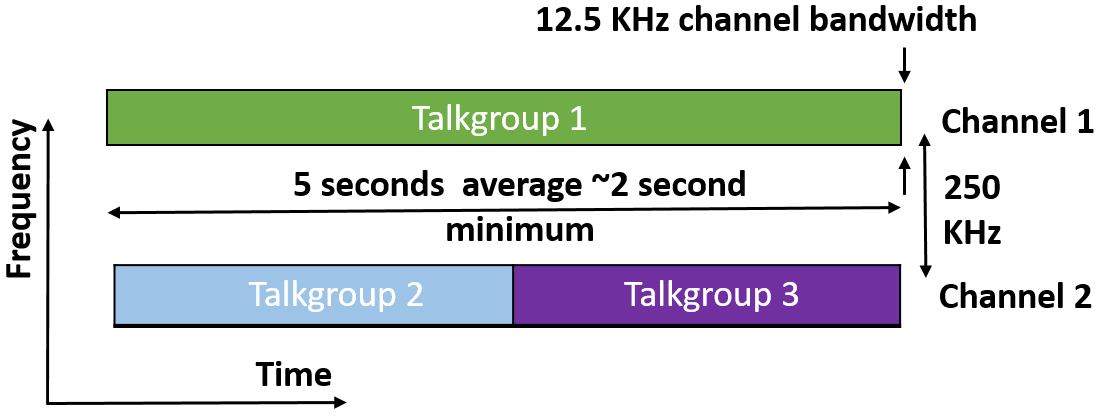}
\caption{APCO-25 channel configuration.}
\label{fig:APCO-25ChannelConfiguration}
\end{figure}

\begin{figure}
        \begin{subfigure}[b]{0.5\textwidth}
         \centering
                \includegraphics[width=8cm,height=9cm,keepaspectratio]{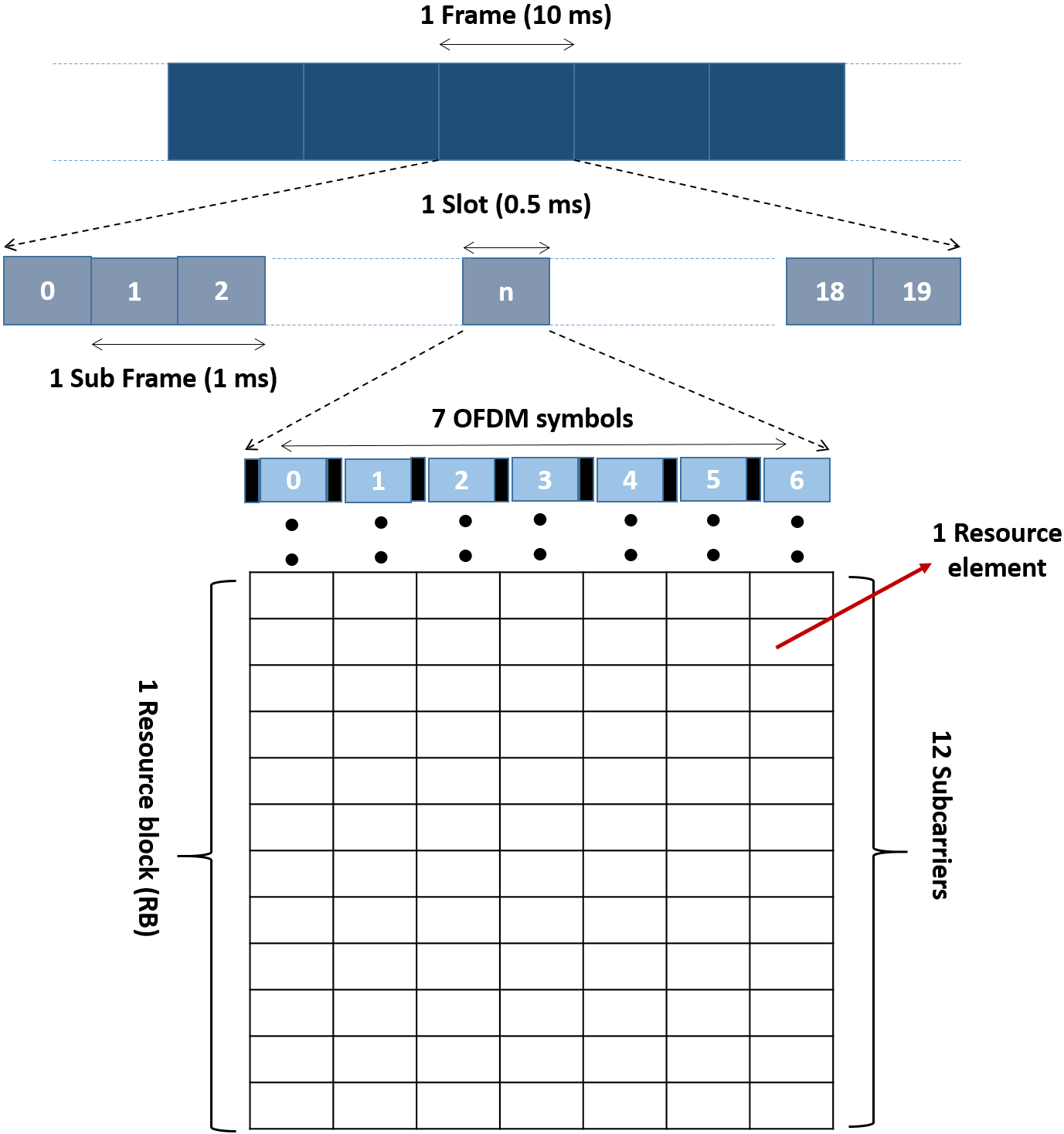}
                \caption{LTE frame structure.}
                \label{fig:LTEframeStructure}
        \end{subfigure}

        \begin{subfigure}[b]{0.51\textwidth}
        \centering
                \includegraphics[width=8cm,height=9cm,keepaspectratio]{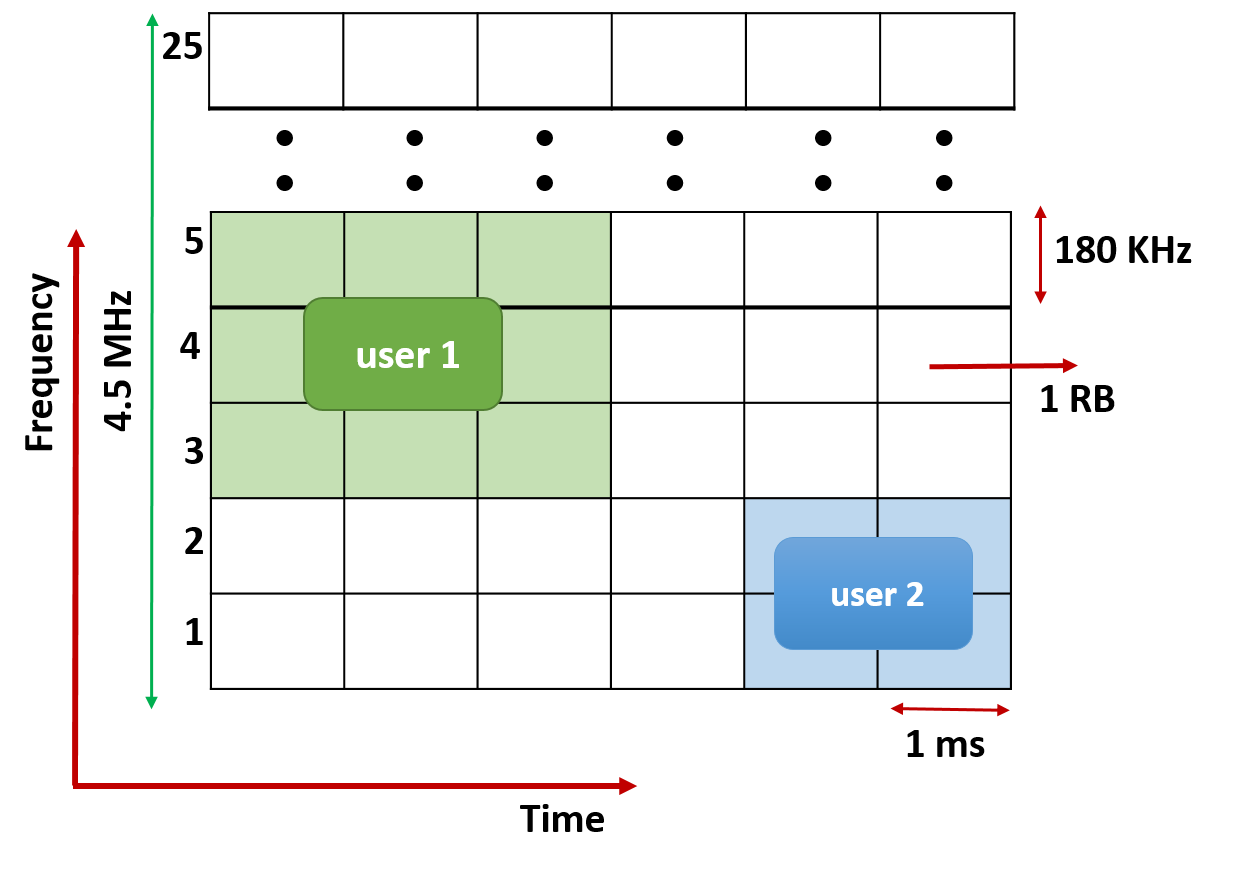}
                \caption{LTE channel configuration. Each block represents a RB, an UE can be scheduled in one or more RBs. LTE FirstNet gives more flexibility in scheduling users to their best channels, and hence enables better QoS.}
                \label{fig:LTEresourceBlock}
        \end{subfigure}
        \caption{LTE channel configuration.}
        \label{LTEchannelConfiguration}
\end{figure}

\subsection{Channel Configuration and Frame Structure}
Channel configuration for the LMRS is shown in Fig.~\ref{fig:APCO-25ChannelConfiguration}. In an LMRS the supported bandwidth are 6.25~KHz, 12.5~KHz, and 30~KHz, wherein 12.5~KHz is the standard bandwidth. The separation between two channels is usually 250~KHz or higher. The talkgroups channel assigned during a transmission can last for a duration between 2 to 5 seconds. The data channel has bandwidth of 12.5 KHz and is FDMA and/or TDMA channel. The data channel is designed with an aim to provide functionality like over the air rekeying, mission-critical voice etc,. On the other hand, the control channel is used solely for resource control. The main task for control channel is to allocate resources, digital communication message bearer, and message handler between RF subsystem and SU. The APCO-25 control channel uses C4FM modulation scheme and supports a baud rate of $9600$ bits/s \cite{P25ControlChannel}.

The LTE frame structure as seen in Fig.~\ref{fig:LTEframeStructure}, has one radio frame of 10 ms duration, 1 sub-frame 1 ms duration, and 1 slot of 0.5 ms duration. Each slot comprises of 7 OFDM symbols. The LTE UL and DL transmission is scheduled by resource blocks (RB). Each resource block is one slot of duration 0.5 ms, or 180 KHz, and composed of 12 sub-carriers. Furthermore, one RB is made up of one slot in the time domain and 12 sub-carriers in the frequency domain. The smallest defined unit is a resource element, which consists of one OFDM subcarrier. Each UE can be scheduled in one or more RBs as shown in Fig.~\ref{fig:LTEresourceBlock}. The control channel in LTE is provided to support efficient data transmission, and convey physical layer messages. LTE is provided with three DL physical control channels, i.e., physical control format indicator channel, physical HARQ indicator channel, and physical downlink control channel. Physical uplink control channel is used by UE in UL transmission to transmit necessary control signals only in subframes in which UE has not been allocated any RBs for the physical uplink shared channel~transmission. Physical downlink shared channel, physical broadcast channel, and physical multicast channel are the DL data channels, whereas physical uplink shared channel and physical random access channel are the UL data channels.

Some further differences between the LMRS and the LTE networks are summarized in Table~\ref{Table:Comparison} \cite{compare}.

\begin{table*}[!htbp]
\centering
\caption{Comparison between the LMRS and the LTE  system.}
\begin{tabular}{|p{1.75cm}|p{7.5cm}|p{8cm}|}
\hline \textbf{Parameters} & \textbf{LMRS}  & \textbf{LTE}\\

\hline Applications & Mission-critical voice, data, location services, and text. & Multimedia, location services, text, and real-time voice and data. \\

\hline Data Rates & APCO-25 has a fixed data rate of 4.4 Kbits/s for voice communications and 9.6 to 96 Kbits/s for data only system. &  Depends on various factors such as signal-to-interference-plus-noise-ratio (SINR), bandwidth allocation, symbol modulation, forward error correction code, and  number of UEs attached to the eNodeB. As per 3GPP specifications, LTE-Advanced provides an increased peak data rate of downlink (DL) 3 Gbits/s, and 1.5 Gbits/s in the uplink (UL).\\

\hline SINR & DAQ is Delivered Audio Quality representing the signal quality for digital radios. The DAQ scale ranges from 1 to 5, with 1 being unintelligible to 5 being perfect \cite{DAQ}. Public safety radio uses a DAQ of 3.4, which corresponds to speech that is understandable and rarely needs repetition, i.e., SINR of 17.7 dB. & Does not depend on cell load but relies on factors such as channel quality indicator (CQI) feedback, transmitted by the UE. CQI feedback lets eNodeB select between QPSK, 16QAM, and 64QAM modulation schemes and a wide range of code rates.  A SINR greater than 20 dB denotes excellent signal quality (UE is closer to eNodeB), whereas SINR less than 0 dB indicates poor link quality (UE is located at cell edge).\\

\hline Modulation type & Fixed modulation scheme with 2 bits per symbol. & Based on CQI feedback eNodeB can select between QPSK, 16QAM, and 64QAM modulation scheme where higher modulation scheme needs higher SINR value. \\

\hline Forward Error Correction (FEC)&  FEC rate is fixed as per DAQ3.4. &  FEC rate is variable and dependent on the SINR values. Higher rate codes (less redundant bits added) can be used when SINR is high. \\

\hline
Antenna Configuration & Single input single output (SISO) node with an omni-directional pattern. & MIMO has been an integral part of LTE with a goal of improving data throughput and spectral efficiency.  LTE-Advanced introduced 8x8 MIMO in the DL and 4x4 in the UL.\\

\hline Transmission Power & subscriber unit always transmits at full power. As per \cite{P25TxPower} Subscriber unit transmit power is within range set of 1 - 5 Watts and for base station output power 100 - 125~Watts depending on 700 MHz, 800 MHz, VHF, or UHF frequency. & UE's transmission power can be controlled with a granularity of 1 dB within the range set of -40 dBm to +23 dBm (corresponding to maximum power transmission power of 0.2 Watts). As per 3GPP Release 9 notes, maximum transmission power is 46 dBm for micro base station and 24 dBm for pico base station. \\

\hline Cell range & Based on the antenna height above the ground the maximum transmission range can be up to 43 km \cite{desourdis2002emerging}. & In theory, GSM base station can cover an area with a radius of 35 km. In case of dense urban environment, coverage can drop to area with radius between 3 km to 5 km and less than 15 km in dense rural environment \cite{sauter2010gsm}. Small cells such as microcell, picocell, femtocell have even smaller range and cover up to 2 km, 200 m, and 10 m respectively.\\

\hline Multicast and Unicast & Multicast for group voice calls and unicast for data. & Primarily unicast for data, MBMS can support multicast data. \\

\hline Security & 256-bit AES encryption \cite{compare, compareSecurity}. & Standardized using 128-bit key \cite{solanki2013lte}.\\

\hline

\end{tabular}
\label{Table:Comparison}
\end{table*}

\section{LMRS vs. LTE: Simulation Results}
\label{SectionVII}
The purpose of this section is to simulate LTE band class 14 system and APCO-25 using NS-3 open source tool. The LTE band class~14 simulation is based on LENA LTE model \cite{LENA} whereas LMRS simulation is based on the definition of APCO-25 portrayed in \cite{P25Phase, C4FM, P25ControlChannel, P25RadioWiki}. We chose to use NS-3 simulator and LENA LTE model due to their effective characterization of LTE  protocol details.

In order to be consistent and to provide a better comparative analysis between LMRS and LTE band class 14, we implement a single cell scenario, and similar user density. The main simulation goal is to compute aggregated throughput capacity and signal quality measurement with increasing cell size in case of both LTE and LMRS. Simulation setup, assumptions, and simulation parameters for LTE band class 14 system and LMRS are discussed in Section~\ref{apcoNS3Sim} and Section~\ref{bandClass14NS3Sim}, respectively. Furthermore, we analyze these simulation results in Section~\ref{NS3SimResultAna} to strengthen the conclusion of the comparative study.

\subsection{Setup for LTE Band Class 14}
\label{bandClass14NS3Sim}
\begin{figure} [!htbp]
\centering
\includegraphics[width=1\linewidth]{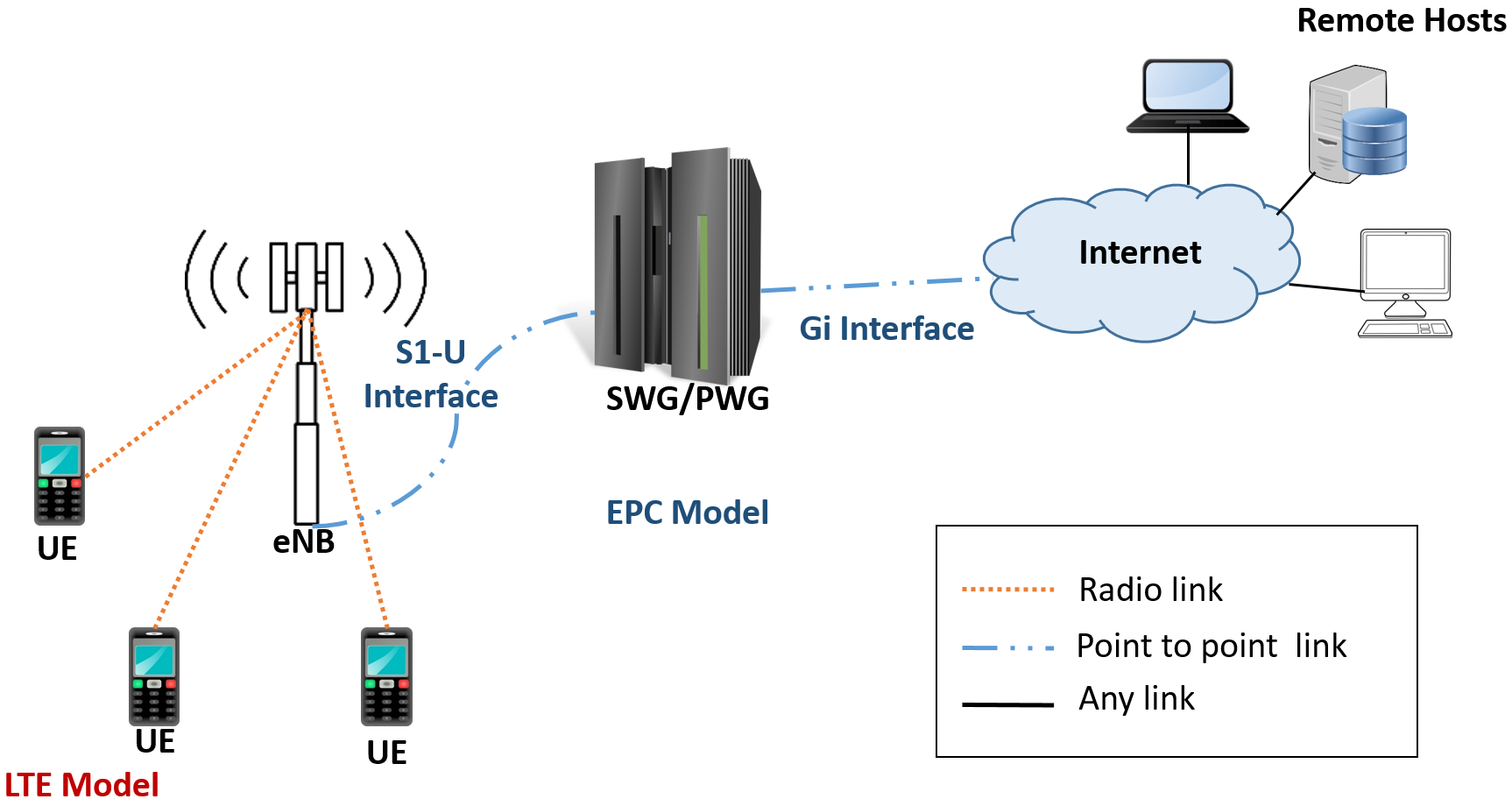}
\caption{LTE model for NS-3 simulations, based on the LENA project \cite{LENA}. All the important details of the LTE PHY and MAC protocols are implemented in the NS-3 simulations.  }
\label{LENASimModel}
\end{figure}

The simulation study of LTE band class 14 includes, throughput computation in uplink and downlink, cell signal quality measurement in terms of SINR, and measuring signal quality in a inter-cell interference scenario. Fig.~\ref{LENASimModel} shows the throughput simulation environment implemented in this paper. Simulation environment consists of randomly placed multiple UEs and one eNodeB (eNB). Other environment entity includes remote host which provides service to end user. Remote host will provide IP-based services to UEs and user datagram protocol (UDP) application is installed on both the remote host and the UEs to send and receive a burst of packets.

\textbf{Simulation scenario-Aggregate throughput computation}: In this simulation scenario, we implement a macro-eNodeB site with a transmission power of $46~dBm$. The simulation comprises $20$ UEs being randomly distributed throughout the cell, which has transmission power of $23~dBm$, and follows random walk 2D mobility model. The round robin MAC scheduler is used to distribute the resources equally to all the users. The communication channel included in the simulation is trace fading loss model in a pedestrian scenario. The LTE band class 14 simulation uses a bandwidth of 10~MHz and the frequency values is shown in Table~\ref{Table:BandClass14FreqParam}.

\begin{figure} [!htbp]
\centering
\includegraphics[width=1\linewidth]{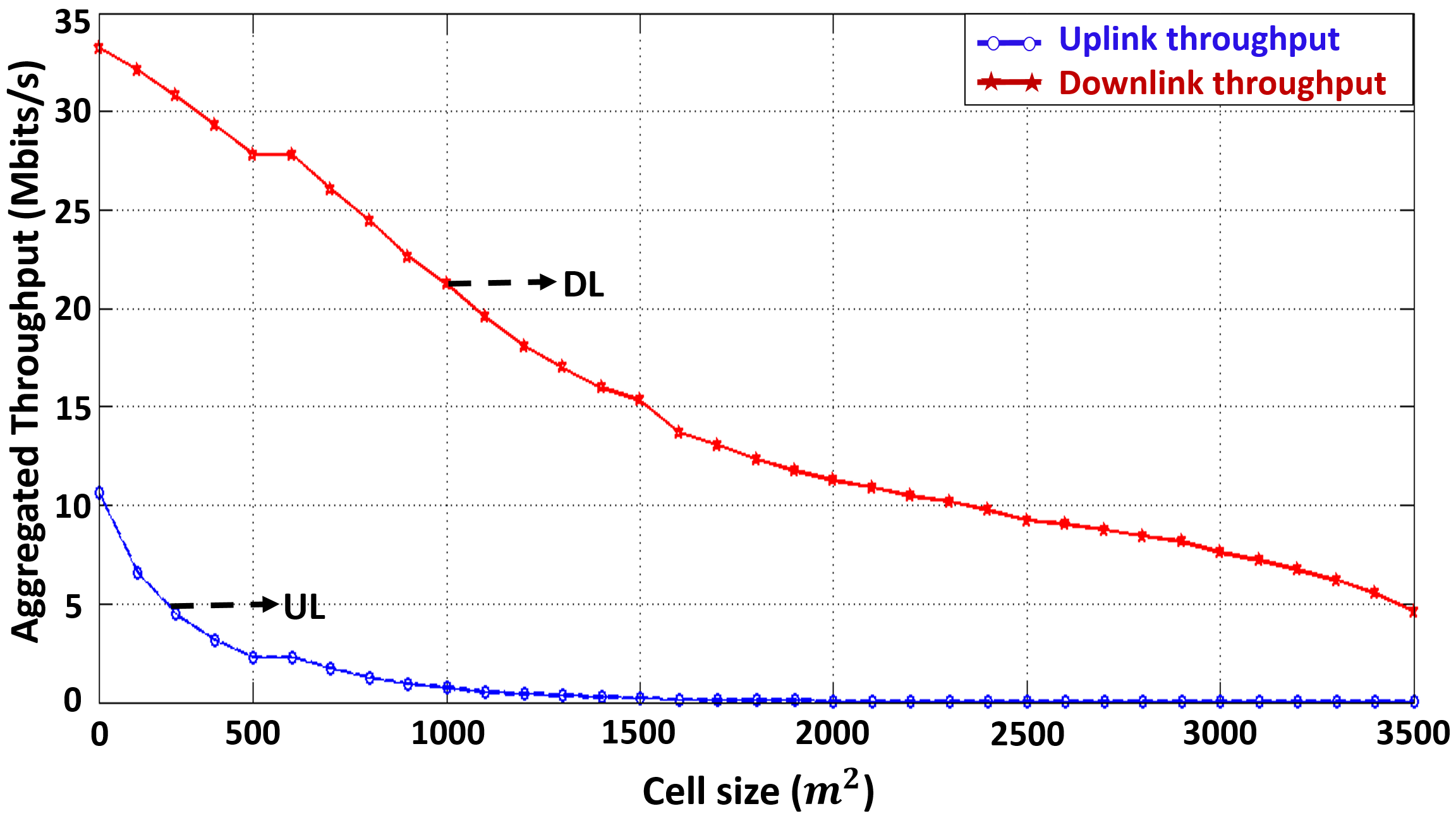}
\caption{Aggregated uplink and downlink throughput observed by 20 UEs in the LTE band class 14. UEs experience better throughput when in close proximity with macro-eNodeB.}
\label{UlAndDlThroughput}
\end{figure}

\textbf{Result observation}: In this simulation scenario for data only transmission and reception, the aggregated DL throughput can be as large as $33$~Mbit/s whereas the aggregated UL throughput can be as large as $11$~Mbit/s as seen in Fig.~\ref{UlAndDlThroughput}. Furthermore, it is also observed that the throughput experience by the UEs is diminished as the distance the between the UE and macro-eNodeB increases.

\begin{table}[!htbp]
\centering
\caption{LTE RF condition classification in \cite{sinr}.}
\begin{tabular}{|c|c|c|c|}
\hline & \textbf{RSRP (dBm)} & \textbf{RSRQ (dBm)} & \textbf{SINR (dB)}\\
\hline Excellent & $\ge$-80  & $\ge$-10 & $\ge$20\\
\hline Good & -80 to -90 & -10 to -15 & 13 to 20\\
\hline Mid-cell & -90 to -100 & -15 to -20 & 0 to 13\\
\hline Cell-edge  & $\le$-100 & $<$-20 & $\le$0\\
\hline
\end{tabular}
\label{RSSIClassification}
\end{table}

\textbf{Simulation scenario-Signal quality measurement:} In this simulation scenario, the UEs are placed at the cell center, post cell center, pre cell-edge and cell-edge regions of a macro-eNB site. The simulation model uses the simulation parameters and their attributes as shown in Table~\ref{Table:BandClass14FreqParam}. The signal quality is a measurement of three quantities: namely RSRP (reference signal received power), RSRQ (reference signal received quality), and SINR. SINR is a measure of signal quality defined by a UE vendor but never reported to the network. The UEs report the CQI to the network indicating DL transmission rate. The CQI is a 4-bit integer and is based on the observed SINR at the UE. In this simulation scenario, SINR factor is used to measure the link quality and the paper assumes the link quality is poor if the CQI value approaches~0.

\begin{figure} [!htbp]
\centering
\includegraphics[width=1\linewidth]{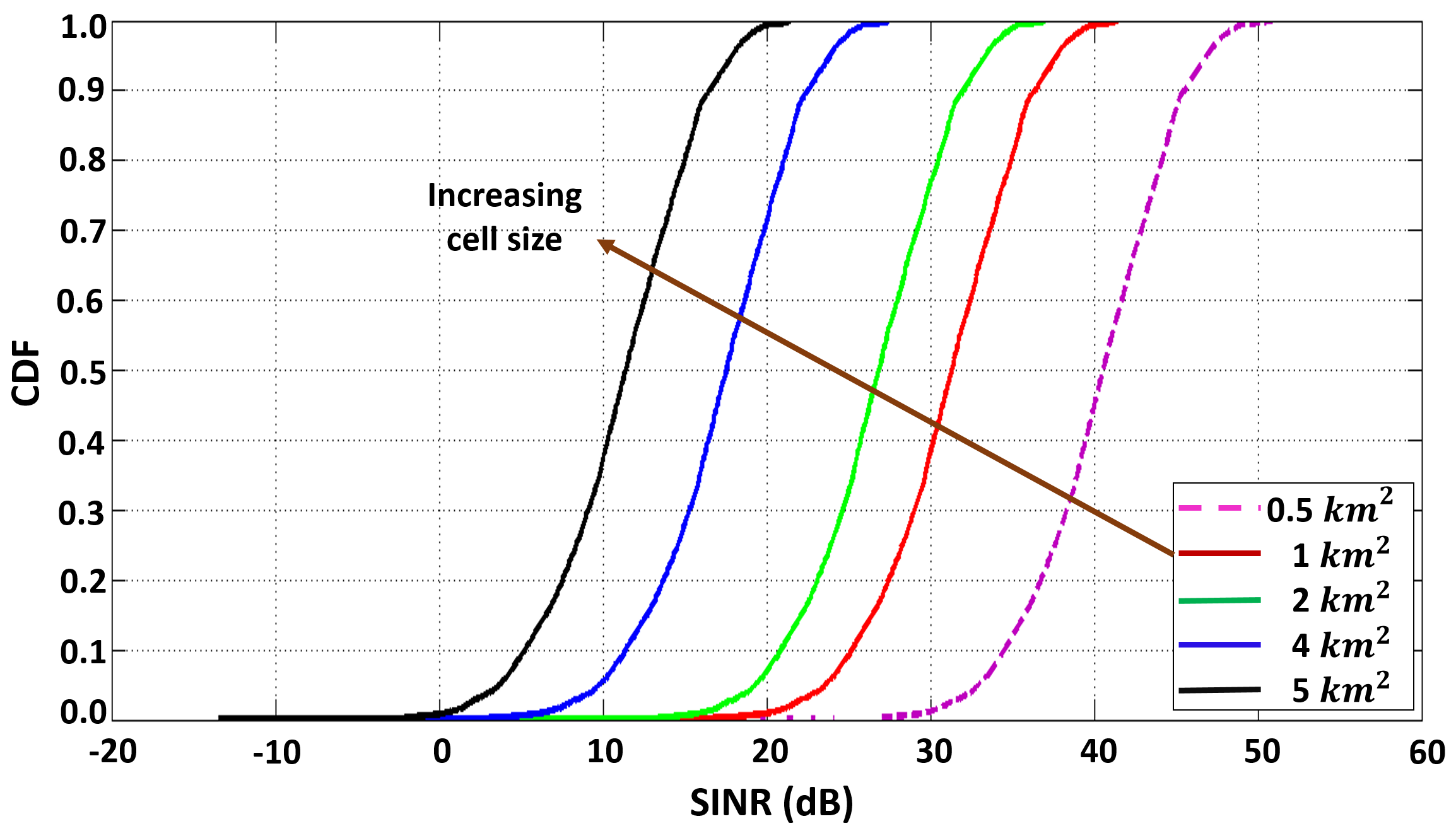}
\caption{The CDF plot of DL SINR values measured by the UEs placed at cell center, post cell center, pre cell-edge, and cell-edge regions.}
\label{fig:CellSINR}
\end{figure}

\textbf{Result observation:} The simulation cell size varies between $0.5~km^2$ to $5~km^2$. As the distance between the eNB and the UE gradually increases, a change in signal quality is observed.  Fig.~\ref{fig:CellSINR} shows the CDF plots of DL SINR values measured by the UEs throughout the cell, which range from $-10$ dB to $50$ dB. These SINR values recorded by the UEs fit between the range of excellent to cell-edge RF conditions as specified in Table~\ref{RSSIClassification}. The UEs when positioned nearby to the base station they experience better signal quality.

\begin{figure} [!htbp]
\centering
\includegraphics[width=8cm,height=14cm,keepaspectratio]{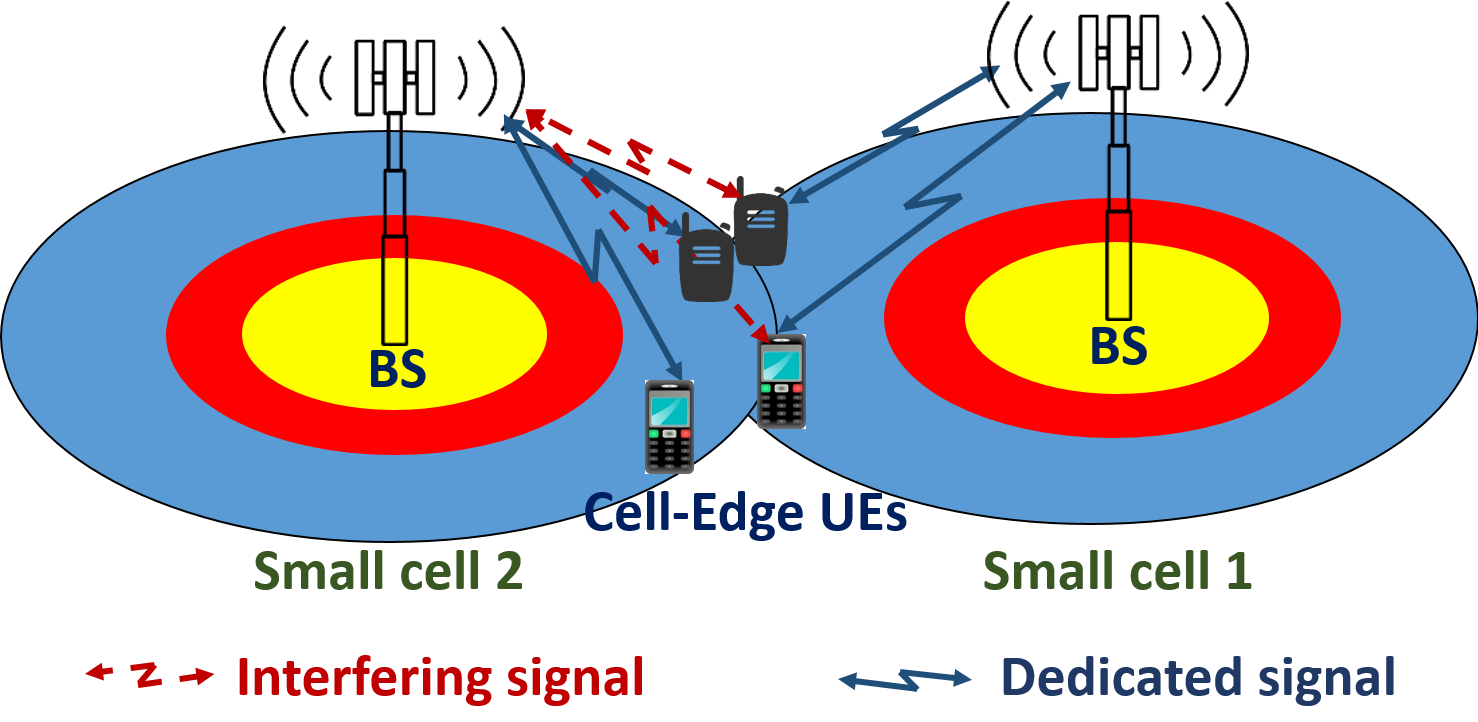}
\caption{Inter-cell interference simulation scenario with cell-edge UEs with interference from neighboring cell. Such a scenario may e.g. correspond to two fire trucks or police cars that utilize LTE small cells, and are parked next to each other during an emergency incident.}
\label{ICIScenario}
\end{figure}

\textbf{Simulation scenario-Inter-cell interference measurement:} The inter-cell interference in our simulation occurs in a homogeneous network with eNodeBs having the same capability, and same transmit power class. The inter-cell interference between two adjacent cell results in signal degradation and the UEs experience poor link quality at the cell-edge \cite{ici1,ici2, ici3}. In the simulation model of the inter-cell interference measurement, 20 UEs are placed beyond the mid-cell region, and along the cell-edge region. The simulation model includes, two homogeneous eNodeBs with fixed location and transmission power of 23 dBm, 20 UEs, round robin MAC scheduler, pedestrian fading model, MIMO 2-layer spatial multiplexing, a bandwidth of 10~MHz, and the other simulation parameters defined in Table~\ref{Table:BandClass14FreqParam}. The simulation scenario is as shown in Fig.~\ref{ICIScenario}, which shows two adjacent homogeneous small cells and cell-edge UEs. The cell-edge UEs in cell 1 observe interference from the neighboring cell 2. The main goal of inter-cell interference simulation is to measure the signal degradation observed by UEs at cell-edge in cell 1.

\begin{figure} [!htbp]
\centering
\includegraphics[width=1\linewidth]{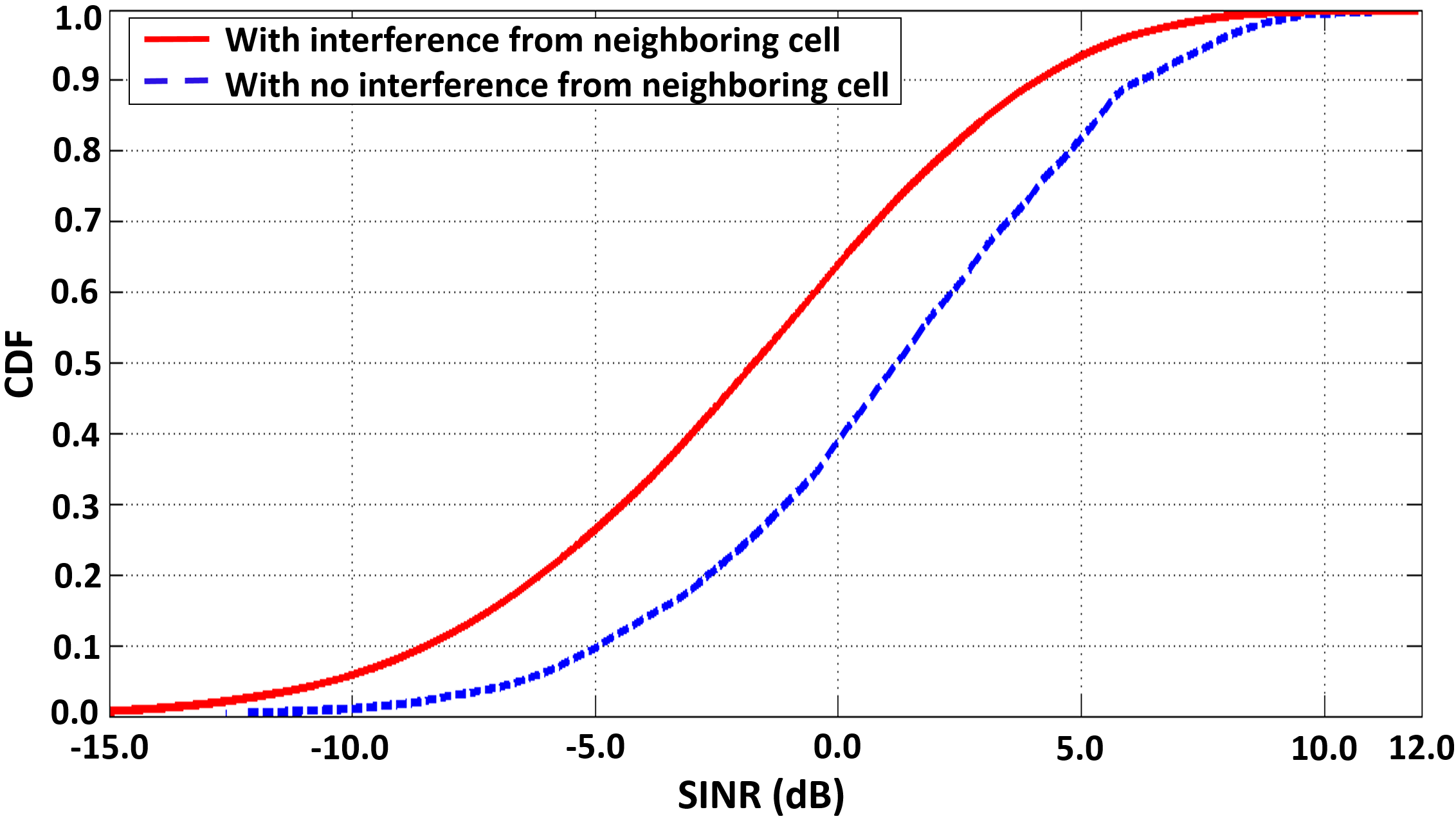}
\caption{The CDF plot of DL SINR values measured by the UEs placed on the edge of cell 1 of Fig.~\ref{ICIScenario}.}
\label{ICISimResult}
\end{figure}

\textbf{Result observation:} The intuitive analysis of Fig.~\ref{ICISimResult} conclude that the cell-edge UEs in Fig.~\ref{ICIScenario} observe signal degradation and experience poor link quality when compared to cell-edge UEs with no interference. The SINR values measured by the cell-edged UEs in both the scenario are plotted in Fig.~\ref{ICISimResult} in form of a CDF. The DL SINR values experienced by the UEs range from $-10$ dB to $10$ dB in the cell with no interference. On the other hand, the SINR values measured by the UEs in cell 1 fall in the range of $-15$ dB to $10$ dB due to the signal degradation caused by the adjacent cell.

\begin{figure}
        \centering
        \begin{subfigure}[b]{0.5\textwidth}
		        \centering
                \includegraphics[width=6cm,height=10cm,keepaspectratio]{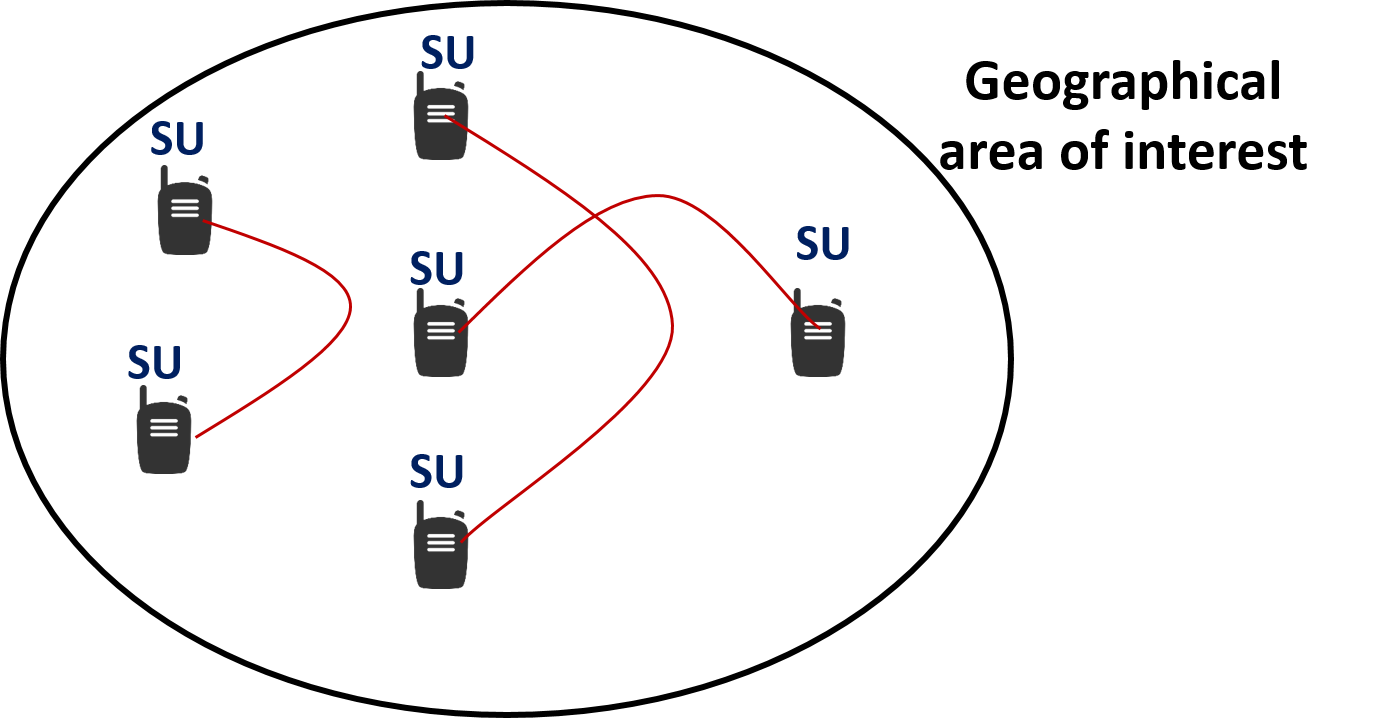}
                \caption{The lines indicate direct communication mode between the subscriber units.}
                \label{fig:DirectMode}
        \end{subfigure}

        \begin{subfigure}[b]{0.50\textwidth}
                \centering
                \includegraphics[width=6cm,height=10cm,keepaspectratio]{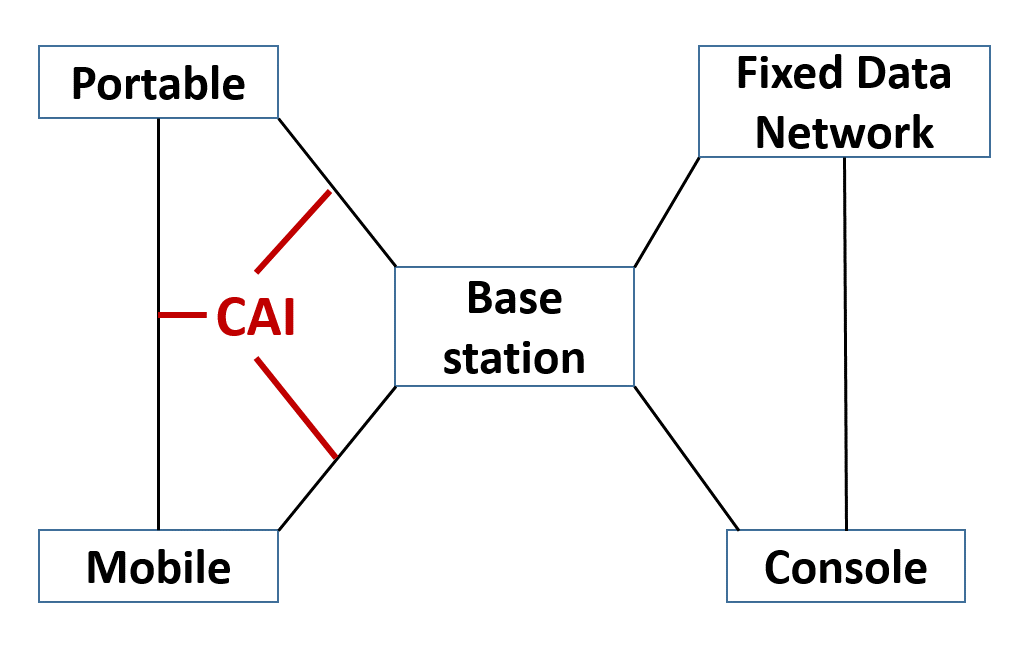}
                \caption{APCO-25 CAI digital voice modulation standard in a radio system.}
                \label{fig:CAI}
        \end{subfigure}
        \caption{ APCO-25 suite of standards for public safety communication referenced in NS-3 simulation.}
        \label{fig:APCO-25}
\end{figure}

\subsection{Setup for APCO-25}
\label{apcoNS3Sim}
The main  goal of APCO-25 simulation study is to compute the throughput, and the SINR distribution. Fig.~\ref{fig:APCO-25} shows the APCO-25 simulation environment implemented in this paper. The subscriber units in the simulation environment operate in direct mode, also known as talk-around, which enables EFR to have radio-to-radio direct communications by selecting a channel that bypasses their repeater/base systems. However this direct mode of communication limits the distance between mobile/portable subscriber units and spectrum reuse \cite{talkaround,P25RadioWiki}. A portable subscriber units can be described as a hand held device, while mobile subscriber units are mounted in EFR vehicle for mission-critical communication, to operates vehicle siren and lights. The APCO-25 Phase I and Phase II both can employ direct mode operations. Phase II TDMA systems require timing reference from the base station via system control channel and traffic channel for slot synchronization. The direct mode operation is challenging in case of Phase II TDMA due to the unavailability of master timing reference from the base station or a repeater \cite{P25RadioWiki}. In direct operation mode, as the distance between the subscriber units increases, the SINR value decreases.

The APCO-25 common air interface (CAI) is a specified standard for digital voice modulation used by the compliant radios as shown in Fig.~\ref{fig:CAI}. Public safety devices form different vendors using APCO-25 CAI can communicate with each other. APCO-25 CAI has a control channel with a rate of 9600 bps and it uses improved multi-band excitation (IMBE) for voice digitization. The IMBE voice encoder-decoder (vocoder) samples the audio input into digital stream for transmission. At the receiver, vocoder produces synthetic equivalent audio from the digital stream \cite{P25RadioWiki, cai}.

\textbf{Simulation scenario-APCO-25:} The NS-3 simulation for APCO-25, consist a square simulation area and with a uniform random distribution of 20 subscriber units across the cell. These 20 subscriber units in direct communication mode use APCO-25 CAI for digital voice modulation and Friis propagation model. The portable subscriber units have transmission power of 5 Watts, mobile subscriber units have transmission power of 10 Watts.

\begin{figure} [!htbp]
\centering
\includegraphics[width=8cm,height=14cm,keepaspectratio]{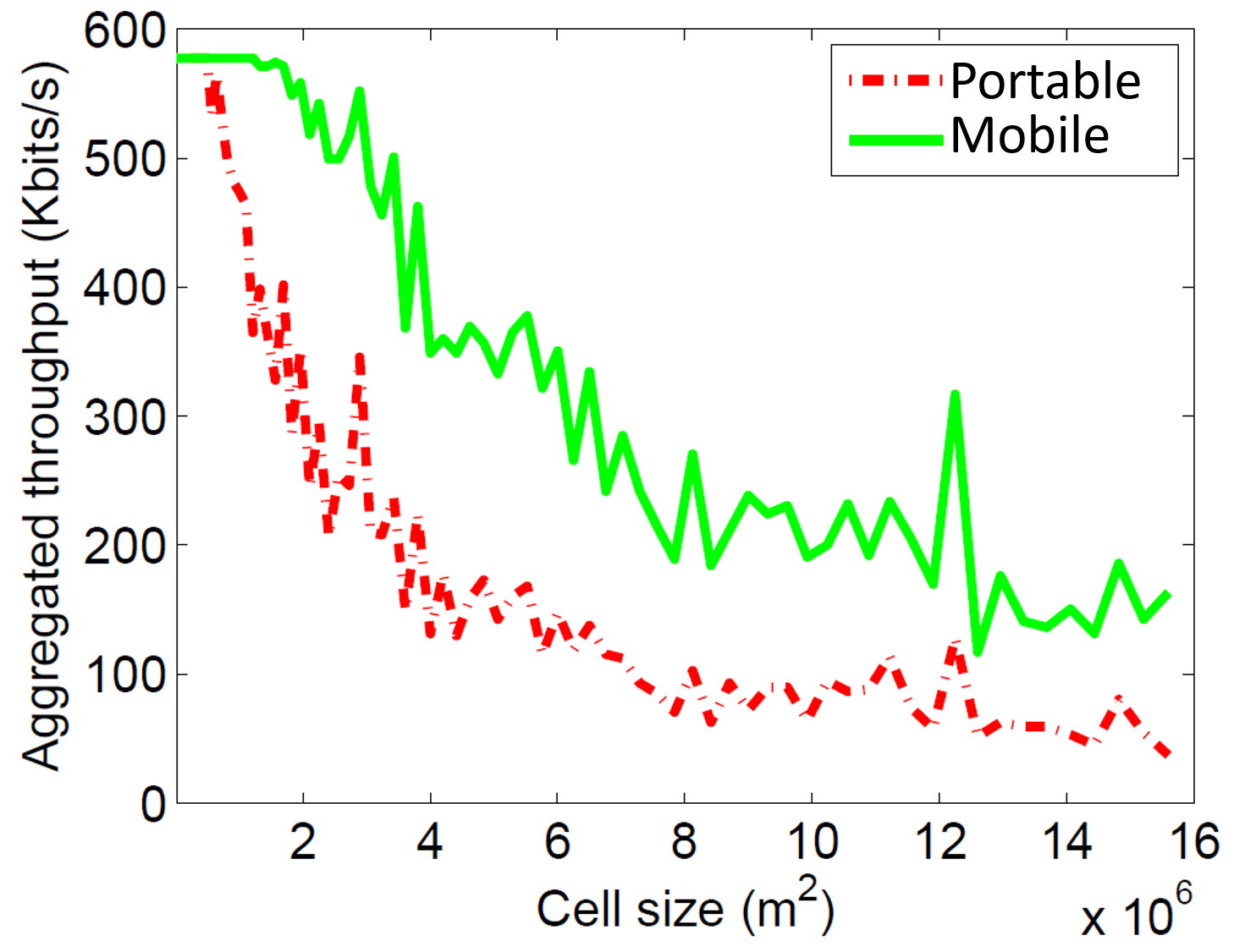}
\caption{The maximum observed throughput in case of data communication is approximately $580$ Kbits/s for both portable and mobile public safety devices.}
\label{LMRSThroughput}
\end{figure}

\begin{figure} [!htbp]
\centering
\includegraphics[width=8cm,height=14cm,keepaspectratio]{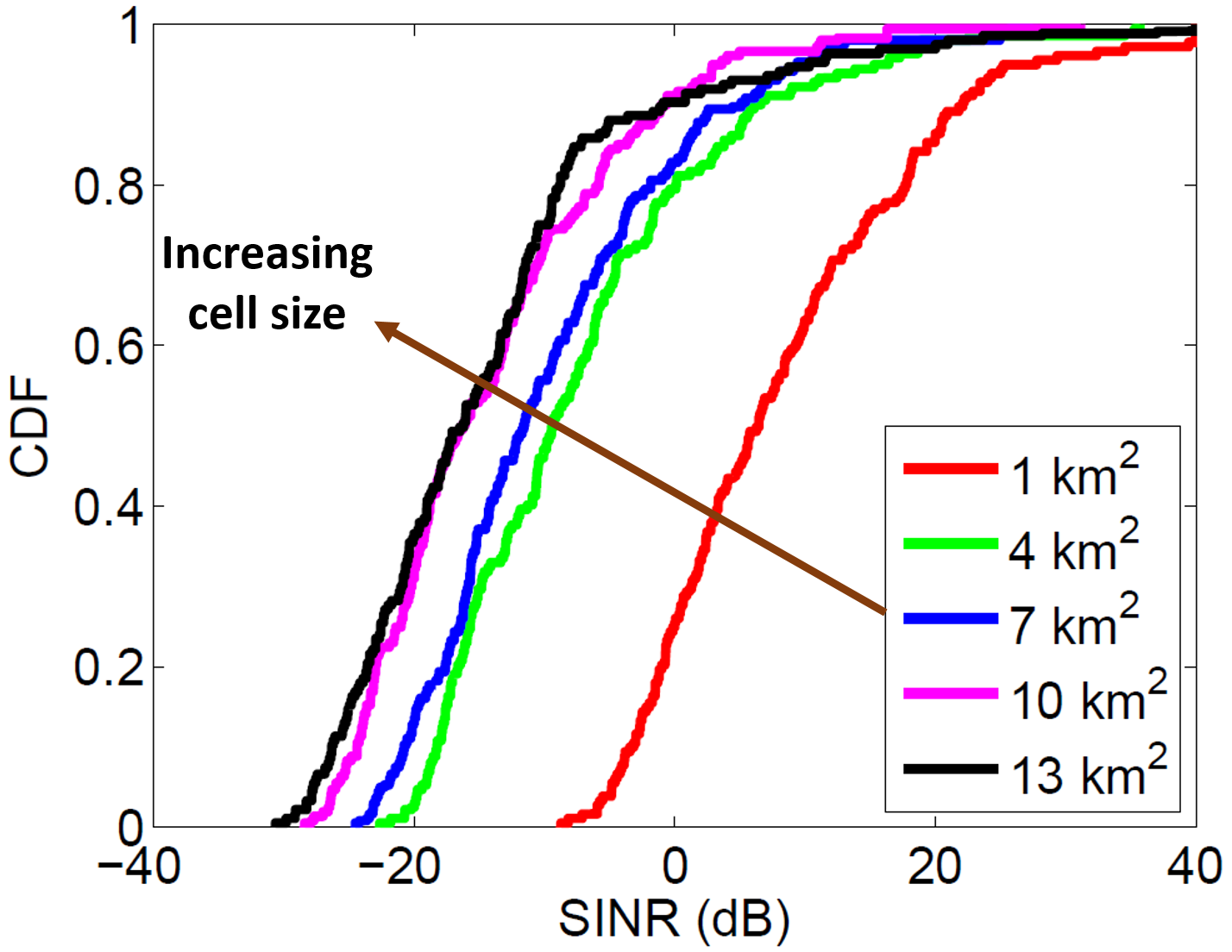}
\caption{SINR CDF plots for APCO-25 portable subscriber units, with increasing distance between the subscriber units.}
\label{portablesinr}
\end{figure}

\begin{figure} [!htbp]
\centering
\includegraphics[width=8cm,height=13cm,keepaspectratio]{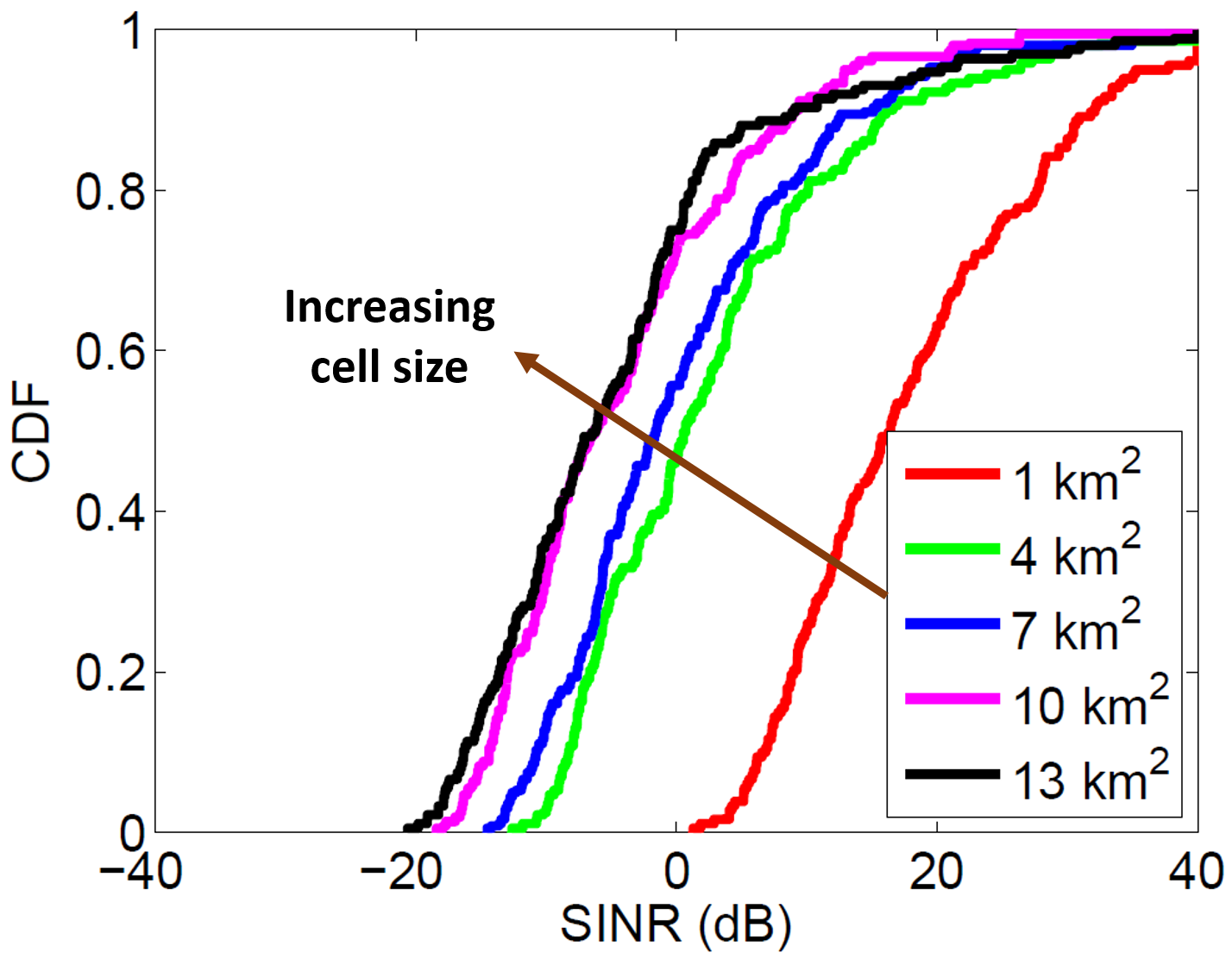}
\caption{SINR CDF plots for APCO-25 mobile subscriber units, with increasing distance between the subscriber units.}
\label{mobilesinr}
\end{figure}

\textbf{Result observation:} In the direct communication mode, maximum throughput is possible when subscriber units are in proximity.  As the geographical distance between the portable and mobile public safety devices increases, the throughput decreases as shown in Fig.~\ref{LMRSThroughput}. The simulation of APCO-25 for data only system and direct communication mode demonstrates an aggregate throughput of $580$~Kbits/s as shown in Fig.~\ref{LMRSThroughput}. Whereas, Fig.~\ref{portablesinr} and Fig.~\ref{mobilesinr} showcases the SINR observed by the portable and mobile public safety devices in the geographical area of interest. From Fig.~\ref{portablesinr} and Fig.~\ref{mobilesinr} it is observed that the SINR values decreases with the increasing distance between two portable and mobile subscriber units. The observed SINR values range from $-30$~dB to $40$~dB for portable devices, whereas mobile devices observe SINR value between $-20$ dB and $40$~dB.

Since the SINR experienced by the communicating subscriber units depends on the proximity between the subscriber units, it can be concluded that the SINR value decreases with increasing remoteness between the communicating subscriber units.  Furthermore, it can also be also noticed, the increasing distance and poor link quality lead to dropped packet which results in lower aggregated cell throughput.  The mobile subscriber units have relatively higher transmission power compared to portable subscriber units. Therefore Fig.~\ref{LMRSThroughput} also conclude that the mobile subscriber units demonstrate relatively higher throughput when compared to portable subscriber units. Furthermore, mobile subscriber units experience better SINR compared to portable subscriber units as shown in Fig.~\ref{portablesinr} and Fig.~\ref{mobilesinr}.

\begin{table*}[!htbp]
\centering
\caption{Results summary of NS-3 simulation for LTE band class~14 and APCO-25.}
\begin{tabular}{|p{5.0cm}|p{12.4cm}|}
\hline \textbf{Simulation scenario} &\textbf{Results observation} \\
\hline Band class~14: aggregate throughput computation & The main goal of this simulation scenario is to compute aggregated data throughput for DL and UL. As observed, the peak aggregated throughout for DL is $33~Mbit/s$ and for UL is $11~Mbit/s$ as seen in Fig.~\ref{UlAndDlThroughput}.\\
\hline Band class~14: signal quality measurement & The main goal of this simulation scenario is to measure SINR experienced by UEs in a cell. As observed in Fig.~\ref{fig:CellSINR}, the signal quality degrades with increasing distance between eNB and UE. The UEs experience SINR value between $-10~dB$ to $50~dB$.\\
\hline Band class~14: inter-cell interference measurement & The main goal of this simulation is to calculate the SINR experienced by the UE in cell-edge region with and without interference from the adjacent cell. The SINR experienced by UEs range from $-10~dB$ to $10~dB$ when no interference present. Whereas, in case of interference from adjacent cell the UEs experience lower link quality and range from $-15~dB$ to $10~dB$ as seen in Fig.~\ref{ICIScenario}.\\
\hline APCO-25: aggregate throughput computation & The main goal of this simulation scenario is to compute aggregated data throughput for portable and mobile subscriber units in direct communication mode. A peak throughput of $580~Kbits/s$ is observed in case of both portable and mobile subscriber units as seen in Fig.~\ref{LMRSThroughput}.\\
\hline APCO-25: signal quality measurement & The main goal of this simulation scenario is to measure SINR experienced by portable and mobile subscriber units in a cell. As observed in Figs.~\ref{portablesinr} and \ref{mobilesinr}, the signal quality degrades with increasing distance between eNB and UE. The portable devices experience SINR values ranging from $-30$~dB to $40$~dB, whereas mobile devices observe SINR value between $-20$ dB to $40$~dB.\\  
\hline
\end{tabular}
\label{table:ns3SimulationSummary}
\end{table*}

\subsection{Result Analysis}
\label{NS3SimResultAna}
The key factors for successful PSC can be broadly classified into mission-critical voice communication and wider coverage for public safety devices. From the comparison of Figs.~\ref{fig:CellSINR},~\ref{portablesinr},~and~\ref{mobilesinr}, it can be deduced that LMRS public safety devices experience larger cell coverage when compared to the band class 14 LTE devices. This volume of cell coverage is critical during the emergency scenario and is necessary to cover the maximum amount emergency prone area. Another rising factor in mission-critical PSC is real-time video communications, high data rate is needed to support real-time multimedia applications \cite{IMS}.  The aggregated throughput of LMRS infrastructure is less when compared to the aggregated throughput of band class 14 infrastructure and is shown in Fig.~\ref{UlAndDlThroughput} and Fig.~\ref{LMRSThroughput}. These data rates experienced by the LTE UEs can definitely support the much needed real-time video communications in several of the PSC scenarios.

An adaptation of successful edition of mission-critical PTT over LTE would take some time \cite{whenMcLTE}. In the meantime, applying the individual strength of LMRS and LTE into a converged public safety device to mission-critical voice and much needed mission-critical real-time data support can be beneficial to PSC.

\section{LMRS vs. LTE: SDR Experiments}
\label{SectionVIII}
This section discusses the experiments conducted using SDR receiver such as the RTL-SDR RTL2832U \cite{rtl2832u} and HackRF \cite{hackrf}. SDR is an implementation of a radio communication system where the components that have normally implemented in hardware such as amplifiers, mixers, filters, modulators/demodulators, and detectors are instead implemented by means of software \cite{sdr1,sdr2,sdr3}. Most SDR receivers use a variable-frequency oscillator, mixer, and filter. The flexible SDR receiver helps in tuning to the desired frequency or a baseband \cite{sdr1}. These flexible tuning characteristics can be used to tune into public safety signals.

SDR provides a low-cost infrastructure and more or less computer-driven environment which makes it an integral part of the public safety research, development, and test activities \cite{sdrpsc,sdrpsc1}. The continual development of SDR related hardware and software have assisted PSC researchers in gaining a better understanding of public safety signals. Furthermore, in this section we capture and analyze LMRS and public safety LTE signals and is shown via examples.

\begin{figure} [!htbp]
\centering
\includegraphics[width=1\linewidth]{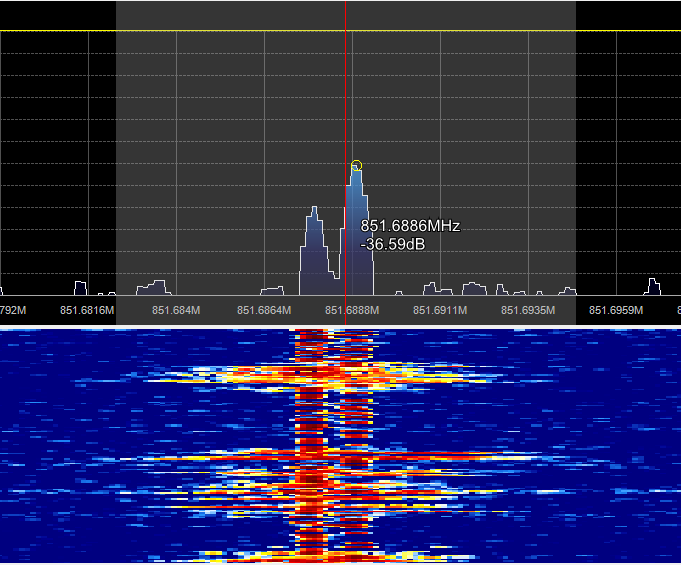}
\caption{Broward County, FL, APCO-25 spectrum monitored using SDR\# software and captured using a HackRF receiver \cite{sdr4,hackrf}.}
\label{fig:CoralSpringsP25Spectrum}
\end{figure}

\subsection{APCO-25}
The setup for the detection of APCO-25 signals consists HackRF connected to a personal computer equipped with a sound card and installed with SDRSharp freeware tool \cite{sdr4, sdr41}. This setup is used as radio scanner to observe unencrypted APCO-P25 digital radio voice spectrum using the instructions provided in \cite{sdr5}. Fig.~\ref{fig:CoralSpringsP25Spectrum} shows the observed Broward County public safety radio spectrum for frequency of 851.68 MHz with signal strength of $-36.59$ dB. The second embedded window is a spectrogram which is a visual representation of the frequency spectrum of the received signal. HackRF has three receiver gain i.e., RF amplifier between $0$ dB or $14$ dB, IF low noise amplifier $0$ dB to $40$ dB, and baseband variable gain amplifier between $0$ dB to $62$ dB. RF amplifier gain is set high to receive string signal, while keeping noise floor as low as possible. In this experiment the RF amplifier gain is set to 9 dB. The two primary factors that describes window function are width of the main lobe and attenuation of the side lobes. 4-term Blackman-Harris window is better suited for  spectral analysis \cite{harris1978use} and therefore used in the current APCO-25 SDR experiments.

\begin{figure}[!htbp]
\centering
\includegraphics[width=1\linewidth]{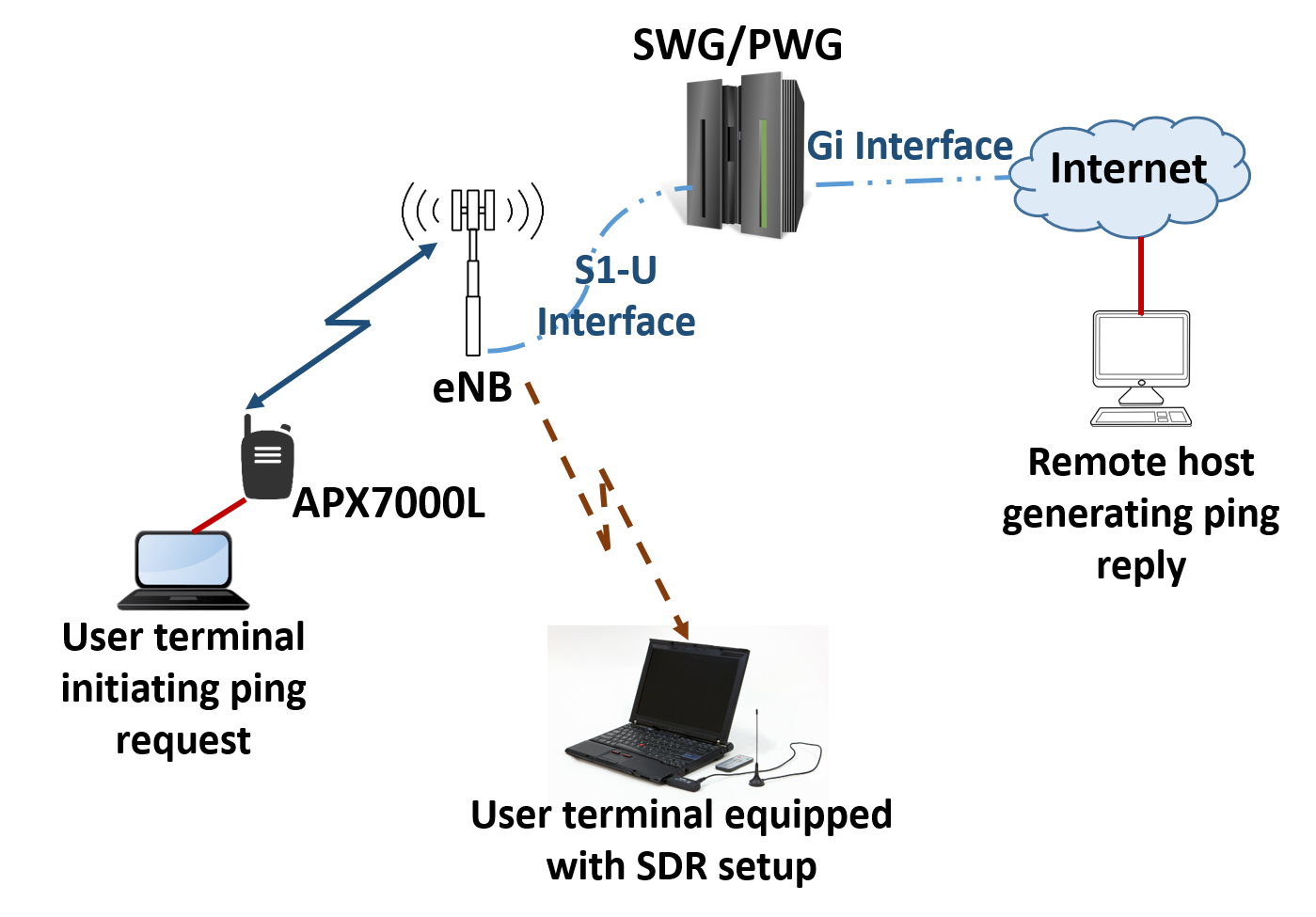}
\caption{Setup for analyzing band class 14 spectrum.}
\label{SetupBandclass14Spectrum}
\end{figure}

\begin{figure}[!htbp]
\centering
\includegraphics[width=1\linewidth]{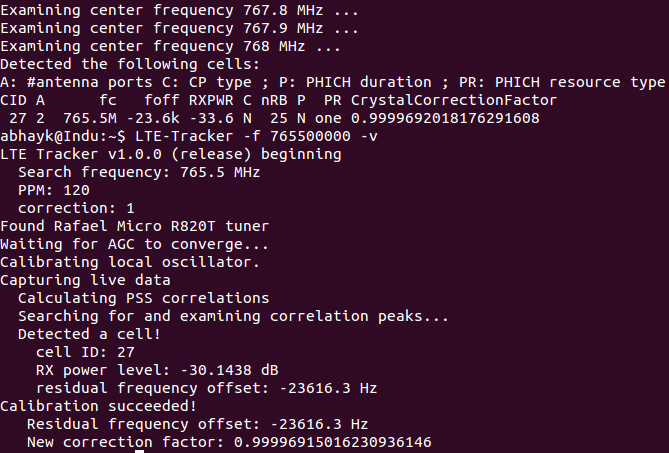}
\caption{Band class 14 cell detection using LTE-Tracker open source tool and RTL-SDR \cite{rtl2832u, sdr6}.}
\label{SigDectBandclass14Spectrum}
\end{figure}

\begin{figure}
        \centering
        \begin{subfigure}[b]{0.5\textwidth}
                \includegraphics[width=1\linewidth]{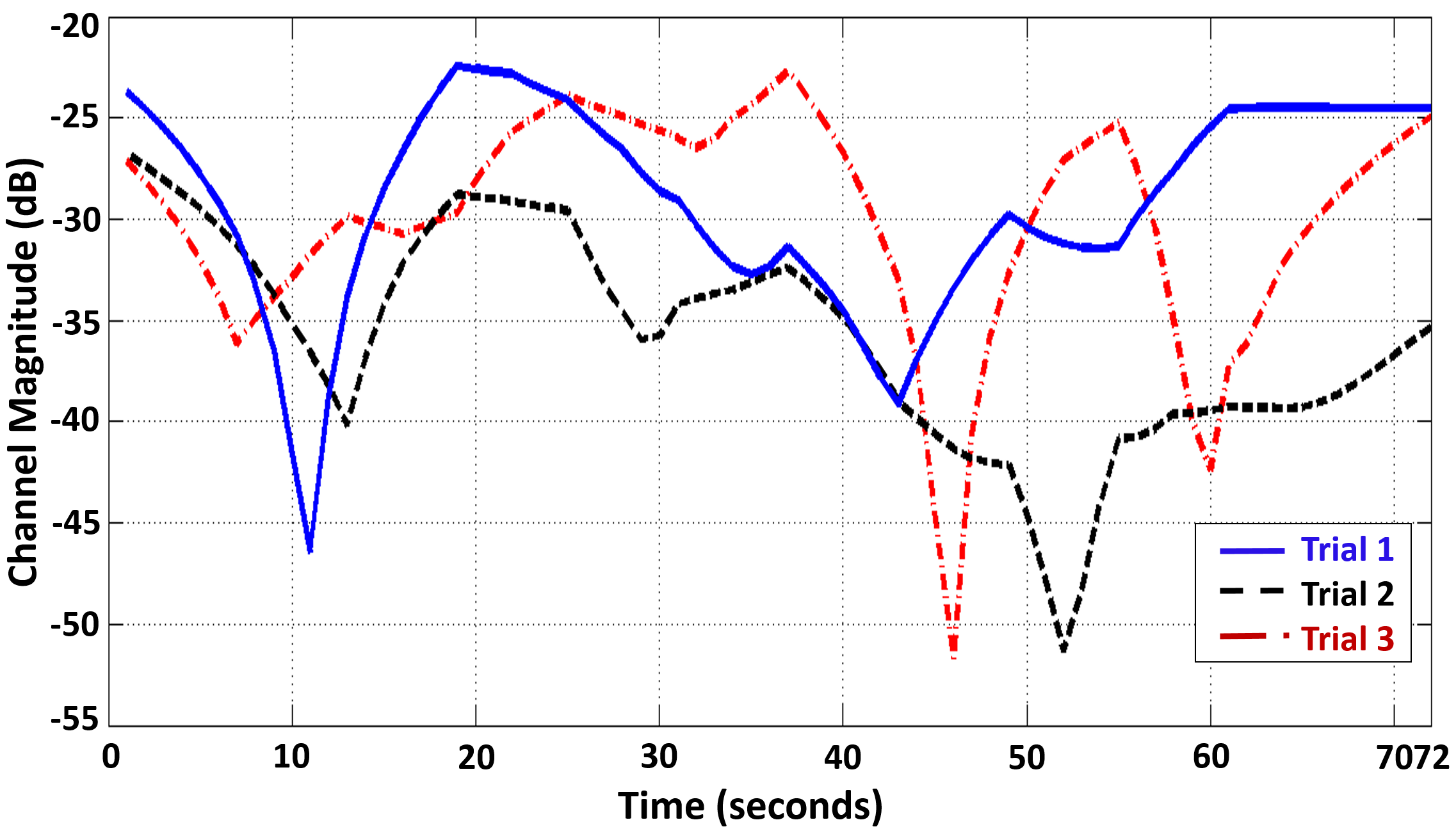}
                \caption{Magnitude plot for the detected cell.}
                \label{fig:magnitude}
        \end{subfigure}

        \begin{subfigure}[b]{0.5\textwidth}
                \includegraphics[width=1\linewidth]{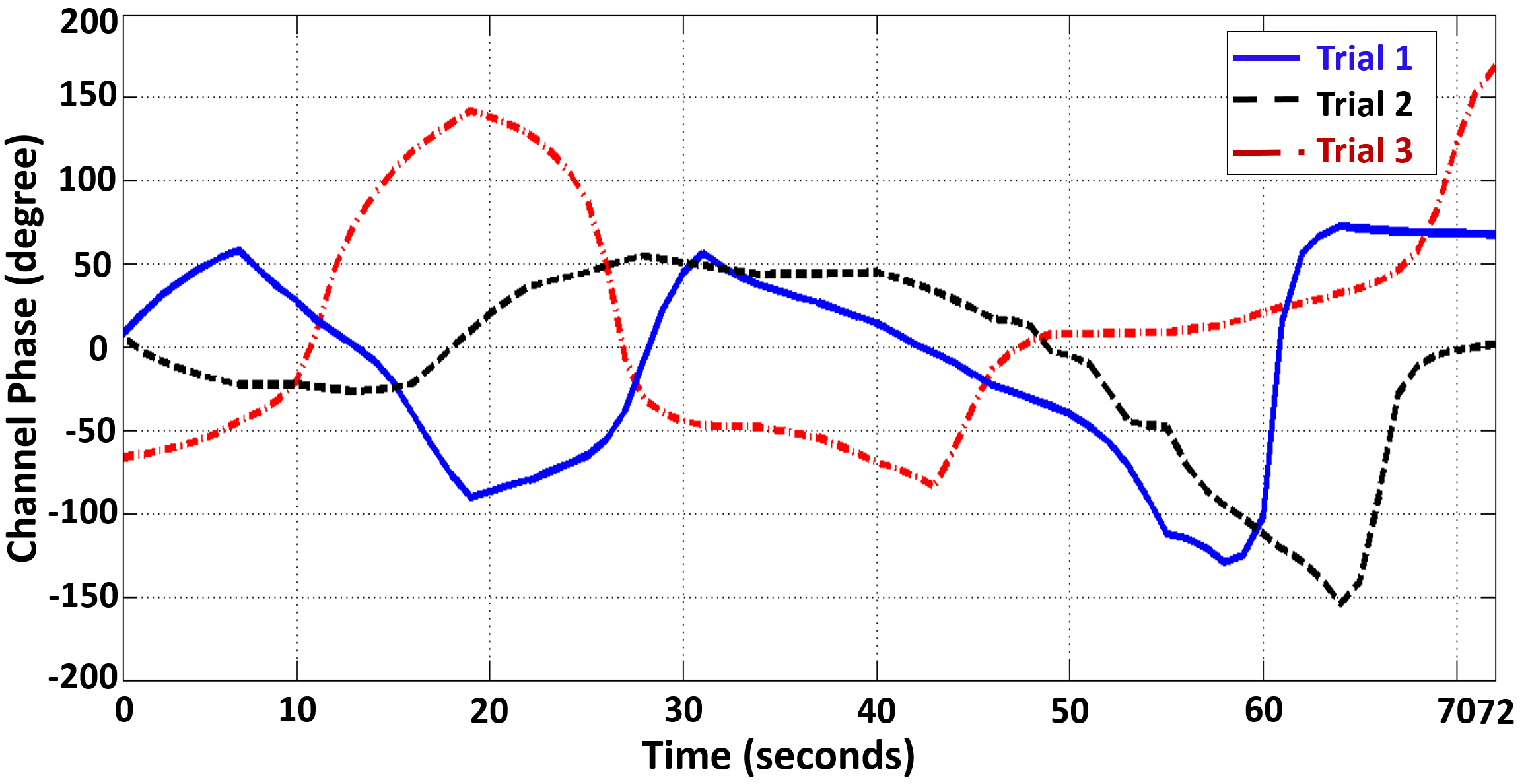}
                \caption{Phase plot for the detected cell.}
                \label{fig:phase}
        \end{subfigure}
        \caption{LTE cell tracker transfer function observed during various RTL-SDR experimentations. The variation in the channel magnitude and phase is observed due to the mobility of user terminal equipped with SDR setup.}\label{bandClass14TF}
\end{figure}

\subsection{Band Class 14}
The SDR experiment for analysis of band class 14 signals is conducted using the Motorola Solutions band class 14 infrastructure. The band class 14 base station in this scenario has DL frequency range of 758 MHz to 768 MHz whereas the UL frequency range is 788 MHz to 798 MHz as specified in Fig.~\ref{fig:BandClass14Plan}. Motorola APX7000L, an LTE-LMR converged device is treated as a UE in this scenario. Ping requests are initiated over a user terminal is routed via APX7000L to reach the remote host which generates the ping replies. An another user terminal equipped with SDR setup is used to capture these band class 14 signals as shown in Fig.~\ref{SetupBandclass14Spectrum}. The user terminal equipped with SDR setup is running an open source LTE cell scanner and LTE tracker \cite{sdr6} tool to identify and track the band class 14 cell. Fig.~\ref{SigDectBandclass14Spectrum} shows the identified band class 14 cell with cell ID 27, with two transmit antennas, center frequency of 765.5 MHz, and receive power level of $-33.6$ dB. Overall 25 resource blocks are occupied by the cell (channel bandwidth of 5 MHz), while the PHICH has normal duration and the resource type one. The transfer functions of the wireless channel from that port on the BS to the dongle's antenna are shown in Fig.~\ref{bandClass14TF}. Fig.~\ref{fig:magnitude} shows the instantaneous magnitude transfer function of port 0 of cell 27 whereas Fig.~\ref{fig:phase} shows the phase plot.

These analyses of LTE band class 14 signals can assist PSC researchers in detecting coverage holes in PSN and gain knowledge on the quality of service experienced by the UEs.

\section{Emerging Technologies for PSC}
\label{SectionIX}
LTE technology has been standardized by the 3GPP in 2008, and has been evolving since then. In response to growing commercial market demands, LTE-Advanced was specified as 3GPP Release 10. 3GPP has developed Release 11 and Release 12 with the aim of extending the functionality and raise the performance of LTE-Advanced \cite{LTETimeline}.
With a vision of 2020, currently LTE has become the target platform for machine-to-machine communications, PSC, and device-to-device services \cite{mallinson2020}. Since the deployment of LTE, 4G is reaching its maturity and relatively incremental improvements are to be expected. The 5G technology is expected to have fundamental technological components that will transform the capabilities of broadband networks \cite{5G}. For example, full-fledged efforts from researchers at the University of Surrey's 5G Innovation Centre managed to attain one terabit per second (Tbps) of data speed \cite{5GTbps}. The three big technologies described in \cite{5G} that will be shaping 5G are ultra-densification of base station deployments, millimeter wave (mmWave) communications, and massive MIMO.

This section of paper is an attempt to emphasize the potential of the technological advancement achieved through LTE, LTE-Advanced, and 5G for transforming the PSC capabilities. In particular, use of  eMBMS, millimeter wave, massive MIMO techniques, small cells, unmanned aerial vehicles (UAVs), LTE-based V2X, unlicensed operation of LTE in the context of PSC, cognitive radio, wireless sensor networks, internet of things, and cybersecurity for PSN will be reviewed.

\subsection{Multimedia Broadcast Multicast Service (MBMS)}

\begin{figure}[!htbp]
\centering
\includegraphics[width=5cm,height=6cm,keepaspectratio]{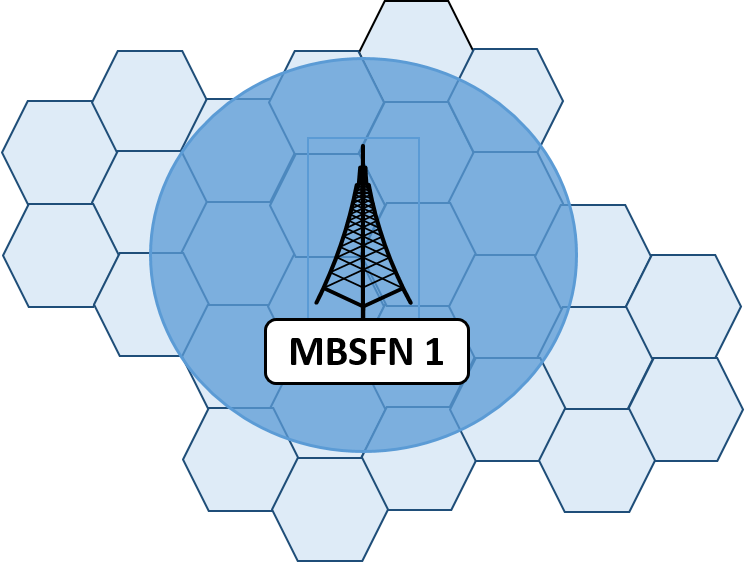}
\caption{In a single frequency network, base stations transmit the same signal at the same time and over the same frequency channel to UE. In this example MBSFN area, the group of cells perform synchronized eMBMS transmission. These transmitted signals appear as multipath components to the UE.}
\label{fig:eMBMS}
\end{figure}

The MBMS is a point-to-point interface specification for upcoming 3GPP cellular networks \cite{mbmsCite, monserrat2012joint,3gpp.22.246}. The MBMS is designed to provide efficient delivery of broadcast and multicast services with a cell and core network. Broadcast transmission across multiple cells is via single-frequency network configuration. The MBMS feature can be divided into the MBMS Bearer Service and the MBMS User Service \cite{monserrat2012joint,3gpp.22.246}. Support for evolved MBMS (eMBMS) will be also included in Release 13, which has important applications for PSC. The eMBMS is designed to improve the efficiency of legacy MBMS. The LTE technology is backward compatible with legacy 3G MBMS and supports new features of the broadcast networks such as digital video broadcast-terrestrial networks.

The eMBMS is a point-to-multipoint interface where multiple LTE cells can be grouped into Multicast/Broadcast Single Frequency Network (MBSFN) and all of the cells in an MBSFN will simulcast the same information for each bearer as shown in Fig.~\ref{fig:eMBMS}. The eMBMS broadcast bearers provide advantages of unlimited group sizes within a cell, better cell edge performance due to simulcast, and is independent of the talkgroups size. The eMBMS would provide PSC services with an ability to use a single set of resources for the traffic destined to multiple public safety devices. The public safety eMBMS broadcast applications would be group calls, PTT, real-time emergency video broadcast, surveillance video broadcast, and non-real-time multimedia services. The eMBMS broadcast coverage improvement would be similar to that of LMRS simulcast\cite{mbms2,mbms3}.

\subsection{Millimeter Wave (mmWave)}
\label{mmWave}
The severe spectrum shortage in conventional cellular bands has attracted attention towards mmWave frequencies that occupy the frequency spectrum for 30 GHz to 300 GHz. It is also considered as a possible candidate from next-generation small-cellular networks \cite{akdeniz2014millimeter,mmWave}. The availability of huge bandwidth in the mmWave spectrum can alleviate the concerns of wireless traffic congestion which is observed in conventional cellular networks \cite{mmWave1,niu2015survey}. The mmWave transceivers can cover up to about a kilometer communication range in some scenarios \cite{mmWave1}. Moreover, densely laid out mmWave small cells in congested areas can expand the data capacity and provide a backhaul alternative to cable \cite{bleicher2013millimeter}. The mmWave's ability opens up possibilities for new indoor and outdoor wireless services \cite{mmWave}.

However, the key limitation of the mmWave technology is its limited range, due to the excessively absorbed or scattered signal by the atmosphere, rain, and vegetation. This limitation causes the mmWave signals to suffer very high signal attenuation, and to have reduced transmission distance. The researchers and scientists have overcome this loss with good receiver sensitivity, high transmit power, beamforming, and high antenna gains \cite{akdeniz2014millimeter, mmWave, bleicher2013millimeter, mmWave1, mmWave2, deng201528, sulyman2014radio, kuttybeamforming}.

A possible solution for extending the mmWave range is proposed in \cite{thomas2015millimeter}. Due to limited propagation range of mmWave, a reliable communication over mmWave requires a substantial number of access points. The authors discuss the transmission of radio frequency signals between the mmWave small cell base stations and radio access points using optical fibers to increase the coverage. This radio over fiber communication is possible by applying advanced optical upconversion techniques. In \cite{kuttybeamforming}, various beamforming techniques are introduced for indoor mmWave communication.

A well augmented mmWave system can position itself as a reasonable solution for PSC needs, by addressing the concerns and requirements such as network congestion and high data rates. A densely populated mmWave small-cell networks can also address the coverage needs during emergency situations.

\subsection{Massive MIMO}
\label{massiveMIMO}
MIMO is an antenna technology that can be used to reduce communication errors  and scale the capacity of a radio link using multiple transmit and receive antennas. Recently massive MIMO is an emerging technology that uses large number of service antennas when compared to current MIMO technology \cite{larsson2013massiveMimo}. Since the introduction of massive MIMO in \cite{marzetta2010noncooperative}, some researchers have shifted focus on solving related traditional problems \cite{jose2011channel, larsson2013massiveMimo, gopalakrishnan2011analysis, mopidevi2011compact, elijahcomprehensive}.

Massive Massive MIMO is typically considered in time-division duplex systems to exploit channel reciprocity since it eliminates the need for channel feedback \cite{hoydis2011massive,elijahcomprehensive}. However, the massive MIMO system is limited by pilot contamination, which is an impairment that is observed during channel estimation stage in TDD networks. The pilot contamination is caused due to the non-orthogonal nature transmission of pilots from neighboring cells \cite{elijahcomprehensive,hoydis2011massive}. Nevertheless, the effects of pilot contamination can be reduced by using mitigation techniques such as pilot-based and subspace-based estimation approaches \cite{elijahcomprehensive}.

Overall, massive MIMO can provide benefits such as 100$\times$ radiated energy efficiency, 10$\times$ capacity increase over traditional MIMO, extensive use of inexpensive low-power components, reduced latency, simplification of the medium access control, spectrum efficiency, reliability, and robustness \cite{larsson2013massiveMimo, gopalakrishnan2011analysis, elijahcomprehensive}.

Massive MIMO can be used for achieving high throughput communications in an emergency situation. For example, real-time situational awareness via high definition multimedia can be enabled via massive MIMO techniques. Massive MIMO can also provide better penetration into buildings due to directional transmissions. This ability of deep penetration can be utilized by the EFR during emergency scenario such as building fire.

\subsection{Small Cells}
Heterogeneous network (HetNet) topology blends macrocells, picocells, and femtocells. HetNet topology aims at providing uniform broadband experience to all users ubiquitously in the network \cite{smallCell1,smallcell2,smallCell3GPP,sui2015interference}. Shrinking cell size benefits by reuse of spectrum across the geographical area and eventually reduction in number of users competing for resources at each base station. Small cells can be deployed both with standalone or within macro coverage, and they can be sparsely or densely distributed  in indoor or outdoor environments\cite{smallCell1}. They transmit at substantially lower power than macro cells and can be deployed to improve the capacity in hotpots, eliminate coverage holes, and ease the traffic load at the macro base station \cite{smallcell2}.

\begin{figure} [!htbp]
\centering
\includegraphics[width=7cm,height=8cm,keepaspectratio]{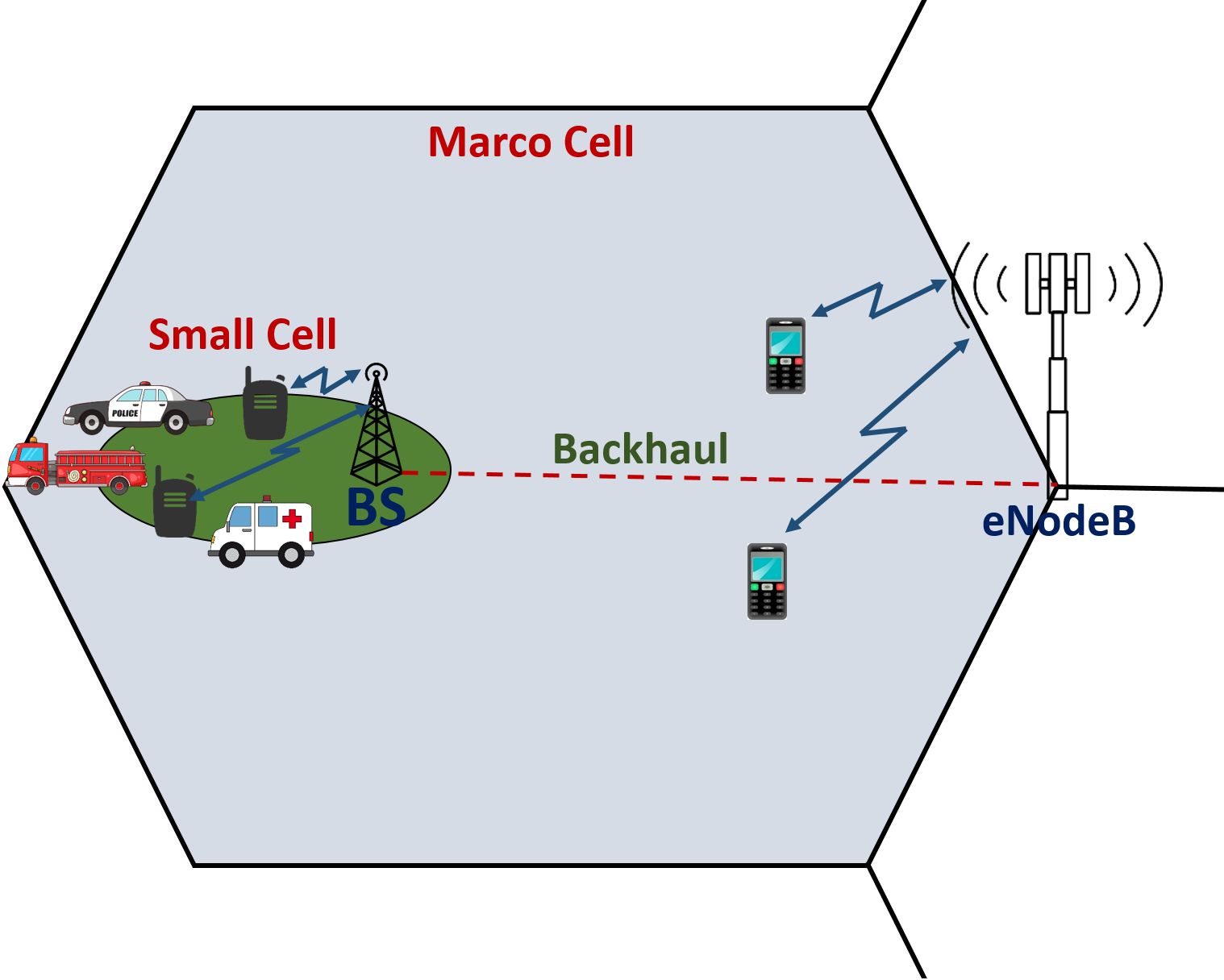}
\caption{Small cell example for PSC during an emergency scene.}
\label{fig:SmallCellPSC}
\end{figure}

LTE based small cells can be deployed in an emergency scenario to improve PSC between EFR as shown in Fig.~\ref{fig:SmallCellPSC}. This hotspot will improve the capacity of PSC, and as a result EFR do not have to compete with general public for eNodeB resources. The base stations used as small cells in such scenario can be portable/deployable systems. The backhual to these base stations can be via the closest eNodeB, via satellite link, or via its own dedicated backhaul. A flexible and low cost PSC deployment strategy using small cells can be effective and enhance the ability of the EFR to save lives.

The \textit{indoor LTE coverage} is becoming a pressing concern as LTE deployments evolve. For instance, EFR would need communication coverage in various parts of a building such as the basement. However, building basements may have weak coverage since the existing LTE deployments does not always consider it crucial. This issue can be resolved using LTE mini-towers to boost indoor communication coverage. Nevertheless, indoor coverage remains ones of the pitfalls for LTE and needs to be addressed for an effective broadband PSN \cite{LTEIndoorCoverageIssue}.

\subsection{Unmanned Aerial Vehicles (UAVs)}
In the United States, Hurricane Sandy affected 24 states, and particularly severe damage was observed in NJ and NY. Electrical power, terrestrial communication systems, and utilities cease to exist due to devastation caused by the Hurricane Sandy. Along with disruption of PSC and commercial wireless services, high winds caused damage to point-to-point backhaul links and fiber-optic cables were cut. The service disruption was observed over at least 3-5 days before any services began to restore. Hurricane Sandy exposed flaw in public-safety LTE plan as discussed in \cite{hurricaneSandyPSLTEflaw}, due to loss of commercial cell sites and no backup coverage from commercial LTE carriers.

The deployment of LTE-equipped UAVs in such an emergency scenario would provide necessary voice and data coverage \cite{merwaday2015uav, gomez2013realistic, arvindUavPS}. In \cite{UAVintoPSLTE}, a similar use case is discussed where LTE-equipped UAVs is deployed as a relay node. UAV will share the same bandwidth allocated for public safety broadband network and PSN could also backhaul the traffic from UAV. The concept of using UAV from wireless communication has been a proven technology. UAV are easy to deploy and can quickly restore critical communications during an emergency operation.  Alcatel-Lucent conducted a test with an ad-hoc LTE network using 4G smartphone and USB Modem. The UAV flew one kilometer while streaming 720p video \cite{UAVLTEVideo}. Higher heights may occasionally result in line-of-sight and stronger connectivity. UAVs have limited battery powers, and therefore may have limited flight times, which may introduce operational constraints.

The use of UAVs in PSC networks pose several challenges that need to be resolved. For example, mobility model, topology formation, energy constraints, and cooperation between UAVs in a multi-UAV network are surveyed in \cite{guptasurvey}. Furthermore, spectrum scarcity specific challenges and integration of cognitive radio into UAVs are addressed in \cite{saleem2015integration}.

UAVs can complement the capabilities of PSN and can operate in the same public safety spectrum block. They can also achieve a radio link with nearest functioning eNodeB \cite{UAVintoPSLTE}. They can be deployed in high-risk areas to assist EFR and used as a relay node to extend network coverage, as discussed in\cite{daniel2009airshield, DronePS, gomez2013realistic,DusanUavPsc}. With clearer rulemaking and policies on usage of UAVs in civilian space \cite{bennett2014civilian}, UAVs can be inducted into PSN and making it a viable solution in filling coverage holes.

\subsection{LTE-based V2X Communication}
LTE-based V2X communication involves the communication of information from vehicle into devices of interest, and vice versa. The V2X communication has many applications such as navigation, driver assistance, travel information, congestion avoidance, and fleet management. It can occur in the form of vehicle to vehicle (V2V), vehicle to infrastructure (V2I), and vehicle to pedestrian (V2P) communication \cite{V2X_nokia,V2X1,V2X2,V2X3,chen2015vehicular}. The ProSe can find good application in V2X communication. In Release 13, the ProSe has been designed for pedestrian mobility speed, and further research and standardization is needed for extensions to be used directly for V2X \cite{V2X3}.

V2X communication can also find application in PSC. In an emergency situation Firefighters, police cars, ambulances vehicles can broadcast/multicast information to other traffic elements and pave way to the EFR vehicles and avoid any traffic congestion. The V2X communications can also be used during disaster management such as city evacuation. The vehicular ecological system can gather information such as accident points and traffic patterns from multiple sources. Furthermore, V2X communication can also broadcast this information to other vehicular nodes. The effective V2X communication can play critical role during disaster management and minimize loss to human life \cite{alazawi2011intelligent,eltoweissy2010towards}.

\subsection{License Assisted Access (LAA)}
With the limited spectrum availability and ever increasing traffic demands, researcher and cellular companies are motivated to find complementary solutions to the licensed spectrum.
Although the licensed spectrum is preferable to provide better user experience, exploring LTE in unlicensed spectrum can lead to higher spectral efficiency \cite{LAA_1}.
LTE in the unlicensed band (LTE-U) can complement the services provided by the LTE in licensed spectrum. As per the regulation, LTE signals transmitted in unlicensed spectrum must be accompanied by a licensed carrier. Primary carrier always uses licensed spectrum (either TDD or FDD) for both control signals and user signals. The secondary carrier would use the unlicensed spectrum for downlink only or both uplink and downlink \cite{LAA1, LAA2, rupasinghe2014licensed, rupasinghe2015reinforcement,ferraos2015mobile}. 5~GHz unlicensed spectrum is a top candidate for the operation that is globally available while it should be ensured that LTE-U design is forward-compatible with upcoming 3GPP specifications \cite{LAA2}. A robust design solutions for unlicensed bands, clearer guidelines for operators, and coexistence framework would be necessary, if LTE and Wi-Fi have to coexist on fair share basis \cite{LAAEricsson, LAA2, LAA3, LAA4}.

LAA is an aggregation of LTE in the licensed and unlicensed spectrum. The main goal of LTE and LTE-Advanced for a licensed-assisted access to unlicensed spectrum would be to achieve better traffic offloading, achieve spectral efficiency, and provide a higher user experience in a cost-effective manner \cite{LAA_1, ULS}. The Wi-Fi performance is maintained while coexisting with licensed-assisted access to unlicensed spectrum as discussed in \cite{LAAEricsson,rupasinghe2014licensed,DusanUavPsc}. The LAA benefits to all user, by increasing the capacity to support the indoor applications, providing seamless indoor/outdoor mobility, better coverage experience, efficient spectrum usage, and better network performance as outlined by the technical report \cite{LAAEricsson}. However, operating on unlicensed spectrum may have design implications on LTE that are not yet well understood. In particular, co-existence between multiple LTE operators sharing the unlicensed band and co-existence with WLAN should be studied and addressed \cite{LAA_9, LAA_8}.

PSC examples discussed earlier under Section~\ref{mmWave} and Section~\ref{massiveMIMO} can benefit from licensed-assisted access to unlicensed spectrum. In Fig.~\ref{fig:SmallCellPSC}, the small cell operating in unlicensed spectrum can be deployed outdoor and indoor. Extending the benefits of small cells with LTE-U would lead to better network performance, enhanced user experience by supplementing the downlink, and provide seamless mobility \cite{LAASmallCell}. These benefits can enhance the EFR capabilities, provide much needed mission-critical communication, and improved data capacity to support real-time data/multimedia/voice \cite{LAAEricsson}. In an emergency scenario, as discussed in Section IX.C where all the terrestrial infrastructure has been devastated. LTE-U based UAV operation can be considered \cite{wolfefeasibility, gomez2013performance}. In \cite{wolfefeasibility}, the feasibility of using UAV in unlicensed for the purpose of search and rescue operations of avalanche victims has been discussed. Based on the simulation results, it can be deduced that UAV operations in unlicensed spectrum can be used by EFR during search and rescue missions.

\subsection{Spectrum sharing and cognitive radio}
\label{spectrumScarcity}
With the explosion of various wireless technologies such as 5G, Internet of Things, home automation, and connected cars. The finite radio spectrum, must accommodate cellular calls and data traffic from various wireless devices. As a result, the radio spectrum is quickly becoming a scarce commodity. For example, according to FCC, United States will be soon running out of radio spectrum \cite{SpectrumScarcityProb}.

As the spectrum gets increasingly scarce, public safety agencies have to compete with commercial entities for the radio spectrum resources. If not properly addressed, problem of spectrum scarcity can result PSN to be less dependable in the future, and may face interoperability challenges \cite{yuksel2013fostering, merwaday2014incentivizing}. In \cite{SpectrumScarcityAfter}, a group of experts from Australia and Europe provide opinions on recent research and policy options about spectrum management, including new approaches to valuation and sharing. Furthermore, effective detection and utilization of white spaces in PSC spectrum are crucial for minimizing outages and maximizing throughput \cite{akhtar2016white}. In \cite{guvencSpectrum2016}, pervasive spectrum sharing is envisioned as an enabler for PSNs with broadband communication capabilities.

Cognitive radio technology presents a viable solution to use spectrum more efficiently, which can potentially address the spectrum scarcity problem. A cognitive radio senses and learns the radio frequency spectrum in order to operate in unused portions of licensed spectrum or whitespaces without causing interference to primary licensed users \cite{SpectrumScarcity1,ghafoor2014cognitive, hossain2009dynamic}. Furthermore, a cognitive radio network offers support for heterogeneity, reconfigurability, self-organization, and interoperability with existing networks as discussed in \cite{ghafoor2014cognitive, hossain2007cognitive}. These ability of cognitive radio can be instrumental during disaster scenarios, with partially/fully destroyed networks. Moreover, the convergence of cognitive radio technology and device-to-device communication can improve the spectrum utilization \cite{sakr2015cognitive} and also benefit PSC by mitigating spectrum scarcity.

\subsection{Wireless Sensor Networks}
\label{subsection::SN}
A sensor network is composed of spatially distributed autonomous sensors with the ability to monitor and control physical environment. A typical sensor network has many sensor nodes and a gateway sensor node. Modern day sensor networks are usually bi-directional which cooperatively pass their data through the networks to the main location. Lately, sensor network received extensive interest for their use cases in PSN \cite{gomez2009secure, gomez2010secure, akan2009cognitive}.

A sensor network is composed of spatially distributed autonomous sensors with the ability to monitor and control physical environment. A typical sensor network has many sensor nodes and a gateway sensor node. Modern day sensor networks are usually bi-directional which cooperatively pass their data through the networks to the main location. Lately, sensor networks received extensive interest for their use cases in PSN \cite{gomez2009secure, gomez2010secure, akan2009cognitive}.

Wireless sensor networks (WSN) have the ability for situational awareness of a large physical environment, and this ability can support the decision makers with early detection of a possible catastrophic event \cite{gomez2009secure}. Furthermore, the localization algorithms can provide key support for many location-aware protocols and enhance the situational awareness \cite{hansurvey}. In a large-scale WSN deployment for situational awareness, the WSN is vulnerable to traffic patterns and node convergence as discussed in \cite{gu2015evolution}. Further analyzes present the sink mobility as a potential solution to enhance the network performance in large-scale WSN with varying traffic patterns.

A traditional multi-hop WSN is resources-constrained and uses fixed spectrum allocation. Therefore, by equipping a sensor node with cognitive radio capabilities, a WSN can reap the benefits of dynamic spectrum access as discussed in \cite{akan2009cognitive}. Furthermore, applying channel bonding schemes to cognitive-WSNs can result in better bandwidth utilization and higher channel capacity as surveyed in \cite{bukhari2015survey}.

The deployment of large-scale WSNs into PSN, alongside the LTE-based PSN, have important advantages. With sensor gateway node connected to LTE-based PSN, data can seamlessly flow into public safety data centers. This real-time data can raise the situational awareness of the public safety agencies and reduces the EFR response time.

However, the integration of WSN into IP-based network makes it vulnerable to various threats as discussed in Section~IX-K. The various security aspects of WSN in PSN are discussed in \cite{gomez2009secure, gomez2010secure}.

\subsection{Internet of Things (IoT)}
The IoT revolves around the pervasive presence of objects such as radio-frequency identification tags, sensors, actuators, and mobile phones having machine-to-machine (M2M) communication either through a virtual or instantaneous connection. The IoT infrastructure is built using cloud computing services and networks of data-gathering sensors. The IoT offers an ability to measure, infer, and understand the surrounding environment and delicate ecologies \cite{iot1,iot2,iot3}. Ubiquitous sensing enabled by the wireless sensor networks also cuts across PSC frontiers. The M2M and connected devices can collect real-time data from EFR devices, such as body-worn cameras, sensors attached to gun holsters, and magazine/baton/radio pouches \cite{iotps1}. This ecosystem of EFR's personal area network can send the real-time alert from these IoT devices to increase the situational awareness. With the availability of broadband services, such as LTE and WiFi, this real-time data can be uploaded and stored using cloud services. The real-time analysis of this data at crime centers using artificial intelligence and decision-making algorithms can be used to respond easily in any emergency situations \cite{iotPsSec}. Furthermore, measures can be taken to prevent a situation turning into an emergency \cite{iotps,iotps2,iot4}.

The IoT can change the way EFR execute their missions, helping them to be more effective and responsive. The real-time tracking of EFR using IoT, can relay the right information to the right person at the right time. Furthermore, this information will enhance real-time decision making capabilities of EFR and avoid any potential disasters. However, as these devices become more integrated into cloud services, they are exposed to vulnerabilities that pose risk to public safety mission-critical data. The important major challenge involves managing the complex threats created by the IoT, while taking advantage of newly discovered capabilities of IoT to enhance PSC \cite{iotPsSec1}.

\subsection{Cybersecurity Enhancements and Data Analytics for PSNs}
\label{PSNSecurity}
The public safety data available from public safety agents such as IoT, sensor nodes, and specialized public safety devices can play a crucial role in disaster management. With the help of intelligent learning techniques, the available data can be analyzed to understand the dynamics of disaster prone areas and plan the rescue operation. Moreover, the real-time data flowing in from different public safety agents can be analyzed to predict and issue an early warning in case of natural or man-made disaster. This data analysis can enhance the capabilities of PSN. The importance of data analytics and related case studies using big data have been surveyed in article \cite{wangbig}.

Furthermore, the public safety entities access public safety data such as medical records, site information, and other video and data information useful to disaster management and emergency response. Such critical information is transmitted in the form of voice, data, or multimedia using wireless laptops, handheld public safety devices, and mobile video cameras. However, this public safety data can be exploited by criminals with proper technology. Therefore, PSN infrastructure must protect the integrity of all the public safety data that is being accessed and transmitted. With extensive elevated network privileges given to EFR during an emergency scenario, having a reliable communication becomes of critical importance \cite{mcgee2012public, ghafghazi2014classification,EdwinUavSecurity}.

In \cite{mcgee2012public}, authors discuss various PSN security threats such as, protection against evesdropping, protection against corrupt information, securing network interfaces between network elements, protection against data exfiltration, and protection against denial of service attacks. Identity management of every public safety device and EFR accessing the critical information is an absolute necessity. As discussed in \cite{Hastings2015,mcgee2012public,ghafghazi2014classification} identity management can be characterized as identification, authentication, and authorization associated with individuals, public safety devices, or public safety applications. There are several proven security methods/techniques for identification, authentication, and authorization. However, the public safety agencies need to develop the technical and policy requirements for use of security techniques in PSN  by considering the scope and context of their applicability to PSC.

\section{Public safety broadband deployments in other regions of the world.}
\label{section:wrc15}
Broadband deployment for PSC has been grabbing the spotlight in different parts of the world. The global standards collaboration emergency communications task force is investigating standards for a globally coordinated approach for PSC. The world radio conference (WRC-15) held in Geneva in November 2015, provided a platform to address public protection and disaster relief  (PPDR) requirements. The WRC-15 \textit{Resolution 646} is an international agreement between UN and ITU. This agreement encourages the public safety entities to use frequency range 694-894 MHz, when outlining a draft for nationwide broadband PSN for PPDR applications \cite{WRC15}. The public safety entities across different countries have different operational frequency range and spectrum requirements.

\begin{table}[!htbp]
\centering
\caption{PDPR identified bands/ranges by international telecommunication union (ITU) \cite{ituFreq}.}
\begin{tabular}{|p{1.5cm}|p{2.70cm}|p{3.3cm}|}
\hline ITU Region 1 & Europe (including Russia and Middle East) and Africa  &  380-470 MHz (380-385/390-395 MHz)\\
\hline ITU Region 2 & Americas  &  746-806 MHz, 806-869 MHz, 4940-4990 MHz\\
\hline ITU Region 3 & Asia-Pacific  &  406.1-430 MHz, 440-470 MHz, 806-824/851-869 MHz, 4940-4990 MHz, and 5850-5925 MHz\\
\hline
\end{tabular}
\label{table:ituRegiontable}
\end{table}

Table \ref{table:ituRegiontable} shows the frequency allocation for PPDR in various parts of the world. The amount of spectrum needed for PSC by different EFR agencies and applications on a daily basis differs significantly. To enable spectrum harmonization for nationwide and cross-border operations would need interoperability between various EFR systems used for PPDR.  As a result, having a dedicated broadband spectrum for PPDR has been one of the WRC-15 agenda items \cite{wrcReport}.

\emph{Europe:} The \textit{ECC Report 199} \cite{EuBbPsc} addresses the broadband spectrum in Europe for PSC networks, which considers the 400 MHz and 700 MHz frequency bands. The report also concludes that 10 MHz of spectrum for the uplink and another 10 MHz for the downlink would provide sufficient capacity to meet the core requirements of PSC.

\emph{United Kingdom:} The public safety agencies in the United Kingdom use narrowband TETRA PSN for data and mission-critical voice services. Furthermore, the United Kingdom also plans on building emergency services mobile communications programme (ESMCP), also referred as the emergency service network (ESN), which will deliver future mobile communication for the country's emergency services using 4G LTE network \cite{ESMCP}. The United Kingdom is preparing for an orderly transition from TETRA to LTE-based ESN by the year 2020 \cite{UKEsn}. However, there is no spectrum available in 700 MHz at-least until 2019. Therefore, the ESN will latch onto 800 MHz to implement voice calls over LTE, PTT capabilities, and satellite backhaul in hard-to-reach areas \cite{UKESNCurrentStatus}.

\emph{Canada:} Industry Canada has initiated a framework and policy for public safety broadband spectrum in the bands 758-763 MHz and 788-793 MHz, i.e., band class 14 as illustrated in Fig.~\ref{fig:BandClass14Plan}. Canadian public safety band plan in upper 700 MHz is shown in \cite{CanadaBbPsc} and it corresponds to FirstNet's band class 14. Recently, the EFR in the City of Calgary, have been testing the first public safety LTE network in Canada \cite{CanadaFirstNet}. With the help of high-speed broadband capabilities Calgary Police Services were able to quickly access data, analyze the information, and securely multicast data to EFR. The Calgary public safety LTE network is composed of cellular sites and uses LTE public safety devices such as VML750 and LEX mission-critical handhelds \cite{CanadaFirstNet}.

\emph{Australia and China:} Similarly, in countries like Australia and China, LTE technology is fast becoming a primary choice for professional digital trunking command and dispatching services. The main goal is to build a rapid, flexible, and effective wireless multimedia communications network, which can be used to improve PDPR applications \cite{ituFreq,OtherBbNwk}.

\section{Issues and Open Research Directions}
\label{issuesFutureResearchDirections}
While the technologies for PSC are ever evolving, given the \textit{discovery-delivery} gap, not all advanced aspects of the public safety technology are commercially available. In particular, a full-fledged deployment of mission-critical LTE-based PSN is unattainable in the short term. This has resulted in the amalgamation of legacy and emerging public safety technologies. The convergence of LMRS-LTE public safety technologies can provide mission-critical voice and broadband data. However, designing and optimizing LMRS-LTE converged devices which can support mature 3GPP techniques such as MBMS is an open challenge for the engineers \cite{converge0,converge1}.

The integration and optimization of 3GPP Rel.~12/13 enhancements such as eMBMS, HetNets, LAA, and LTE-based V2X for PSC is an open challenge to both academic and engineering community. Currently, the MBSFN area and subframes in eMBMS are static and cannot be adjusted according to user distribution. Therefore, design and optimization of flexible MBSFN resource structures that can accommodate different user distributions have become a challenge and an important area of research. The integration of Rel. 13 enhancements for D2D/ProSe framework into MBMS/eMBMS is an another open area of research \cite{monserrat2012joint}. Small cells need better assessment of attributes such as grid configuration, alignment with macro network to avoid interference, handover analysis, and backhauling to meet the capacity requirements of cell-edge users. Furthermore, a systematic convergence of mmWave, massive MIMO, and UAVs with small cells are an open areas of research \cite{smallCell1,smallcell2,sui2015interference}. LTE-U and LAA coexisting fairly in the unlicensed spectrum with other broadband technologies has become an intense debate. Therefore, having policies to ensure fair access to all technologies is increasingly becoming a popular area of investigation. Furthermore, designing protocols for carrier aggregation of licensed and unlicensed bands is an another important area of research \cite{LAAEricsson,rupasinghe2014licensed}. The integration of ProSe and LTE-based broadcast services into LTE-based V2X can ensure connectivity between EFR vehicles, roadside infrastructure, and people in PSN. This open area of research can help to further strengthen the mobility management, public warning systems, and disaster management. However, resource allocation, latency, interference management and mobility management are important issues in V2X communication that need some study \cite{alazawi2011intelligent,eltoweissy2010towards}.

The mmWave, massive MIMO, and UAVs are advanced technologies  with potential applications for PSC. Each of these technologies have their own metrics such as throughput provided, latency, the cost of providing higher throughput, ease of deployment, and commercial viability for assessing success in PSN. By addressing challenges such as interference management, spatial reuse, and modulation and coding schemes for dynamically changing channel user states,  mmWave can result in more practical and affordable system for the development of cellular and PSN networks \cite{bleicher2013millimeter,mmWave2,sulyman2014radio,niu2015survey}. Similarly, efficient model of massive MIMO in PSN needs more accurate analysis of computational complexity, processing algorithms, synchronization of antenna units, and channel model \cite{mopidevi2011compact}. The application of UAVs in PSN is shrouded by public’s privacy concerns and lack of comprehensive policies, regulations, and governance for UAVs \cite{EdwinUavSecurity}. Investigating the role of mmWave and massive MIMO with UAVs for higher throughput gains for robust UAV deployment \cite{NadisankaUav}. Furthermore, developing new propagation model for better autonomy and quicker deployment of UAVS as aerial base stations is an open research direction \cite{merwaday2015uav,DronePS,gomez2013realistic,UAVintoPSLTE,wolfefeasibility,gomez2013performance, gomez2015capacity, WahabUavUWB,arvindUavPS,DusanUavPsc}.

SDR and cognitive radio technologies provide additional flexibility and efficiency for overall spectrum use. These technologies can be combined or deployed autonomously for PSC. However, PSN using SDR or cognitive radio technologies must operate in compliance with radio and spectrum regulations. Furthermore, a comprehensive implementation of these technologies for PSC present technical and operational challenges. An SDR for PSC, which is interoperable across vendor infrastructures, frameworks, and radio bands is an area for future research \cite{sdr5,sdrpsc1,sdrpsc}. Similarly, cognitive radio technology needs extensive research in areas of energy efficient spectrum sensing and sharing to be more effective. Extensive research over application of LAA,  database-assisted spectrum sharing, and prioritized spectrum access in cognitive radio technology can further enhance the spectrum efficiency \cite{ferraos2015mobile, rupasinghe2015reinforcement,rupasinghe2014licensed,DusanUavPsc}. 

Public safety wearables have become increasingly popular amongst EFR and are an important area of research. Tailoring public safety wearables and wireless sensors into IoT can help in real-time data aggregation and situation analysis. Therefore, WSNs and IoT needs to evolve within the wider context of PSN and address issues related to infrastructure, design, cost, interoperability, data aggregation, regulations, policies, and information security \cite{gomez2009secure,gomez2010secure,akan2009cognitive}. Furthermore constructing robust models for multi-hop synchronization and tethering real-time wireless sensor data attached to EFR equipment have become a contemporary area of research \cite{iotps,iotps2}. 

The EFR access public safety data such as medical records, site information, multimedia, and other information useful for disaster management and emergency response. The perpetrators with proper technology can exploit this public safety data. Therefore, given the scope of next generation PSN, securing PSN infrastructures, public safety device integrity, and public safety data is an important issue and open area of research within cybersecurity space \cite{ghafghazi2014classification, mcgee2012public, iotPsSec,EdwinUavSecurity}. Furthermore, the policy requirements for the use of public safety devices and information operating in PSN is an another major issue and area of research for the public safety agencies and federal regulatory bodies \cite{Hastings2015}.

\section{Concluding Remarks} \label{sec:conclusion}
\label{SectionXII}
In this paper, an overview of legacy and emerging public safety communication technologies is presented along with the spectrum allocation for public safety usage across all the frequency bands in the United States. The challenges involved with PSN operations are described, and the benefits of LTE-based PSN over LMRS are also discussed. Simulation results of LTE band class 14 and LMRS, convey that the LTE band class 14 devices experience significantly higher throughput when compared with LMRS subscriber units. These observed throughput values  can support the mission-critical broadband data services. A far-reaching cell coverage is one of the key factors in PSC and the simulation results show that the LMRS subscriber units experience better cell coverage and range when compared to LTE band class 14 UEs. In the upcoming years, LMRS is bound to stay as the primary PSC solution for mission-critical voice connectivity, alongside LTE providing the much need mission-critical real-time data. The technological advancement achieved through LTE, LTE-Advanced, and 5G will continue to enhance and transform the PSC capabilities in the future.

\section*{Appendix}
Definitions of the acronyms used in this survey article are listed in Table~\ref{Table:Abbreviation}.
\begin{table}[!htbp]
\centering
\caption{List of acronyms.}
\begin{tabular}{|c|p{6cm}|}
\hline 3GPP & 3rd generation partnership project \\
\hline C4FM & Continuous 4-level frequency modulation \\
\hline CAI & Common air interface \\
\hline CDF & Cumulative distribution function \\
\hline CN & Core network \\
\hline COLTS & Cells on light trucks \\
\hline COW & Cells on wheels \\
\hline CQI & Channel quality indicator \\
\hline DACA & Deployable aerial communication architecture \\
\hline DL & Downlink \\
\hline Earfcn & E-UTRA absolute radio frequency channel number \\
\hline EFR & Emergency first responders \\
\hline eNB & eNodeB \\
\hline EMS & Emergency medical services \\
\hline ESN & Emergency service network \\
\hline EPC & Evolved packet core \\
\hline ESMCP & Emergency services mobile communications programme \\
\hline FCC & Federal communication commissions \\
\hline FEC & Forward error correction \\
\hline GB & Guard bands \\
\hline HetNet & Heterogeneous network \\
\hline LAA & Licensed assisted access \\
\hline LMRS & Land mobile radio system \\
\hline LTE & Long term evolution \\
\hline LTE-U & LTE in unlicensed band \\
\hline MIMO & Multiple-input and multiple-output \\
\hline MME & Mobility management entity \\
\hline NPSBN & Nationwide public safety broadband network \\
\hline PPDR & Public protection and disaster relief \\
\hline ProSe & Proximity services \\
\hline PGW & Public data network gateway \\
\hline PSC & Public safety communications \\
\hline PSN & Public safety network \\
\hline RAN & Radio access network \\
\hline RB & Resource block \\
\hline RSRP & Reference signal received power \\
\hline RSRQ & Reference signal received quality \\
\hline SAE & System architecture evolution \\
\hline SDR & Software-defined radio \\
\hline SGW & Serving gateway \\
\hline SINR & Signal-to-interference-plus-noise ratio \\
\hline SOW & System on wheels \\
\hline UAV & Unmanned aerial vehicle \\
\hline UE & User equipment \\
\hline UL & Uplink \\
\hline VNS & Vehicle network system\\
\hline WSN & Wireless sensor network \\
\hline
\end{tabular}
\label{Table:Abbreviation}
\end{table}

\bibliographystyle{IEEEtran}
\bibliography{Citations}

\begin{IEEEbiography}{Abhaykumar Kumbhar}
received his B.E in Electronics and Communication Engineering from Visvesvaraya Technological University, India and Master degree in Computer Science from Villanova University.
Currently, he is pursuing his Ph.D degree in Electrical Engineering from Florida International University. He joined Motorola Solutions in 2012 as Senior Firmware Engineer and since then he has been developing embedded software for public safety devices.
\end{IEEEbiography}

\begin{IEEEbiography}{Farshad Koohifar}
received his B.S. in Electrical Engineering, Iran University of Science and Technology and M.Sc. in Electrical Engineering, Sharif University of Technology. Currently, he is pursuing his Ph.D degree in Electrical Engineering from Florida International University.
\end{IEEEbiography}

\begin{IEEEbiography}{{I}smail~G\"uven\c{c}}
(senior member, IEEE) received his Ph.D. degree in electrical engineering from University of South Florida in 2006, with an outstanding dissertation award. He was with Mitsubishi Electric Research Labs during 2005, and with DOCOMO Innovations Inc. between 2006-2012, working as a research engineer. Since August 2012, he has been an assistant professor with Florida International University.
His recent research interests include heterogeneous wireless networks and future radio access beyond 4G wireless systems. He has published more than 100 conference/journal papers and book chapters, and several standardization contributions. He co-authored/co-edited three books for Cambridge University Press, served as an editor for IEEE Communications Letters (2010-2015) and IEEE Wireless Communications Letters (2011-present), and as a guest editor for several other journals. Dr. Guvenc is an inventor/coinventor in 23 U.S. patents, and has another 4 pending U.S. patent applications. He is a recipient of the 2014 Ralph E. Powe Junior Faculty Enhancement Award and 2015 NSF CAREER Award.
\end{IEEEbiography}

\begin{IEEEbiography}{Bruce Mueller}
received his Bachelor degree in Electrical Engineering from Rose-Hulman Institute of Technology and Master degree in Electrical Engineering from the University of Michigan. He joined Motorola in year 1989 and since then he was worked several broadband wireless systems and technologies. Currently, he is the research director for Government and Public Safety Business, Motorola Solutions developing the wireless and research capability.
\end{IEEEbiography}

\end{document}